\documentclass[letter]{article}

\usepackage[english]{babel}
\usepackage[utf8]{inputenc}
\usepackage[T1]{fontenc}

\usepackage[letterpaper,top=3cm,bottom=2cm,left=3cm,right=3cm,marginparwidth=1.75cm]{geometry}

\usepackage{amsmath}
\usepackage{graphicx}

\usepackage[colorlinks=true, allcolors=blue]{hyperref}
\usepackage{stackengine}
\usepackage{physics}
\usepackage{amssymb} 
\usepackage{mathrsfs}
\usepackage{amsfonts}
\usepackage{amssymb}
\usepackage[useregional]{datetime2}
\usepackage[left]{lineno}
\usepackage{longtable}
\usepackage{booktabs}
\usepackage[framemethod=TikZ]{mdframed}
\usepackage{tcolorbox}
\tcbuselibrary{breakable}
\usepackage{csquotes}
\usepackage{pdflscape}
\usepackage{tabularx}
\usepackage{longtable}
\usepackage{makecell}
\usepackage{adjustbox}
\usepackage{array}
\usepackage{csquotes}
\usepackage{blindtext}
\usepackage[affil-it]{authblk}

\usepackage[backend=biber, maxcitenames=2, uniquename=false, style=apa, sorting=nyt,doi=false,isbn=false,url=false, natbib=true]{biblatex}

\DeclareNameFormat{labelname:poss}{% Based on labelname from biblatex.def
  \nameparts{#1}% Not needed if using Biblatex 3.4
  \ifcase\value{uniquename}%
    \usebibmacro{name:family}{\namepartfamily}{\namepartgiven}{\namepartprefix}{\namepartsuffix}%
  \or
    \ifuseprefix
      {\usebibmacro{name:first-last}{\namepartfamily}{\namepartgiveni}{\namepartprefix}{\namepartsuffixi}}
      {\usebibmacro{name:first-last}{\namepartfamily}{\namepartgiveni}{\namepartprefixi}{\namepartsuffixi}}%
  \or
    \usebibmacro{name:first-last}{\namepartfamily}{\namepartgiven}{\namepartprefix}{\namepartsuffix}%
  \fi
  \usebibmacro{name:andothers}%
  \ifnumequal{\value{listcount}}{\value{liststop}}{'s}{}}
\DeclareFieldFormat{shorthand:poss}{%
  \ifnameundef{labelname}{#1's}{#1}}
\DeclareFieldFormat{citetitle:poss}{\mkbibemph{#1}'s}
\DeclareFieldFormat{label:poss}{#1's}
\newrobustcmd*{\posscitealias}{%
  \AtNextCite{%
    \DeclareNameAlias{labelname}{labelname:poss}%
    \DeclareFieldAlias{shorthand}{shorthand:poss}%
    \DeclareFieldAlias{citetitle}{citetitle:poss}%
    \DeclareFieldAlias{label}{label:poss}}}
\newrobustcmd*{\posscite}{%
  \posscitealias%
  \textcite}
\newrobustcmd*{\Posscite}{\bibsentence\posscite}
\newrobustcmd*{\posscites}{%
  \posscitealias%
  \textcites}
 
\newcommand{\E}[2]{\mathbb{E}_{#1} \negmedspace \left[ #2 \right]}
\newcommand{\Cov}[3]{\mathrm{Cov}_{#1} \negmedspace \left( #2, #3 \right)}
\newcommand{\Var}[2]{\mathrm{Var}_{#1} \negmedspace \left( #2\right)}
\renewcommand{\eqref}[1]{Eq.\ref{#1}}
\title{Resolving conceptual issues in Modern Coexistence Theory}
\date{\today}
\author[1,2,*]{Evan C. Johnson}
\author[1]{Alan Hastings}

\affil[1,]{Department of Environmental Science and Policy; University of California Davis; Davis, California 95616 USA}
\affil[2,]{Center for Population Biology; University of California Davis; Davis, California 95616 USA}
\affil[*]{Corresponding author: Evan Johnson, evcjohnson@ucdavis.edu}

\begin{document}
\maketitle
\clearpage

\section*{Abstract} 

In this paper, we discuss the conceptual underpinnings of Modern Coexistence Theory (MCT), a quantitative framework for understanding ecological coexistence. In order to use MCT to infer how species are coexisting, one must relate a complex model (which simulates coexistence in the real world) to simple models in which previously proposed explanations for coexistence have been codified.  This can be accomplished in three steps: 1) relating the construct of coexistence to invasion growth rates, 2) mathematically partitioning the invasion growth rates into coexistence mechanisms (i.e., classes of explanations for coexistence), and 3) relating coexistence mechanisms to simple explanations for coexistence. Previous research has primarily focused on step 2. Here, we discuss the other crucial steps and their implications for inferring the mechanisms of coexistence in real communities. 

Our discussion of step 3 --- relating coexistence mechanisms to simple explanations for coexistence --- serves a heuristic guide for hypothesizing about the causes of coexistence in new models; but also addresses misconceptions about coexistence mechanisms. For example, the storage effect has little to do with bet-hedging or "storage" via a robust life-history stage; relative nonlinearity is more likely to promote coexistence than originally thought; and fitness-density covariance is an amalgam of a large number of previously proposed explanations for coexistence (e.g., the competition--colonization trade-off, heteromyopia, spatially-varying resource supply ratios). Additionally, we review a number of topics in MCT, including the role of "scaling factors"; whether coexistence mechanisms are approximations; whether the magnitude or sign of invasion growth rates matters more; whether Hutchinson solved the paradox of the plankton; the scale-dependence of coexistence mechanisms; and much more.
\newline 
\newline
Keywords: modern coexistence theory, environmental variation, spatiotemporal, stabilizing mechanisms, invasion growth rate, storage effect, relative nonlinearity, fitness-density covariance, coexistence mechanisms, scale-dependence 

\tableofcontents

\newpage

\section{Introduction}
\label{Introduction}

The competitive exclusion principle (\cite{volterra1926variationsST}, \cite{Lotka1932TheSupply}, \cite{GauzeG.F.GeorgiĭFrant︠s︡evich1934TsfeST}; \cite{levin1970community}) states that no more than $L$ species can coexist by partitioning $L$ resources. Taking the competitive exclusion principle to heart, G.E. Hutchinson published "The paradox of the plankton" (\citeyear{hutchinson1961paradox}), which asked how dozens of phytoplankton could coexist in a single lake, despite there being only a handful of limiting nutrients. The well-mixed, homogeneous nature of the epilimnion (the top-most thermal stratum of a lake, where most phytoplankton reside) meant that the paradox of the plankton could not be resolved with appeals to spatial variation, which had long been thought (e.g., \cite{grinnell1904originST}) to permit coexistence. Instead, resolutions to the paradox featured temporal variation. Note that throughout this paper, we use 'variation' and 'fluctuations' interchangeably.

One resolution to the paradox of the plankton is \textit{relative nonlinearity}, an explanation for coexistence in which species specialize on resource variance (\cite{levins1979coexistenceST}; \cite{tilman1982resourceST}) rather than mean resource levels. The requisite temporal variation can be generated exogenously --- for instance, via lake turnover or seasonal run-off (\cite[p.~220-223]{WetzelRobertG2001L:la}) --- or endogenously, via  resource-consumer dynamics (\cite{armstrong1976coexistence}, \citeyear{armstrong1980competitive}). Another resolution to the paradox is \textit{the storage effect}, an explanation for coexistence in which species specialize on different aspects of a variable environment (\cite{chesson1981environmentalST}). In the jargon of community ecology, both relative nonlinearity and the storage effect are known as \textit{fluctuation-dependent coexistence mechanisms}.

Relative nonlinearity can theoretically support a number of species that grows quadratically with the number of discrete resources (\cite[p.~253]{Chesson1994}), and the storage effect can theoretically support an arbitrary number of species (\cite[p.~259]{Chesson1994}). Therefore, fluctuation-dependent coexistence mechanisms resolved the paradox of the plankton, but imposed additional problems. Ecology has explained how species can coexist, but how do species actually coexist? If there are no theoretical limits to biodiversity, why are the number of coexisting species that which we observe? Answering these questions will require a quantitative framework that is capable of measuring the relative importance of different explanations for coexistence.

Community ecologists have put forward many explanations for coexistence, the most prominent of which are are specialized natural enemies (\cite{nicholson1937role}; \cite{holt1977predationST}; \cite{holt1994simpleST}; \cite{holt1994ecological}), a trade-off between competition and colonization (\cite{levins1971regional}; \cite{Sousa1979}; \cite{hastings1980disturbance}; \cite{tilman1994competition}), the Janzen-Connell Hypothesis (\cite{janzen1970herbivores}; \cite{connell1971role}; \cite{stump2015distance}); the partitioning of resources across space (\cite{macarthur1958population};\cite{hutchinson1961paradox}; \cite{tilman1982resourceST}; \cite{holt1984spatial}); opportunist-gleaner tradeoffs (\cite{fredrickson1981microbial}); seasonal variation in resource supply (\cite{stewart1973partitioning}; \cite{grover1997resource}) or endogenously cyclical resource-consumer dynamics (\cite{armstrong1976coexistence}; \citeyear{armstrong1980competitive}), temporal partitioning of the environment (\cite{loreau1989coexistence}; \cite{loreau1992time}; \cite{klausmeier2010successional}), the storage effect (\cite{chesson1981environmentalST}, \cite{chesson2003quantifying}), and neutral theory (cite: \cite{caswell1976community}; \cite{hubbell2001unified}, \cite{kalyuzhny2015neutral}). Each explanation, having emerged from simple models (e.g., \cite{levins1971regional}), experiments (e.g., \cite{paine1966food}), or curious patterns in field data (e.g., \cite{janzen1970herbivores}), are certainly \textit{partial explanations}. Real ecological communities are complex, and many of the above phenomena may be operating at once. Spectacularly, each of these partial explanations can be grouped into natural categories called \textit{coexistence mechanisms} and assigned a measure of relative importance. \textit{Modern Coexistence Theory} (MCT) is the framework that makes this possible.

MCT has been widely successful. It has been the basis of important conceptual and theoretical advances (e.g., \cite{chesson1997roles}; \cite{stump2015distance}; \cite{snyder2003local}; \cite{chesson2008interaction}; \cite{chesson2010storage}; \cite{schreiber2021positively}), and several attempts to infer the mechanisms of coexistence in real communities (\cite{caceres1997temporal}; \cite{venable1993diversity}; \cite{pake1995coexistence}; \cite{pake1996seed}; \cite{adler2006climate}; \cite{sears2007new}; \cite{descamps2005stable}; \cite{facelli2005differences}; \cite{angert2009functional}; \cite{Adler2010}; \cite{usinowicz2012coexistence}; \cite{chesson2012storage}; \cite{chu2015large}; \cite{usinowicz2017temporal}; \cite{ignace2018role}; \cite{hallett2019rainfall}; \cite{armitage2019negative}; \cite{armitage2020coexistence}; \cite{zepeda2019fluctuation}; \cite{zepeda2019fluctuation}; \cite{towers2020requirements}; \cite{holt2014variation}; \cite{ellner2016quantify}) or laboratory microcosms (\cite{Jiang2007}; \cite{letten2018species}). Additionally, MCT unifies seemingly dissimilar explanations for coexistence through categorization into coexistence mechanisms, thus organizing a scattered literature and highlighting similarities, such as the symmetrical role (with regards to coexistence) of resource specialization and specialist predators (\cite{chesson2008interaction}).

\begin{landscape}
\begin{figure}[h]
      \includegraphics[scale = 0.68]{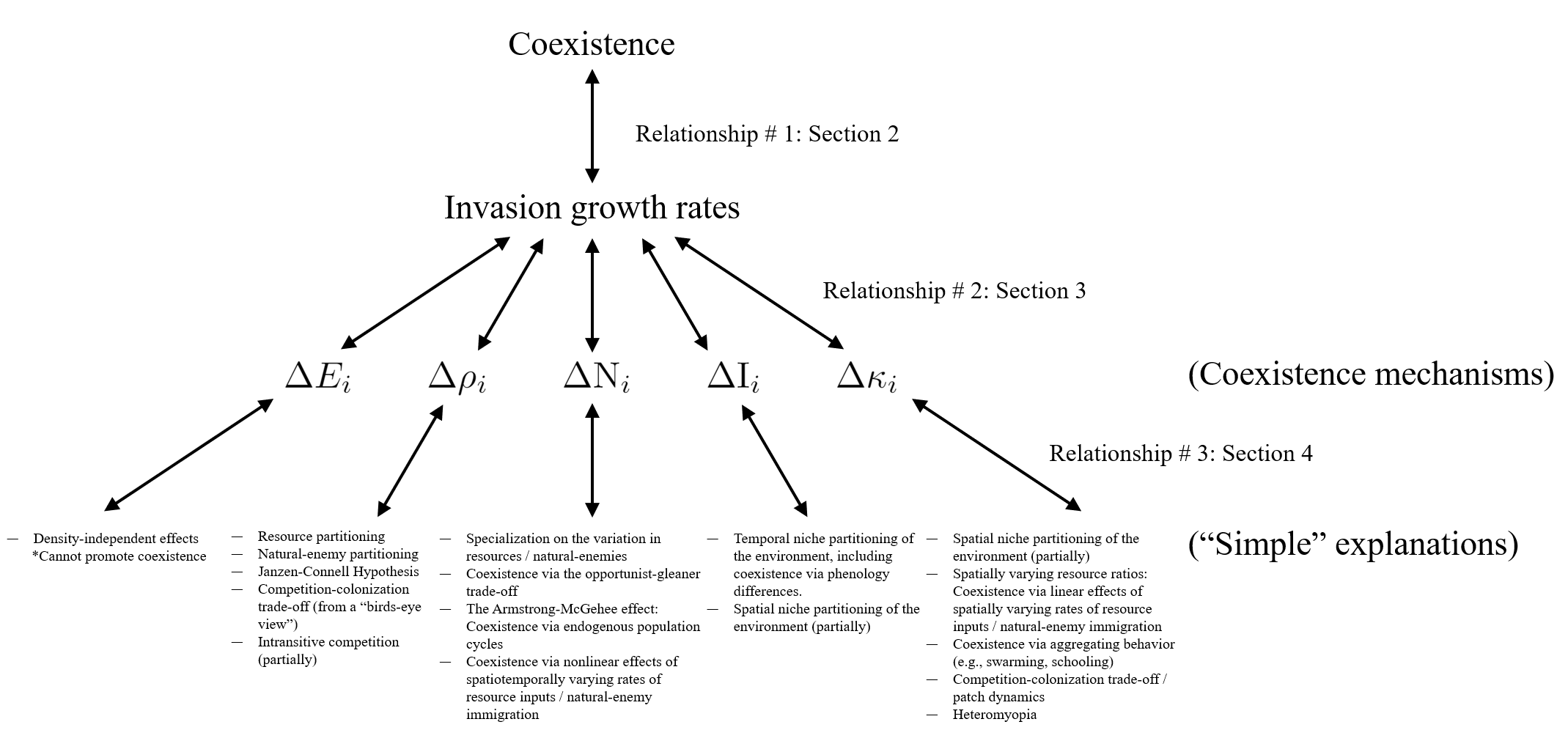}
  \caption{The goal of Spatiotemporal Modern Coexistence Theory: connecting \textit{actual coexistence} to \textit{simple explanations for coexistence}}
  \label{fig:edifice}
\end{figure}
\end{landscape}

The main innovation of MCT is the partitioning of invasion growth rates into coexistence mechanisms. However, we perceive of MCT as a broad edifice (Fig. \ref{fig:edifice}) that relates real coexistence (i.e., coexistence in real ecological communities) to simple yet incomplete explanations for coexistence (i.e., simple models in which coexistence has been demonstrated). The edifice is composed of three kinds of relationships, each corresponding to a level of arrows in Figure 1: 1) the relationship between coexistence and invasion growth rates 
(Section \ref{sec:The relationship between coexistence and invasion growth rates}), 2) the relationships between the invasion growth rate and coexistence mechanisms (Section \ref{sec:Spatiotemporal coexistence mechanisms}), and 3) the relationship between coexistence mechanisms and simple explanations for coexistence (Section \ref{sec:Interpreting coexistence mechanisms}).

\newpage

\newpage

\section{The relationship between coexistence and invasion growth rates}
\label{sec:The relationship between coexistence and invasion growth rates}

\subsection{Overview}
\label{sec:The relationship between coexistence and invasion growth rates:Overview}

The focal object of MCT is the invasion growth rate, the long-term per capita growth rate of a rare species ("the invader") where the remaining species ("the residents") have attained their limiting dynamics. But how do invasion growth rates relate to coexistence? Here is the basic idea: 

For a species to persist, it must be able to increase when it becomes rare; some degree of rarity is inevitable, as environmental fluctuations will eventually draw populations away from their typical abundances. If all species in a community can recover from rarity, then all species can be said to coexist with one another. An invasion growth rate measures the tendency of a rare species to increase, so following intuition, one might claim that coexistence occurs if each species in a community has a positive invasion growth rate in the sub-community of $S-1$ resident species (where $S$ is the total number of species). This is known the \textit{mutual invasibility criterion for coexistence} (\cite{turelli1978reexamination}; \cite{chesson2000general}; \cite{Chesson1989}; \cite{Grainger2019TheResearch}).

The mutual invasibility criterion is intuitive but imperfect. The upshot is that the mutual invasibility criterion is neither necessary nor sufficient for coexistence, but that invasion growth rates can be used heuristically, or through an alternative coexistence criterion. We first turn to a brief discussion of invasion growth rates. 

\subsection{Invasion growth rates}
\label{sec:The relationship between coexistence and invasion growth rates:Invasion growth rates}
The 'long-term per capita growth rate of a rare species' can be more precisely understood as the dominant Lyapunov exponent (\cite{Metz1992}) or the dominant stochastic Lyapunov exponent (\cite{crutchfield1982fluctuations}; \cite{Dennis2003}) of a dynamical system that represents an ecological community. The lypaunov exponent is the fundamental mathematical object which governs invasion: the geometric mean of growth multipliers in scalar populations (\cite{Lewontin1969}; \cite{Stearns2000}), Floquet multipliers in a periodic environments (\cite{Klausmeier2008}), and eigenvalues in stage-structured models (\cite{CaswellHal2001Mpm:}) are all just special cases of the dominant Lyapunov exponent.

Implicit in the calculation of invasion growth rates is the assumption that populations have an infinite number of individuals. An invader with an infinite number of individuals can lose an arbitrary number of individuals without going extinct. Because the invader cannot go extinct, it will eventually experience a perfectly representative collection of environmental states and resident densities. This has several effects: 1) the initial conditions of the invader's environment are washed-out over time; 2) the invader's environment (which includes resident species) has plenty of time to "settle down" to its limiting dynamics; and 3) when the invader's environment is stationary (\cite[p.~236]{Chesson1994}; \cite{Nisbet1982Modelling}), the invasion growth rate can be computed by integrating per capita growth rates across the stationary distribution of environmental states.

For the invader, an infinite-population model will approximate a finite-population when the invader's density is low enough to not affect any species' per capita growth rate, but high enough that demographic stochasticity (and thus stochastic extinction) can be ignored. In such a scenario, the sign of the invasion growth rate would perfectly predict whether or not the invader will recover in the short-term (\cite{jansen1998shaken}).  What happens in the very long term cannot be determined with a single invasion growth rate; For instance, an initially successful invader may by eventually excluded by the residents, a scenario that has been called "the resident strikes back" (\cite{Mylius2001TheAttractor}; \cite{Geritz2002}; or see Fig. \ref{fig:phase} c). Species at extremely low density (e.g., extirpated then reintroduced) are affected by demographic stochasticity and are subject to stochastic extinction. Even so, for these species a positive invasion growth rate is necessary for a non-zero probability of invasion (\cite{Schreiber2011}).

The invasion growth rate is always calculated in the context of the \textit{limiting dynamics of the invader's environment}, which typically includes internal variables (e.g., resident densities, predator densities, resource concentrations) and external variables (e.g., temperature, disturbances). "The limiting dynamics" can be more precisely understood as a unique, long-term (asymptotic) joint frequency-distribution of such variables. Therefore, invasion growth rates cannot be measured in models where resident sub-communities have alternative state states (the invasion growth rate is not unique) and models with strong unidirectional environmental change (the frequency distribution does not converge). 

However, MCT can still be used when unidirectional environmental change is considerably slower than demographic change. For example, even though lake communities are affected by climate change, the time-scale of phytoplankton invasion (likely no more than a couple years, considering fast generation times) is short compared the decadal time-scale on which climate-change has appreciable effects on phytoplankton dynamics (\cite{izmest2011long}). Therefore, it may be reasonable to not incorporate climate change projections into one's model of phytoplankton dynamics, with the understanding that the validity of one's inferences regarding coexistence only extends so far into the future. 

One ought to be wary of calculating invasion growth rates over long time-scales, especially when environmental change is much slower than demographic change. For example, the invasion growth rate of a maple tree species may converge after 500,000 years, perhaps after the effects of anthropogenic climate change have been attenuated by several Milankovitch cycles. However, our models of contemporary population dynamics will certainly be poor representations of population dynamics in the far future. In addition, the infinite population assumption breaks-down: In theory, long periods of unfavorable conditions can be offset by sufficiently favorable conditions; in reality, long periods of unfavorable conditions lead to extinction. 

Finally, the invasion growth rate is measured when the "internal structure" of the invader's population has reached a quasi-steady state. For age/stage structured population, the relevant mathematical object is the \textit{stable-age distribution} (\cite{CaswellHal2001Mpm:}). For spatially-structured populations, the relevant mathematical object is the quasi-steady spatial distribution (\cite{chesson2000general}; \cite{ferriere2001invasion}; \cite{Stump2018}). Ones hope that the internal structure of the invader's population evolves much faster than invader's density, such that the invasion growth rate can be uniquely determined while the invader is still rare. In concrete models, this assumption can be checked with simulation data: plot summary statistics of population structure (e.g., pair correlations, the fraction of juveniles) against total density (for an example, see \cite{le2003adaptive}, Fig. 7). More generally, there are a number of tricks that can aid in the computation of invasion growth rates in particular classes of models; for more information, see \cite{johnson2022coexistence}, Section 3.

\subsection{Inferring coexistence from invasion growth rates: The mutual invasibility criterion}
\label{sec:The relationship between coexistence and invasion growth rates:Inferring coexistence from invasion growth rates: The mutual invasibility criterion}

\begin{figure}[h] 
      \includegraphics[scale = 0.45]{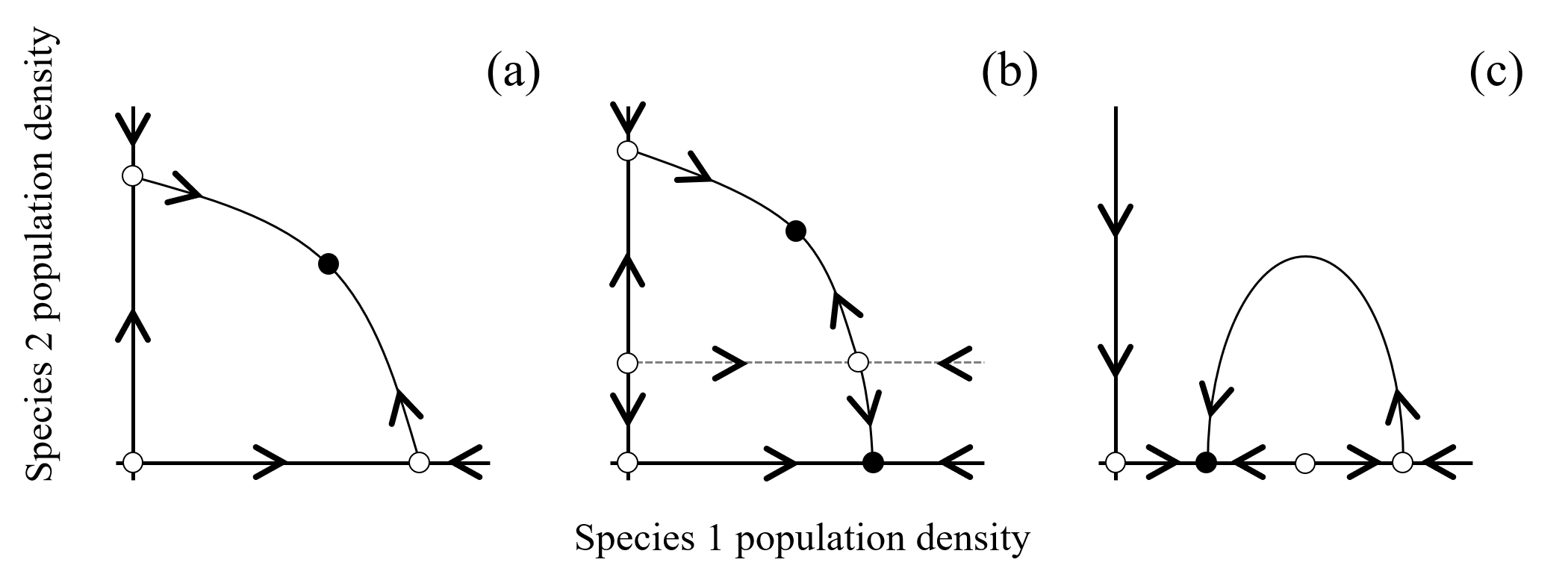}
  \caption{Mutual invasibility in phase space. \textit{(a)} Mutual invasibility is sometimes indicative of coexistence. \textit{(b)} Mutual invasibility is not necessary for coexistence in the case of a strong allee effect. \textit{(c)} Mutual invasibility is not sufficient for coexistence in this case because of a phenomenon known as the "resident strikes back". Closed circles and open circles are stable and unstable equilibria, respectively. Lines and arrows in phase space denote stable and unstable manifolds. The dotted line in panel \textit{b} denotes species 2's allee threshold.}
  \label{fig:phase}
\end{figure} 

\newpage

\newcolumntype{R}[2]{%
    >{\adjustbox{angle=#1,lap=\width-(#2)}\bgroup}%
    l%
    <{\egroup}%
}
\newcommand*\rot{\multicolumn{1}{R{30}{1em}}}% no optional argument here, please!
\begin{landscape}
\begin{table}
%\centering
\caption{\label{tab:co_def} \textbf{The properties of various coexistence criteria}.  \textit{Deterministic skeleton needed} = \textit{Yes} means that the criterion only works if we assume small perturbations from deterministic dynamics, and therefore, that one must either start with a deterministic model or extract a \textit{deterministic skeleton} from a stochastic model (\cite{Coulson2004}). Columns 6--9 show whether a criterion is \textit{Necessary} and/or \textit{Sufficient} for coexistence, operationalized here as each species spending very little time at low density in a single instantiation of a stochastic model (as in  \cite{Schreiber2011}, Theorem 1). \textit{Small, frequent perturbations} is in reference to a mathematical construct: arbitrarily small perturbations to population density, which transform population-dynamical trajectories into \textit{$\epsilon$-chains} (\cite{hofbauer1980dynamical}). \textit{Large, infrequent perturbations}  may place population densities arbitrarily close to (but not exactly at) zero density. \textit{Robust} = yes means that if the criterion is satisfied, then it will also be satisfied for small $C^1$ perturbations (\cite[p.~38]{GuckenheimerJohn1983Nods}) to the vector field of the model (or the model's 'deterministic skeleton'). This table does not include various types of \textit{structural stability} (e.g., \cite{meszena2006competitive}; \cite{saavedra2017structural}; \cite{Song2020})}

\newcolumntype{L}[1]{>{\raggedright\let\newline\\\arraybackslash\hspace{0pt}}m{#1}}

\begin{tabularx}{2.5\textwidth}{ L{0.15\linewidth} L{0.32\linewidth}  l l l l l l l l X}
Criterion &
Description &
\rot{Designed to work with complex dynamics} &
\rot{Designed to work with large perturbations} &
\rot{Deterministic skeleton needed} &
\rot{Necessary in the face of small, frequent perturbations} &
\rot{Sufficient in the face of small, frequent perturbations} &
\rot{Necessary in the face of large, infrequent perturbations} &
\rot{Sufficient in the face of large, infrequent perturbations} &
\rot{Robust} &
relevant references \\
\toprule
Feasibility & There is at least one equilibrium point (not necessarily stable) where all species have positive density. & No & No & Yes & No & No & Yes & No & No & \cite{Grilli2017}; \newline
\cite{saavedra2017structural} \\
\midrule 
Positive, stable equilibrium & There is a feasible equilibrium which is also asymptotically, Lyapunov stable. & No & No & Yes & No & Yes & No & No & Yes & \cite{May1974}; \newline \cite{hastings1988food}; \newline \cite{Logofet2005}  \\
\midrule
Mutual invasibility & In an $S$-species community, each species has a positive invasion growth rate in the sub-community defined by the limiting dynamics of $S-1$ resident species. & Yes & Yes & No & No & Yes & No & Yes & No & \cite{turelli1981niche}; \newline \cite{Chesson1989}; \newline \cite{Siepielski2010} \\
\midrule
Positive attractor & There is an attracting set that only includes positive species densities. & Yes & No & Yes & Yes & Yes & Yes & No & Yes & \cite{Schreiber2006}; \newline \cite{armstrong1980competitive}  \\
\midrule
Hofbauer's permanence criterion & Weighted averages of invasion growth rates must be positive; see \eqref{SPT} in the main text. &  Yes & Yes & No & No & Yes & No & Yes & No & \cite{schuster1979dynamical}; \newline \cite{jansen1998shaken}; \newline \cite{Schreiber2000}; \newline \cite{Benaim2019} \\
\bottomrule
\end{tabularx}
\label{tab:coexistence criteria}
\end{table}
%\end{table}%
\end{landscape}

The \textit{mutual invasibility criterion} claims that coexistence occurs if each species in a community has a positive invasion growth rate in the sub-community of the remaining $S-1$ resident species. As intuitive as this sounds, it is not generally true. Even if we ignore the complications set forth in the previous section (e.g., a sub-community of residents may have alternative stable states, the invader is subject to stochastic extinctions in finite-population models) mutual invasibility is neither a necessary nor sufficient condition for coexistence (Fig. \ref{tab:coexistence criteria}).

To see why mutual invasibility is not necessary, consider a single-species community where the  solitary species experiences a strong allee effect. This species can persist under normal conditions, but will not be able to recover if perturbed to low density.  Another example, which proceeds with much the same reasoning, is a two-species community in which both species are obligate mutualists. The same basic idea caries over to more complex settings: notably, intransitive competition in purely competitive communities can generate emergent allee effects (\cite{barabas2018chesson}).

To see why mutual invasibility is not sufficient for coexistence, consider the case of "the resident strikes back" (Fig. \ref{fig:phase}c;  \cite{Mylius2001TheAttractor}; \cite{Geritz2002}; Fig. \ref{fig:phase}c), wherein the invader is initially successful but eventually excluded as the resident sub-community relaxes to a alternative stable state. This phenomenon arises in a variety of models (\cite{case1995surprising}; \cite{doebeli1998invasion}; \cite{parvinen1999evolution}; \cite{sachdeva2017divergence}). 

Another problem with the mutual invasibility criterion can be demonstrated with the classic Rock-Paper-Scissors model (\cite{gilpin1975limit}; \cite{May1975}). When the rock species is perturbed to low density, 'scissors beats paper', leaving a community with only the scissors species. The rock species can invade this community, but since 'rock beats scissors', the end result is a community with only the rock species. The mutual invasibility criterion cannot be used here, because the $S-1$ resident species cannot persist when the invader is perturbed to low density. The criterion is uninformative in this context because the complete community may or may not coexist, despite the fact that the invasion growth rates in all stable sub-communities (i.e., those that contain a single resident) are positive.

The mutual invasibility criterion works well in a narrow set of cases. It is a sufficient condition for coexistence in two-species communities where per capita growth rates are decreasing functions of population size (this functionally excludes allee effects, mutualisms, and alternative stable states; see \cite{Chesson1989}), and in $S$-species communities in which all $S-1$ subsets of species are able to coexist (\cite{Case2000}). The mutual invasibility criterion works well for 2-species communities because an invasion analysis in a two-species system produces a single resident, and it is natural for a solitary species to persist in the absence of any of its competitors. By contrast, an invasion analysis in a three-species system produces two residents whose coexistence cannot be taken for granted. Speciose communities constructed with random parameters typically have some sub-communities of $S-1$ residents that can not coexist (\cite{case1990invasion}; \cite{Capitan2017}; \cite{servan2018coexistence}), but certain special arrangements of parameters, notably \textit{diffuse competition} (\cite{Stump2017}) and \textit{Volterra dissipative} (\cite{volterra1937principes}; \cite{LogofetD.O.DmitriĭOlegovich1993Mag:}) arrangements, will allow for the coexistence of all sub-communities.

\subsection{Alternative coexistence criteria}
\label{sec:The relationship between coexistence and invasion growth rates:Alternative coexistence criteria}

Historically, many coexistence criteria have been use, each with strengths and weaknesses (Table \ref{tab:coexistence criteria}). Given the shortcomings of the mutual invasibility criterion, why should should we not consign it to the dustbin of history? There are at least two answers. 1) Much of theoretical ecology aims to develop understanding via heuristics: statements that are evocative yet not universally true. For example, the statement "coexistence occurs when intraspecific competition is greater than interspecific competition" is only generally true in two-species Lotka-Volterra models (\cite{May1975}; \cite{saavedra2016nested}) but may nevertheless predict coexistence in reality (\cite{Godoy2017},  \cite{Friedman2017CommunityMicrocosms}). Even if such heuristics are not useful in a pragmatic sense, they seem to clarify ecological concepts in a way that is satisfying to many ecologists. Thus, at the very least, the applied value of the mutual invasibility criterion does not bear on its ability to produce insights and publications. 2) Alternative coexistence criteria are even worse. Historically, theoretical work has focused on the local stability of a positive equilibrium. This criterion is simultaneously too strong, because it automatically excludes communities with complex (e.g., cyclical or chaotic) dynamics, and too weak, because it only considers the consequences of perturbations that are infinitesimally small (\cite{hastings1988food}). In empirical work, coexistence has been most often conflated with co-occurrence (i.e., species occupying the same habitat) and a supplementary just-so story about life-history differences (\cite{Siepielski2010}; \cite{schoener1982controversy}). Co-occurrence is not an appropriate criterion for coexistence, since an observed population may be supported by a population located elsewhere via migration (i.e., source-sink dynamics), or may be in the process of being competitively excluded. 

Unbeknownst to many theoretical ecologists, mathematical biologists have been hard at work, developing a coexistence theory that can handle large perturbations and intransitive rock-paper-scissor type dynamics (\cite{schuster1979dynamical}; \cite{hofbauer1981general}; \cite{Schreiber2000}; \cite{garay2003robust}; \cite{Schreiber2011}; \cite{Schreiber2012}; \cite{Roth2014}; \cite{hening2018coexistence}; \cite{Benaim2019}). Of particular interest is a notion of global stability called  \textit{permanence} or \textit{uniform persistence}: a tendency for all species' densities to be bounded from above and below, uniformly in initial conditions. \textcite{hofbauer1981general} developed a sufficient criteria for permanence, which is satisfied by picking weights $p_j$ (i.e., positive constants that sum to one) such that the weighted sum of $\E{t}{r_j}$ (the invasion growth rate), is positive for each and every sub-community $\mu$ in which one or more species is missing. Or, in mathematical form,

\begin{equation} \label{SPT}
    \sum \limits_j p_j r_j(\mu) > 0, \quad \text{for all} \; \mu.
\end{equation}

The Hofbauer criterion has been refined and extended for several different kinds of models (\cite{Schreiber2000}; \cite{Roth2014}; \cite{Schreiber2011}; \cite{Benaim2019}). The centrality of invasion growth rates suggests that the Hofbauer criterion could be integrated with MCT's partition of invasion growth rates into coexistence mechanisms. While it is always possible to use MCT to partition invasion growth rates, it is unclear how we may interpret these growth rates when community assembly is complicated. For example, consider the community assembly graph $\emptyset \rightarrow \{1\} \rightarrow \{1,2\} \rightarrow \{1,2,3 \}$. In order for species 2 to coexist, it must be able to invade species 1, \textit{and} it must be able to survive the invasion of species 3, yet species 2's invasion growth rate only captures the former event. Further, it is unclear how the Hofbauer criterion could be used in conjunction to compute community-average coexistence mechanisms (as in \cite{chesson2003quantifying}, \cite{barabas2018chesson}). How would different sub-communities and different species be weighted in this average, keeping in mind the the weights $p_j$ are not necessarily unique? The integration of the Hofbauer criterion and MCT appears to be a fruitful avenue of future research, but such a project is outside the scope of this paper. 

\subsection{What is coexistence?}
\label{sec:The relationship between coexistence and invasion growth rates:Alternative coexistence criteria}

Most ecologists would agree with the following \textit{folk-definition} (which a psychometrician might call a theoretical definition): Coexistence is when interacting species co-occur for a long time. One could operationalize this definition by simulating a stochastic model forward in time and computing the probability of all species persisting past some user-specified time horizon. While this approach seems reasonable at first glance, there are a few things that are unsatisfactory. 1) Choosing a time horizon in order to define "a long time" feels subjective; inferences may be sensitive to this decision. 2) Inferences could be sensitive to the initial state of the invader's environment (again, this includes internal variables like resident densities), but we often do not have access to this information; even if we did, we would like to somehow integrate over possible initial conditions, since we care about what allows species to coexist \textit{in general}, not just in the near future. 3) We often hope to obtain biological insight by expressing the conditions for coexistence in terms of model parameters. However, it is almost always impossible to derive analytical expressions for probability distributions of future population densities, making mathematical analysis largely incompatible with the "simulate and check for coexistence" approach. 

To circumvent these problems, we can use two alternative notions of coexistence: \textit{local stability} and \textit{global stability}. A community exhibits local stability if species return to an attractor (of the underlying \textit{deterministic skeleton} of a stochastic model) following a small perturbation. A community exhibits global stability (also known as permanence, or uniform persistence) when species tend to return to their typical abundances regardless of initial conditions and the size of environmental perturbations. For more rigorous definitions, see \textcite{Schreiber2006}. 

Both local stability and global stability solve the problem of specifying a time horizon by assuming infinite population sizes (thereby precluding stochastic extinctions; see Section \ref{sec:The relationship between coexistence and invasion growth rates:Invasion growth rates}) and particular kinds of environmental perturbations. In the case of local stability, perturbations are so small so that they cannot overstep the community's basin of attraction. In the case of global stability, perturbations can be arbitrarily large cannot cause instantaneous extinction (this is because perturbations directly affect per capita growth rates, not population size). Local stability sidesteps the issue of initial conditions by focusing on the deterministic skeleton and assuming that the community starts close to a dynamical attractor (if one exists), whereas the irrelevancy of initial conditions is baked into the definition of global stability. Finally, local and global stability are mathematically tractable: in simple models, linear stability analysis can be used to test for a stable equilibrium point, mutual invasibility, or the Hofbauer conditions (Table \ref{tab:coexistence criteria}).

Indeed, the historical coexistence criteria of Table \ref{tab:coexistence criteria} are all criteria for either local stability or global stability. A positive, stable equilibrium is a special case of local stability; a positive attractor is the more general test for local stability. Feasibility is a necessary condition for global stability (\cite{aubin1988permanence}; \cite{jansen1998shaken}), but not local stability (for an example of coexistence without feasibility see \cite{armstrong1980competitive}). The Hofbauer condition (\eqref{SPT}) is a sufficient condition for global stability, and the mutual invasibility criterion is the special case of the Hofbauer condition for two species, where the weights $p_j$ (which are necessary for satisfying the Hofbauer criterion, \eqref{SPT}) can be selected arbitrarily (see \cite{Schreiber2011}, sec. 4). 

It is important to remember that coexistence criteria give us limited information about local stability and global stability (Table \ref{tab:coexistence criteria}), which themselves are imperfect operationalizations of the construct of coexistence. Local stability is too weak because it only handles the case of infinitesimal perturbations, but global stability is too strong because it requires robustness to arbitrarily large perturbations. Nevertheless, by circumventing the arbitrariness of choosing time horizons and initial conditions, the coexistence criteria and underlying stability concepts give us a systematic way to analyze and compare models.

\section{Spatiotemporal coexistence mechanisms}
\label{sec:Spatiotemporal coexistence mechanisms}

In this section, we describe how the invasion growth rate is partitioned into coexistence mechanisms, following the derivation of \cite{johnson2022coexistence}. Variables and notation are explained on-the-fly, but one may also refer to Table \ref{tab:symbols}.

The general strategy is to "decompose and compare" (\cite{Ellner2019}). \textit{Decompose} each species' long-term average per capita growth rate into additive terms using a Taylor series. Then, \textit{compare} the terms corresponding to the invader to like-terms corresponding the resident species. Note that the coexistence mechanisms are species-level quantities. To learn how a coexistence mechanism affects coexistence in general, we can compute community-average coexistence mechanisms (read on to Section \ref{Community-average coexistence mechanisms}).

\paragraph{Derivation of coexistence mechanisms}

Consider a community with discrete-time dynamics, with spatial structure but no age/stage structure. The local finite rate of increase of species $j$ at time $t$ and patch $x$ is defined as $\lambda_j(x,t) = n_j(x,t+1) / n_j(x,t)$, where $n_j(t)$ is population density. 

\begin{enumerate}

    \item First, express the local finite rate of increase as a function $g_j$ of an environmentally dependent parameter $E_j(x,t)$ and a competition parameter $C_j(x,t)$:
    
    \begin{equation}
        \lambda_j = g_j(E_j, C_j).
    \end{equation}
    
    Note that we have dropped the explicit space and time dependence for notational simplicity. Though $E_j$ is typically a demographic parameter that depends on the abiotic environment (e.g., per capita fecundity depends on degree-days), $E_j$ may more generally represent the effects of density-independent factors. The parameter $C_j$ is called the competition parameter, but it may represent the effects of density-dependent factors, including refugia, mutualists, and natural enemies.

    \item  The local finite rate of increase, $\lambda_j$, is expanded with a Taylor series of $g_j$ about the \textit{equilibrium values}, $E_j^*$ and $C_j^*$, chosen so that $g_j(E_j^*, C_j^*) = 1$.
    
\begin{equation} \label{taylor_decomp}
\begin{aligned}
 \left. {\lambda_j}%
_{\stackunder[1pt]{}{}}%
 \right|_{%
 \stackon[1pt]{$\scriptscriptstyle E_j = E_{j}^{*}$}{$\scriptscriptstyle C_j = C_{j}^{*}$}}
 \approx \; & 1 + \alpha_j^{(1)} (E_j - E_{j}^{*}) + \beta_j^{(1)} (C_j - C_{j}^{*}) \\ & + 
\frac{1}{2} \alpha_j^{(2)} (E_j - E_{j}^{*})^{2} + \frac{1}{2} \beta_j^{(2)} (C_j - C_{j}^{*})^{2} + 
\zeta_j  (E_j - E_{j}^{*})   (C_j - C_{j}^{*}),
\end{aligned}
\end{equation}

where coefficients of the Taylor series, 

\begin{equation}  \label{taylor_coef}
\begin{aligned}
 \alpha_j^{(1)} = \pdv{g_j\scriptstyle{(E_j^*, C_j^*)}}{E_j},  \quad
 \beta_j^{(1)} = \pdv{g_j\scriptstyle{(E_j^*, C_j^*)}}{C_j},  \quad
 \alpha_j^{(2)} = \pdv[2]{g_j\scriptstyle{(E_j^*, C_j^*)}}{E_j},  \quad
 \beta_j^{(2)} = \pdv[2]{g_j\scriptstyle{(E_j^*, C_j^*)}}{C_j,}  \quad
 \zeta_j = \pdv{g_j\scriptstyle{(E_j^*, C_j^*)}}{E_j}{C_j},  \quad
\end{aligned}
\end{equation}

are all evaluated at $E_j = E_j^*$ and $C_j = C_j^*$, as implied by the notation. The truncated Taylor series will lead to an accurate approximation of the invasion growth rate if environmental fluctuations are small compared to other model parameters (details in \cite{Chesson1994}; \cite{chesson2000general}). However, one can still measure coexistence mechanisms when these \textit{small-noise assumptions} are not met.

    \item The appropriate spatial and temporal averaging is applied in order to express average growth rates entirely in terms of moments of local growth $\lambda_j$ and \textit{relative density}, $\nu_j = n_j / \E{x}{n_j}$.
    
    The \textit{long-term average per capita growth rate} is the temporal average of the logged metapopulation finite rate of increase, which we denote by $\E{t}{r_j}$. Here, averages, variances, and covariance are denoted $\E{x,t}{\cdot}$, $\Var{x,t}{\cdot}$, and $\Cov{x,t}{\cdot}{\cdot}$ respectively, with the subscripts indicating whether the statistic is computed over space $x$, time $t$, or both. Using another Taylor series, the long-term average growth rate of species $j$ can be approximated as

\begin{equation} \label{time_decomp_2}
\begin{aligned}
   \E{t}{r_j} \approx \E{x,t}{\lambda_j} + \E{t}{\Cov{x}{\nu}{\lambda_j}} - 1 -   \frac{1}{2} \Var{t}{\E{t}{\lambda_j}}.
\end{aligned}
\end{equation}

    \item The Taylor series expansion of $\lambda_j$ (derived in step 1) is substituted into the expression for the average growth rate (\eqref{time_decomp_2}), resulting in a long expression for species $j$'s average growth rate:
    
    \begin{equation} \label{big_decomp}
\begin{aligned}
\E{t}{r_j} \approx \; &  \alpha_j^{(1)} \E{x,t}{(E_j - E_{j}^{*})} + \beta_j^{(1)} \E{x,t}{(C_j - C_{j}^{*})} \\ 
+ & \; \frac{1}{2} \alpha_j^{(2)} \Var{x,t}{E_j} + \frac{1}{2} \beta_j^{(2)} \Var{x,t}{C_j} + \zeta_j  \Cov{x,t}{E_j}{C_j} \\ 
+ & \; \E{t}{\Cov{x}{\nu_j}{ \alpha_j^{(1)} (E_j - E_{j}^{*}) + \beta_j^{(1)} (C_j - C_{j}^{*})}} \\ 
- & \; \frac{1}{2} \alpha_j^{(1)^{2}} \Var{t}{\E{x}{E_j}} - \frac{1}{2} \beta_j^{(1)^{2}} \Var{t}{\E{x}{E_j}} - \alpha_j^{(1)}\beta_j^{(1)} \Cov{t}{\E{x}{E_j}}{\E{x}{C_j}}.
\end{aligned}
\end{equation}

\item The invader is compared to the residents. By definition, no resident species can grow or decline on average (i.e., $\E{t}{r_s} = 0$), so we may subtract a linear combination of the $S-1$ resident species from the invasion growth rate

\begin{equation} \label{inv_res_dif}
    \E{t}{r_i} = \E{t}{r_i} - \frac{1}{S-1} \sum\limits_{s \neq i}^S \frac{a_i}{a_s} \E{t}{r_s},
\end{equation}

without at all altering the invasion growth rate. The constant $a_j$ represents the intrinsic speed of population $j$'s population dynamics, typically operationalized as generation time when all species in the community are at their typical abundances. The quotient $a_i/a_s$ is called a \textit{speed conversion factors} because $a_s$ in the denominator is canceled by a resident's speed, implicit in the resident's growth rate components, leaving only the invader's speed, $a_i$. The speed conversion factors have replaced the \textit{scaling factors} which featured in previous iterations of MCT. Both types of re-scaling are discussed further in Section \ref{Scaling factors}.

The long decomposition of the average growth rate (\eqref{big_decomp}) can be substituted into \eqref{inv_res_dif}, and like-terms can be grouped such that the invasion growth rate is expressed as a sum of invader--resident comparisons. These comparisons are the \textit{coexistence mechanisms}. 
\end{enumerate}

\begin{tcolorbox}[breakable, title = Formulas for small-noise coexistence mechanisms, subtitle style={boxrule=0.4pt, colback=black!30!white}]

\tcbsubtitle{The invasion growth rate}

\begin{flalign} \label{sn_co_mech}
    \E{t}{r_i} \approx \Delta E_{i} + \Delta\rho_i + \Delta\mathrm{N}_i + \Delta\mathrm{I}_i + \Delta\kappa_i
\end{flalign}

\tcbsubtitle{Density independent effects}

\begin{flalign}
    \label{dE}
    \Delta E_{i} = & \left[\alpha_{i}^{(1)} \E{x,t}{E_i - E_{i}^{*}} + \frac{1}{2} \alpha_i^{(2)}\Var{x,t}{E_i} - \frac{1}{2} \alpha_i^{(1)^{2}} \Var{t}{\E{x}{E_i}}\right] \nonumber \\ 
     & - \frac{1}{S-1} \sum\limits_{s \neq i}^S \frac{a_i}{a_s} \left[\alpha_s^{(1)} \E{x,t}{E_s - E_{s}^{*}} + \frac{1}{2} \alpha_s^{(2)}\Var{x,t}{E_s} - \frac{1}{2} \alpha_s^{(1)^{2}} \Var{t}{\E{x}{E_s}}\right] 
\end{flalign}

\tcbsubtitle{Linear density-dependent effects}
      
\begin{flalign} \label{drho}
   \Delta\rho_i = & \beta_i^{(1)} \E{x,t}{C_i - C_{i}^{*}}  - \frac{1}{S-1} \sum\limits_{s \neq i}^S \frac{a_i}{a_s} \beta_s^{(1)} \E{x,t}{C_s - C_{s}^{*}} 
\end{flalign}

\tcbsubtitle{Relative nonlinearity}
      
\begin{flalign} \label{dN}
     \Delta\mathrm{N}_i = & \frac{1}{2} \left[ \beta_i^{(2)}\Var{x,t}{C_i} - \beta_i^{(1)^{2}} \Var{t}{\E{x}{C_i}}\right] \nonumber  \\
    & - \frac{1}{S-1} \sum\limits_{s \neq i}^S \frac{a_i}{a_s} \left[ \beta_s^{(2)}\Var{x,t}{C_s} - \beta_s^{(1)^{2}} \Var{t}{\E{x}{C_s}}\right]
\end{flalign}

\tcbsubtitle{The storage effect}
      
\begin{flalign} \label{dI}
     \Delta\mathrm{I}_i =  & \left[ \zeta_i \Cov{x,t}{E_i}{C_i} - \alpha_i^{(1)}\beta_i^{(1)} \Cov{t}{\E{x}{E_i}}{\E{x}{C_i}}\right] \nonumber  \\
   & - \frac{1}{S-1} \sum\limits_{s \neq i}^S \frac{a_i}{a_s} \left[ \zeta_s \Cov{x,t}{E_s}{C_s} - \alpha_s^{(1)}\beta_s^{(1)} \Cov{t}{\E{x}{E_s}}{\E{x}{C_s}}\right]
\end{flalign} 

\tcbsubtitle{Fitness-density covariance}
      
\begin{flalign} \label{dkappa}
    \Delta\kappa_i = & \E{t}{\Cov{x}{\nu_i}{ \alpha_i^{(1)} E_i + \beta_i^{(1)} C_i}} \nonumber  \\
    & - \frac{1}{S-1} \sum\limits_{s \neq i}^S \frac{a_i}{a_s} \E{t}{\Cov{x}{\nu_s}{ \alpha_s^{(1)} E_s + \beta_s^{(1)} C_s}}
\end{flalign}     
\end{tcolorbox}

%\begin{table}[ht]
%\centering
\begin{longtable}[t]{p{0.3\linewidth}  p{0.7\linewidth}} 
\caption{\label{tab:symbols} The symbols and terminology of Modern Coexistence Theory (MCT). Table modified from \cite{johnson2022coexistence}.} \\
\toprule \\
 & \multicolumn{1}{c}{Description}  \\
\hline \\
\multicolumn{2}{l}{MCT-specific terminology} \\ 
\cmidrule(lr){1-2} \\
invader & a rare species; for mathematical convenience, the per capita growth rate of this species is approximated by perturbing population density to zero.  \\
resident & a common species, more precisely understood as a species at its typical abundances \\
invasion growth rate & the long-term average of the per capita growth rate of an invader \\
partition & a scheme for breaking up an invasion growth rate into a sum of component parts \\
coexistence mechanism & a class of explanations for coexistence; corresponds to a component of the invasion growth rate partition of Spatiotemporal MCT \\
invader--resident comparison & a comparison between an invader and the resident species; measures a rare-species advantage  \\
speed conversion factor & converts the population-dynamical speed of resident species to that of the invader; corrects for average fitness differences in the invader--resident comparison; replaces the \textit{scaling factors}, also known as \textit{comparison quotients}, from previous versions of MCT  \\
\end{longtable}
\begin{longtable}[t]{p{0.1\linewidth}  p{0.8\linewidth}}
\midrule \\ 
\multicolumn{2}{l}{Variable} \\ 
\cmidrule(lr){1-2} \\
$x$ & a location in space \\
$t$ & a point in time \\
$j$ & species index \\
$n_j(x,t)$ & the population density of species $j$ at patch $x$ and time $t$. \\
$\nu_j(x,t)$ & relative density, calculated as local population density divided by the spatial average of population density, i.e., $n_j(x,t) / \E{x}{n_j}$ \\
$\lambda_j$(x,t) & the local finite rate of increase. In non-spatial models, $\lambda_j$ is defined as $n_{j}(x,t+1)/n_{j}(x,t)$. However, in spatial models, $\lambda_j$ is defined as $n_{j}'(x,t)/n_{j}(x,t)$, where $n_{j}'(x,t)$ is the population size after the local growth phase, but before the dispersal phase. \\
$\widetilde{\lambda}_j(t)$ & the metapopulation finite rate of increase, defined as a density-weighted average over patches: $\widetilde{\lambda}_j = \E{x}{(n_j / \E{x}{n_j}) \lambda_j}$ \\
$\E{t}{r_j}$ & The long-term average growth rate; for resident species, this is zero by definition; for invader, this is the invasion growth rate; in discrete-time models with no age/stage-structure, it is equal to $\E{t}{\log(\widetilde{\lambda}_j)}$ \\
$E_j(x,t)$ & the environmental parameter; more generally understood as the effects of density-independent factors \\
$C_j(x,t)$ & the competition parameter; more generally understood as the effects of density-dependent factors \\
$g_j$ & a function that gives the local finite rate of increase: $\lambda_j(x,t) = g_j(E_j(x,t), C_j(x,t))$ \\
$E_j^*$ & the equilibrium environmental parameter, defined so that $g_j(E_j^*, C_j^*) = 1$ \\
$C_j^*$ & the equilibrium competition parameter, defined so that $g_j(E_j^*, C_j^*) = 1$ \\
$\sigma$ & the scale of environmental fluctuations \\
$S$ & the total number of species in the community; $S-1$ is the number of residents \\
$a_j$ & the \textit{speed} of the population dynamics of species $j$; the intrinsic capacity to grow or decline quickly; often operationalized $a_j = 1/GT_j$, where $GT_j$ is the generation time of species $j$  \\
$\frac{a_i}{a_j}$ & speed conversion factors; a constant that effectively converts the population-dynamical speed of species $j$ to that of species $i$ \\
\midrule \\
\multicolumn{2}{l}{Coexistence mechanisms} \\
\cmidrule(lr){1-2} \\
$\Delta E_i$ & Density-independent effects; the degree to which density-independent factors favor the invader \\
$\Delta \rho_i$ & Linear density-dependent effects; specialization on resources and/or natural enemies\\ 
$\Delta N_i$ & Relative nonlinearity; specialization on the spatiotemporal variance of resources and/or natural enemies  \\ 
$\Delta I_i$ & The storage effect; specialization on different states of a spatiotemporally varying environment  \\ 
$\Delta \kappa_i$ & Fitness-density covariance; the differential ability of rare species to end up in locations with high ecological fitness  \\
\midrule \\
\multicolumn{2}{l}{Taylor series coefficients} \\ 
\cmidrule(lr){1-2} \\
$\alpha_{j}^{(1)}$ & the linear effects of fluctuations in $E_j$, defined as $\pdv{g_j}{E_j}\Bigr|_{\substack{E_j = E_{j}^{*}\\C_j = C_{j}^{*}}} = \pdv{g_j\scriptstyle{(E_j^*, C_j^*)}}{E_j}$ \\ 
$\alpha_{j}^{(2)}$ & the nonlinear effects of  of fluctuations in $E_j$, defined as $\pdv[2]{g_j\scriptstyle{(E_j^*, C_j^*)}}{E_j}$  \\ 
$\beta_{j}^{(1)}$ & the linear effects of fluctuations in $C_j$, defined as $\pdv{g_j\scriptstyle{(E_j^*, C_j^*)}}{C_j}$  \\ 
$\beta_{j}^{(2)}$ & the nonlinear effects of fluctuations in $C_j$, defined as $\pdv[2]{g_j\scriptstyle{(E_j^*, C_j^*)}}{C_j}$  \\ 
$\zeta_{j}^{(1)}$ & the non-additive (i.e., interaction) effects of fluctuations in $E_j$ and $C_j$, defined as $\zeta_j = \pdv{g_j\scriptstyle{(E_j^*, C_j^*)}}{E_j}{C_j}$  \\ 
\midrule \\ 
\multicolumn{2}{l}{Superscripts and subscripts} \\ 
\cmidrule(lr){1-2} \\
Subscripts &  \\ 
\cmidrule(lr){1-1} \\
$j$ & index of an arbitrary species   \\
$i$  & index of the invader  \\
$r$ & index of a resident  \\
$x$ & indicates that a summary statistic (e.g., mean, covariance, variance) is calculated by summing across space  \\
$t$ & indicates that a summary statistic (e.g., mean, covariance, variance) is calculated by summing across time \\
\midrule  \\ 
\multicolumn{2}{l}{Operators} \\ 
\cmidrule(lr){1-2} \\
$\E{x,t}{\cdot}$ & The spatiotemporal \textit{sample} arithmetic mean; for a variable $Z$ that varies over $K$ patches and $T$ time points, \\ 
& $\E{x}{Z} = (1/K)\sum_{x = 1}^{K} Z(x,t)$, \\
& $\E{t}{Z} = (1/T)\sum_{t = 1}^{T} Z(x,t)$, and \\
& $\E{x,t}{Z} = (1/(T K)) \sum_{t = 1}^{T} \sum_{x = 1}^{K} Z(x,t)$  \\
$\Var{x,t}{\cdot}$ & The spatiotemporal \textit{sample} variance for a variable $Z$ that varies over $K$ patches and $T$ time points, \\
& $\Var{x}{Z} = (1/K)\sum_{x = 1}^{K} (Z(x,t) - \E{x}{Z})^2$, \\
& $\Var{t}{Z} = (1/T)\sum_{t = 1}^{T} (Z(x,t) - \E{t}{Z})^2$, and \\
& $\Var{x,t}{Z} = (1/(T K)) \sum_{t = 1}^{T} \sum_{x = 1}^{K} (Z(x,t) - \E{x,t}{Z})^2$ \\
$\Cov{x,t}{\cdot}{\cdot}$ & The spatiotemporal \textit{sample} covariance of variables $W$ and $Z$ that vary over $K$ patches and $T$ time points,  \\
& $\Cov{x}{W}{Z} = (1/K)\sum_{x = 1}^{K} (W(x,t) - \E{x}{W})(Z(x,t) - \E{x}{Z})$, \\
& $\Cov{t}{W}{Z} = (1/T)\sum_{t = 1}^{T} (W(x,t) - \E{t}{W})(Z(x,t) - \E{t}{Z})$, and \\
& $\Cov{x,t}{W}{Z} = (1/(T K)) \sum_{t = 1}^{T} \sum_{x = 1}^{K} (W(x,t) - \E{x,t}{W})(Z(x,t) - \E{x,t}{Z})$ \\
\bottomrule \\
\end{longtable}
%\end{table}%

\section{Interpreting coexistence mechanisms}
\label{sec:Interpreting coexistence mechanisms}

\newpage
\begin{table}
%\centering
\caption{\label{tab:co_max} The maximum number of species that can coexist via various coexistence mechanisms, in a system with $L$ discrete resources and $M$ discrete environmental states; Table from \cite{johnson2022coexistence} }

\begin{tabular}{ >{\raggedright}p{0.4 \linewidth} >{\centering\arraybackslash} p{0.1\linewidth} >{\centering\arraybackslash} p{0.2\linewidth} >{\centering\arraybackslash} p{0.15 \linewidth} >{\centering\arraybackslash} p{0.2 \linewidth} }

Coexistence mechanisms & Models with no variation & Models with only spatial variation & Models with only temporal variation & Models with spatiotemporal variation \\
\toprule
$\Delta E$: Density-independent effects & 1 & 1 & 1 & 1 \\
$\Delta \rho$: Linear density-dependent effects & L & $L$ & $L$ & $L$ \\
$\Delta N$: Relative nonlinearity & 0 & $(L(L-1))/2$ & $(L(L-1))/2$ & $L(L-1)$ \\
$\Delta I$: Storage effect & 0 & $L M$ & $L M$ & $2 L M$ \\
$\Delta \kappa$: Fitness-density covariance & 0 & $LM +(L(L-1))/2$ & 0 & $LM +(L(L-1))/2$ \\
\bottomrule
\end{tabular}
\label{tab:max species}
\end{table}
%\end{table}%

Why coexistence mechanisms? Why not use some other scheme for partitioning invasion growth rates? For one, the coexistence mechanisms are demarcated with respect to the absence vs. presence of density-dependence (i.e., $E_j$ vs. $C_j$) and variation (i.e., $\E{x,t}{C_j}$ vs. $\Var{x,t}{C_j}$), two integral concepts in population biology. Second, the coexistence mechanisms are distinct from a historical perspective: nobody discovered two coexistence mechanisms in the same paper (though fitness-density covariance is an amalgam of several previously proposed explanations). Third, each coexistence mechanism has a unique limit to the maximum number of species it can support, given a set number of regulating factors and environmental states (Table \ref{tab:max species}).  Finally, the coexistence mechanisms reveal commonalities between seemingly disparate explanations for coexistence. Herbivores are physically very different from soil nutrients, but both can behave similarly from a population-dynamical perspective (\cite{chesson2008interaction}).

That is not to say that the coexistence mechanisms are the only reasonable way partition the invasion growth rate. \textcite{Ellner2019} give a generic method for calculating unorthodox partitions and provide an example involving species' traits. One may decompose the conventional coexistence mechanisms further into contributions from individual (or subsets of) regulating factors or into contributions from spatial variation and temporal variation (\cite{johnson2022methods}). One may also aggregate coexistence mechanisms to compare the main effect of density-independent factors (i.e., $\Delta E_i$) to the main effect of density-dependent factors (i.e., $\Delta \rho_i + \Delta N_i$); or to compare all spatial mechanisms to all temporal mechanisms (\cite{johnson2022methods}, Section 4.1). 

Causal diagrams (Fig. \ref{fig:Density independent effects}, \ref{fig:Linear density-dependent effects}, \ref{fig:Relative nonlinearity}, \ref{fig:The storage effect}, \ref{fig:Fitness-density covariance}) can be used to demonstrate how each coexistence mechanism operates. With the exception of density-independent effects ($\Delta E_i$), each coexistence mechanism has two features: 1) a negative feedback loop involving population density, and 2) some degree of specialization / exclusivity / ecological differentiation. For example, in the causal diagram for the linear density-dependent effects, $\Delta \rho_i$ (Fig. \ref{fig:Linear density-dependent effects}), species $j$ has a species-specific competition parameter $C_j$, implying that species $j$ specializes on particular resources or natural enemies. In the causal diagram for the storage effect, $\Delta I_i$, (Fig. \ref{fig:The storage effect}), specialization manifests as the species-specific environmental parameter, $E_j$. Note that the causal diagrams are highly stylized; they focus on a feedback loop corresponding to a single species, and therefore only show a small subset of a much larger community-level causal diagram.

There are at least two reasons for relating simple explanations for coexistence to the coexistence mechanisms. First, the resulting taxonomy of models can serve as a heuristic guide for determining the precise causes of coexistence in a new model. Second, we need to multiple, disparate models to give a fully generally interpretation of coexistence mechanisms. As we will see, it is likely that some misconceptions about relative nonlinearity and the storage effect are the consequence of over-generalizing from two highly-similar models (the lottery model and the annual plant model; \cite{Chesson1994}).

\subsection{$\Delta E_{i}$: Density-independent effects}
\label{sec:Interpreting coexistence mechanisms:Density-independent effects}

\begin{figure}[!htb]
 \centering
      \includegraphics[scale = 0.5]{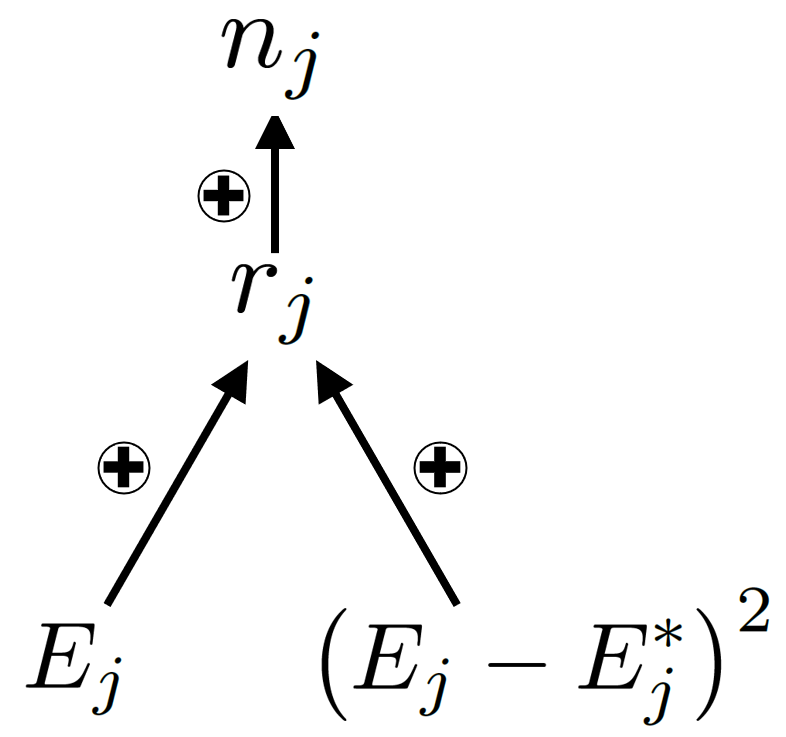}
  \caption{Density-independent effects. The density-independent factors (captured by $E_j$ and $\Var{}{E_j}$) affect growth rates and population density. Note the absence of any feedback loops. The quantity $\left(E_j - E_j^* \right)^2$ becomes $\Var{x,t}{E_j}$ when averaged across space and time; see \eqref{taylor_decomp} and \eqref{big_decomp} in Section \ref{sec:Spatiotemporal coexistence mechanisms}.}.
        \label{fig:Density independent effects}
\end{figure}

The first coexistence mechanism, $\Delta E_{i}$, is termed \textit{density-independent effects} and can be interpreted as the degree to which density-independent factors favor the invader. The value of $\Delta E_{i}$ does not depend on any species' density. This represented by the lack of a feedback loop in  clearly in Figure \ref{fig:Density independent effects}. Consequentially, one species will have the largest $\Delta E_i$, regardless of which species is the invader; if all other terms in the invasion growth rate partition are zero, then all other species in the community will be excluded (\cite{chesson1997roles}). This thought experiment demonstrates 1) that density-dependent factors are necessary for coexistence and therefore $\Delta E_i$ might rightfully not deserve the title of "coexistence mechanism"; and 2) why all the Taylor series terms (of the average growth rate decomposition, \eqref{big_decomp}) containing only $E_j$'s are shunted into $\Delta E_{i}$, while the growth rate components containing only $C_j$'s are split between $\Delta\rho_i$ and $\Delta\mathrm{N}_i$: the density-independent effects are between-species differences that cannot be responsible for coexistence, so it is often uninteresting to partition them further (but see \cite{Ellner2019}). 

\subsection{$\Delta \rho_i$: Linear density-dependent effects}
\label{sec:Interpreting coexistence mechanisms:Linear density-dependent effects}

\begin{figure}[h]
 \centering
      \includegraphics[scale = 0.5]{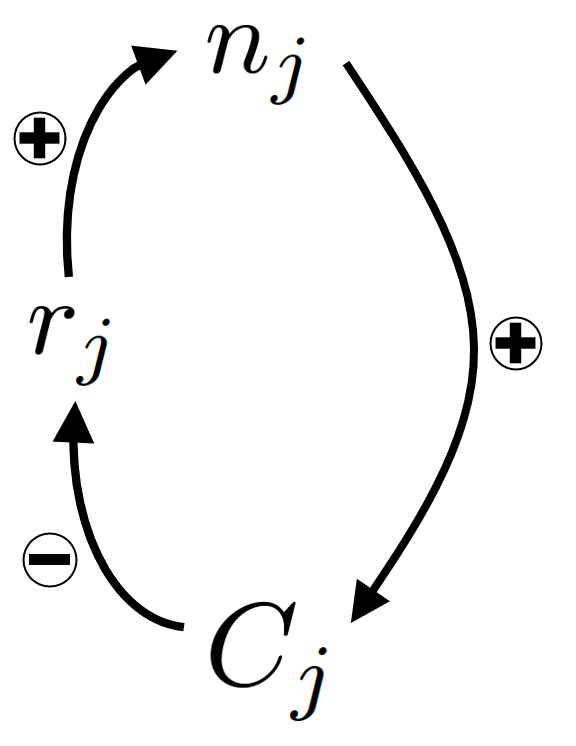}
  \caption{Linear density-dependent effects. The negative feedback loop involves the species-specific competition parameter $C_j$, which includes the effects of resources and natural enemies.} 
        \label{fig:Linear density-dependent effects}
\end{figure}

The second quantity, $\Delta \rho_i$, is called the \textit{linear density-dependent effects}. The linear density-dependent effects is best understood as the class of classic explanations for coexistence:  resource and natural-enemy partitioning. More precisely, $\Delta \rho_i$ is the rare-species advantage resulting from specialization on the mean level of density-dependent factors, which could take the form of mineral nutrients, water, carbon, prey species, light, space, refugia, pathogens, parasites, parasitoids, predators, herbivores, etc. The "specialization" need not be complete in the sense that each species only affects and is affected by a single density-dependent factor. Rather, there is some contingent (i.e., model specific) threshold of specialization needed in order to attain coexistence (\cite{barabas2016effect}). In the two-species Lotka-Volterra model, there is a simple mathematical condition for the "specialization threshold" (i.e., intraspecific competition > interspecific competition). In more speciose communities, such simple equations do not generally exist (\cite{saavedra2017structural}; \cite{LogofetD.O.DmitriĭOlegovich1993Mag:}).

Naturally, coexistence in explicit resource-consumer models can be attributed to $\Delta \rho_i$ (e.g., \cite{Ellner2019}). The same can be said of coexistence in Lotka-Volterra-like models (\cite{volterra1937principes}; \cite{hassell1976discrete}; \cite{walters1999linking};  \cite{dallas2021initial}), where species densities themselves can be treated as density-dependent factors. Lotka-Volterra dynamics are usually viewed as a useful but imperfect simile for the dynamics associated with competition or apparent-competition (\cite{abrams2008competition}; \cite{mayfield2017higher}; \cite{o2018whence}). However, when resource dynamics are fast, a specific form of resource-consumer dynamics are well-approximated by Lotka-Volterra dynamics (\cite{macarthur1970species}; \cite{chesson1990macarthur}).

The linear density-dependent effects encompasses several notable explanations for coexistence. First, $\Delta \rho_i$ captures coexistence mechanisms that operate on finer-grained spatial or temporal scales than that of observation/data-collection (more on this in Section \ref{The scale-dependence of coexistence mechanisms}). For example, the competition--colonization trade-off can be attributed to fitness-density covariance from a "worm's-eye view" (\cite{bolker1999spatial}; \cite{shoemaker2016linking}), but can be attributed to the linear density-dependent effects from a "bird's-eye view". the competition--colonization trade-off (\cite{skellam1951random}; \cite{levins1971regional}) can be attributed to $\Delta \rho_i$. The same is true for related explanations, such as the fecundity-dispersal trade-off (\cite{Yu2001}) and a seed size-number trade-off (\cite{turnbull1999seed}; \cite{muller2010tolerance}). The latter factoid can be verified by looking at the equations in \posscite{levins1971regional} classic paper on the competition--colonization trade-off and using the process of elimination to exclude fluctuation-dependent coexistence mechanisms. There are no patch-level equations, which excludes spatial fluctuation-dependent coexistence mechanisms; and the per capita growth rate equation is linear and deterministic, which excludes temporal fluctuation-dependent coexistence mechanisms.

Second, the Janzen-Connell hypothesis of tropical tree coexistence (\cite{janzen1970herbivores}; \cite{connell1971role}) also falls under the umbrella of $\Delta \rho_i$. Unlike ordinary natural-enemy partitioning, the Janzen-Connell hypothesis posits that coexistence is boosted further by \textit{distance-responsive predation}: parasites and diseases tend to kill seeds and seedlings which are near to their parent trees. However, \textcite{stump2015distance} used MCT to show that distance-responsive predation generally undermines coexistence, thus disproving the Janzen-Connell hypothesis in its most platonic form. 

Finally, coexistence via intransitive competition  (\cite{soliveres2018everything}) is \textit{partially} captured by $\Delta \rho_j$. Intransitive competition means that there is no best competitor in all settings, such that coexistence occurs via indirect effects that span across a network of interspecific interactions. This process is well-caricatured by Rock-Paper-Scissors dynamics, where species A beats B, B beats C, C beats A, and so on (\cite{May1975}, \cite{gilpin1975limit}). Intransitive competition is normally studied with Lotka-Volterra models, it can also arise in more complex, multi-trophic models (\cite{schreiber2004simple}; \cite{schreiber2018evolution}). While it has not been documented empirically or theoretically (to our knowledge), intransitivity can be mediated through other coexistence mechanisms. For instance, intransitivity via relative nonlinearity may occur if species $A$ generates a lot of resource variation, which disproportionately hurts species $B$ via relative nonlinearity, and so on. 

Invasion analysis is generally seen as incompatible with coexistence via intransitive competition. The problem is that in perturbing a species to invader state, intransitive loops (\textit{sensu} \cite{Gallien2017}, Fig. B3) involving the invader are destroyed, which which may cause knock-on extinctions (see Section \ref{sec:The relationship between coexistence and invasion growth rates:Inferring coexistence from invasion growth rates: The mutual invasibility criterion}). However, intransitive loops among the resident species affect the densities of the those species, and thus the level of competition felt by the invader. It is in this sense that $\Delta \rho_i$ partially captures the effects of intransitive competition. \textcite{Gallien2017} suggest a measure of the effects of intransitive competition on coexistence: the difference between invasion growth rate in the real world, and a hypothetical world where an some intransitive loops have been broken by removing a single resident species (this is repeated for each resident, and then averaged). This measure captures the effects of some intransitive loops (i.e., all the loops that pass through a single resident), but not all of them (i.e., the loops among the $S-2$ remaining residents). We submit the following method for measuring the effects of all intransitive loops (excluding those involving the invader) simultaneously: take the difference between the true invasion growth rate, and the invasion growth rate where there is a single resident species; naturally, this is repeated across all resident, and then averaged. 

\subsection{$\Delta\mathrm{N}_i$: Relative nonlinearity}
\label{sec:Interpreting coexistence mechanisms:Relative nonlinearity}

\begin{figure}[h]
 \centering
      \includegraphics[scale = 0.5]{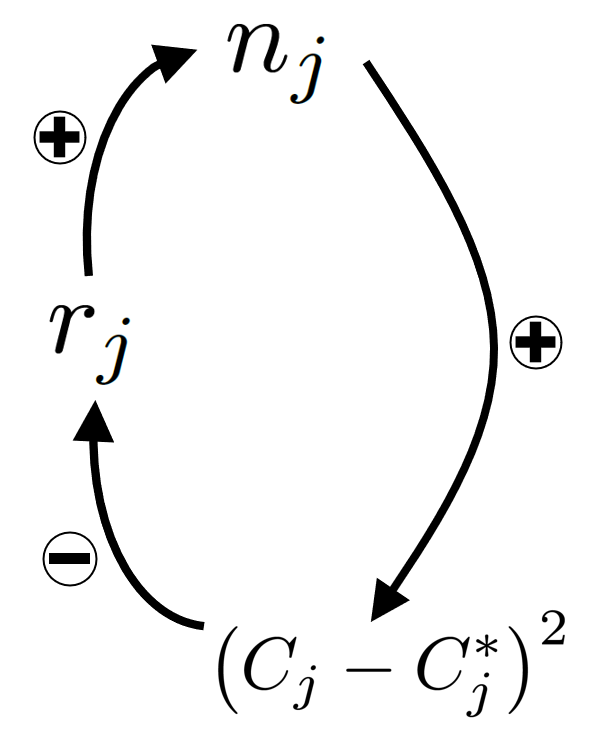}
  \caption{Relative nonlinearity. The negative feedback loop is mediated through the variance of regulating factors. The quantity $\left(C_j - C_j^* \right)^2$ becomes $\Var{x,t}{C_j}$ when averaged across space and time (see \eqref{taylor_decomp} and \eqref{big_decomp} in Section \ref{sec:Spatiotemporal coexistence mechanisms}). Increased population density $n_j$ increases the magnitude of fluctuations in competition, $\left(C_j - C_j^* \right)^2$. These fluctuations may affect the metapopulation growth rate $r_j$ either positively or negatively through the coefficient $\beta_j^{(2)}$ (see \eqref{taylor_coef}, Section \ref{sec:Spatiotemporal coexistence mechanisms}); here we show a negative effect, as this is the case in resource-consumer models where the consumer has a type II functional response and the competition parameter is $C_j = 1/"\text{resource concentration}"$. } 
      \label{fig:Relative nonlinearity}
\end{figure}

The first fluctuation-dependent mechanism, $\Delta\mathrm{N}_i$, is called \textit{relative nonlinearity}. It can be interpreted as the rare-species advantage that results from specialization on the variation in a density-dependent factors (\cite{levins1979coexistenceST}; \cite{tilman1982resourceST}). The variation can be generated endogenously via population dynamics (\cite{armstrong1976coexistence}, \citeyear{armstrong1980competitive}); or exogenously, either directly via a fluctuating resource supply (\cite{stewart1973partitioning}; \cite{hsu1980competition}; \cite{smith1981competitive}; \cite{butler1985mathematical}; \cite{abrams2004does}), or indirectly via environmental fluctuations (\cite{Chesson1994}; \cite{Yuan2015}). 

Compared to other coexistence-promoting mechanisms --- specifically $\Delta \rho_i$ and $\Delta I_i$ --- relative nonlinearity is understudied. The relegation of relative nonlinearity probably has several causes. For one, \textcite{Chesson1994} showed that in systems with a single competitive factor, only one species can coexist via relative nonlinearity, but an unlimited number of species could coexist via the storage effect (but see Section \ref{sec:Interpreting coexistence mechanisms:The storage effect} to see why this result may be misleading). Further, \textcite{chesson2000general} found that there is no relative nonlinearity in the the annual plant model and the lottery model with only spatial variation. In models where relative nonlinearity can arise, \textcite{chesson2000general} writes that "\ldots the limited ability for relative nonlinearity to promote coexistence when acting alone means that it is best viewed as modifying other mechanisms ... by decreasing the degree of dominance of a superior competitor with a relatively concave growth rate ... ". As we will show below, this is only generally true for models where resource variation is driven by environmental stochasticity: fluctuations in the per capita demographic rates of of resources or consumers.

Theoretically, many species \textit{can} coexist via relative nonlinearity when there are many competitive factors. In a system with $L$ regulating factors, there are $L(L+1)$ unique spatial and temporal covariances (\cite{johnson2022coexistence}); treating the covariance between regulating factors as an effective regulating factor and following the mathematics of the competitive exclusion principle (\cite{levin1970community}), we conclude that the maximum number of species that can coexist via relative nonlinearity is $L(L+1)$ (Table \ref{tab:co_max}). However, it is unclear how to devise a concrete model in order to attain this outcome, to say nothing of how representative such a model would be of real-world population dynamics.

\begin{figure}[h]
 \centering
      \includegraphics[scale = 0.5]{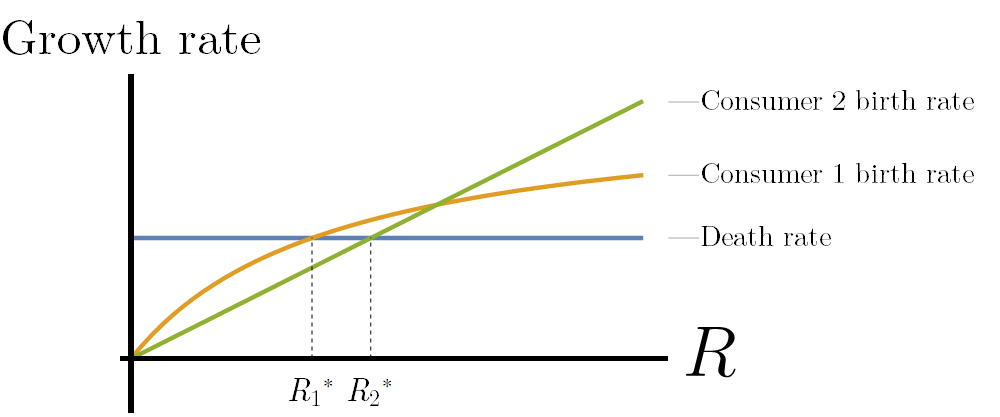}
  \caption{An opportunist-gleaner trade-off. Consumer 1 (the "gleaner") excludes consumer 2 (the "opportunist") in the absence of resource fluctuations (by \posscite{tilman1982resourceST} $R^*$ rule: $R_1^* < R_2^*$), but consumer 1 is hurt more by resource fluctuations (by Jensen's inequality). Coexistence via relative nonlinearity is possible if Consumer 1 increases resource variation when it is abundant.}
      \label{fig:Opportunist Gleaner}
\end{figure}

To better understand the relationship between resource variation and coexistence, we consider a community with a single resource and two consumers that exhibit an  opportunist-gleaner tradeoff (Fig. \ref{fig:Opportunist Gleaner}). We also use a heuristic that originates from Tilman's (\citeyear{tilman1980resources}, \citeyear{tilman1982resourceST}) graphical analysis of resource-consumer models: For coexistence to occur, species must consume proportionately more of that which most limits their own growth. In the example portrayed in Figure \ref{fig:Opportunist Gleaner}, the gleaner is hurt by resource variation, so it is most limited by mean resource levels. Thus, coexistence requires that the gleaner disproportionately decreases mean resource levels, or equivalently, increases resource variation when it is abundant.

The gleaner can increase variation (and thus promote coexistence) by inducing cyclical resource-consumer dynamics (\cite{armstrong1976coexistence}, \citeyear{armstrong1980competitive}). This outcome is contingent upon model parameters, but it is not a quirk: the gleaner has a faster consumption rate (at low resource concentrations), and is therefore inherently more destabilizing than the opportunist.

If resource dynamics are subject to environmental stochasticity, then resource variation should scale monotonically with mean resource levels (according to the small-noise approximation of population dynamics; \cite{gardiner1985handbook}; \cite{lande2003stochastic}). Here, since the gleaner has a lower $R^*$ than the opportunist, the gleaner tends to decrease resource variation, thus undermining coexistence. There is some reason to believe that this outcome is common in the real-world: Taylor's law (\cite{taylor1961aggregation}; \cite{taylor2019taylor}) shows that the aforementioned relationship between the mean and variance is common, at least for biotic resources. Though relative nonlinearity cannot promote coexistence in the case of environmental stochasticity, it is nevertheless important because it can change competitive outcomes (e.g., if resource variation is severe enough, then the opportunist will exclude the gleaner).

When the resource supply rate fluctuate through time, the gleaner tends to increase resource variation, thus promoting coexistence (\cite{hsu1980competition}; \cite{smith1981competitive}). The reason for the increase in variation is related to the different slopes the two consumer's birth-rate curves around their respective equilibrium resource levels (at $R^{*}_1$ and $R^{*}_2$; see Fig. \ref{fig:Opportunist Gleaner}). If the slope is steep, a resource surplus causes a dramatic increase in consumer birth rates; the subsequently large consumer population then reduces resource levels. In other words, resource levels are regulated via a negative feedback loop with consumers, and the strength of this negative feedback is proportional to the slope of the birth rate function. Because the gleaner species necessarily has a shallow slope (than the opportunist), it necessarily increases resource variation. Note here that coexistence is possible but not guaranteed. Experimental work with phytoplankton microcosms supports the idea that fluctuating resource supply changes population dynamics and sometimes causes coexistence (\cite{grover1990resource}; \cite{grover1991resource}; \cite[ch.~5]{grover1997resource}). 

So far in this section, we have discussed temporal relative nonlinearity in the context of resource competition. However, relative nonlinearity should work similarly in the apparent competition module (i.e., two prey, one predator), due to duality between resource concentration and the inverse of predator density ("enemy-free space" \cite{jeffries1984enemy}). Indeed, previous research has already demonstrated that in models with apparent competition, relative nonlinearity via endogenous cycles can promote coexistence (\cite{schreiber2004coexistence}) and that relative nonlinearity via environmental stochasticity does not permit multiple species to coexist (\cite{stump2017optimally}, Appendix D.2).

Spatial relative nonlinearity has only been explicitly studied in a few papers ( \cite{chesson2000general}; \cite{snyder2004spatial}; \cite{stump2018spatial}). It has been suggested that spatial relative nonlinearity should arise less readily than temporal relative nonlinearity (\cite{chesson2000general}; \cite{snyder2004spatial}; \cite{barabas2018chesson}) but this suggestion is clearly an extrapolation from the lottery model and annual plant model with fluctuating fecundity. Within the context of resource-consumer models with opportunist-gleaner trade-offs, we expect that spatial relative nonlinearity behaves similarly to temporal relative nonlinearity (Table \ref{tab:relative nonlinearity}). There is one exception: population cycles are necessarily a temporal phenomenon, so there is no purely spatial analogue of the endogenously generated resource-consumer cycles.

If there is spatial variation in the per capita (or per concentration) parameters of resource dynamics, then we can apply the same argument that we used in the case of temporal environmental stochasticity: resource variation is proportional to mean resource levels, so the dominant competitor in the absence of fluctuations --- the gleaner --- tends to decrease resource variation, thus undermining coexistence. This coexistence-undermining effect is strongest when there is complete local retention (i.e., individuals never disperse away from their home patches). Local retention in spatial models plays a similar role to temporal autocorrelation in temporal models, in the sense that both allow population buildup when/where conditions are favorable (see \cite{johnson2022towards}). As we will see, local retention tends to boost the spatial storage effect and fitness-density covariance.

If there is spatial variation in resource supply rates, then we can apply the same argument that we used in the case of temporally-fluctuating resource supply: The gleaner is less capable of dampening fluctuations in resource concentrations, which increases resource variation (relative to the opportunist), and in-turn promotes coexistence. There is an interesting twist: the coexistence-promoting effect of this mechanism is strongest when there is no local retention of consumers. Local retention strengthens the feedback loop between population density and resource concentration, such that consumers tamp-down resources in good patches (i.e., patches with high resource supply rates). When consumers disperse, good patches lose consumers (on net), which increases the spatial variation in resource concentrations. In Appendix \ref{Spatial variation in resource supply promotes coexistence}, we analyze a model and show -- with simulations and math -- that spatial variation in resource supply can indeed promote coexistence.

In conclusion, whether or not relative nonlinearity promotes coexistence in opportunist-gleaner models depends on the ultimate source of resource variation (Table \ref{tab:relative nonlinearity}). While "it depends" is perhaps an unsatisfactory answer, it is valuable in that it refutes the conventional wisdom that relative nonlinearity simply tweaks growth rates or switches competitive outcomes.

\renewcommand{\arraystretch}{1.25}
\begin{table}
\caption{\label{tab:relative nonlinearity} Does relative nonlinearity promote coexistence in a model with an opportunist gleaner trade-off? Here, \textit{Promotes coexistence} $=$ \textit{Yes} means that it is possible for both species to coexist. \textit{No} means only one species persists.}
\begin{tabular}{ p{0.5 \linewidth} p{0.5\linewidth}}
Source of resource/natural-enemy variation & Promotes coexistence? \\
\toprule
\textit{\underline{Temporal variation}} \\ 
Endogenous population cycles & Yes \\
Environmental stochasticity & No \\
Fluctuating resource supply rate & Yes \\
\textit{\underline{Spatial variation}} & \\ 
Endogenous population cycles & N/A \\
Environmental stochasticity & No \\
Fluctuating resource supply rate & Yes \\
\bottomrule
\end{tabular}
\end{table}

\subsection{$\Delta\mathrm{I}_i$: The storage effect}
\label{sec:Interpreting coexistence mechanisms:The storage effect}

\begin{figure}[h]
 \centering
      \includegraphics[scale = 0.5]{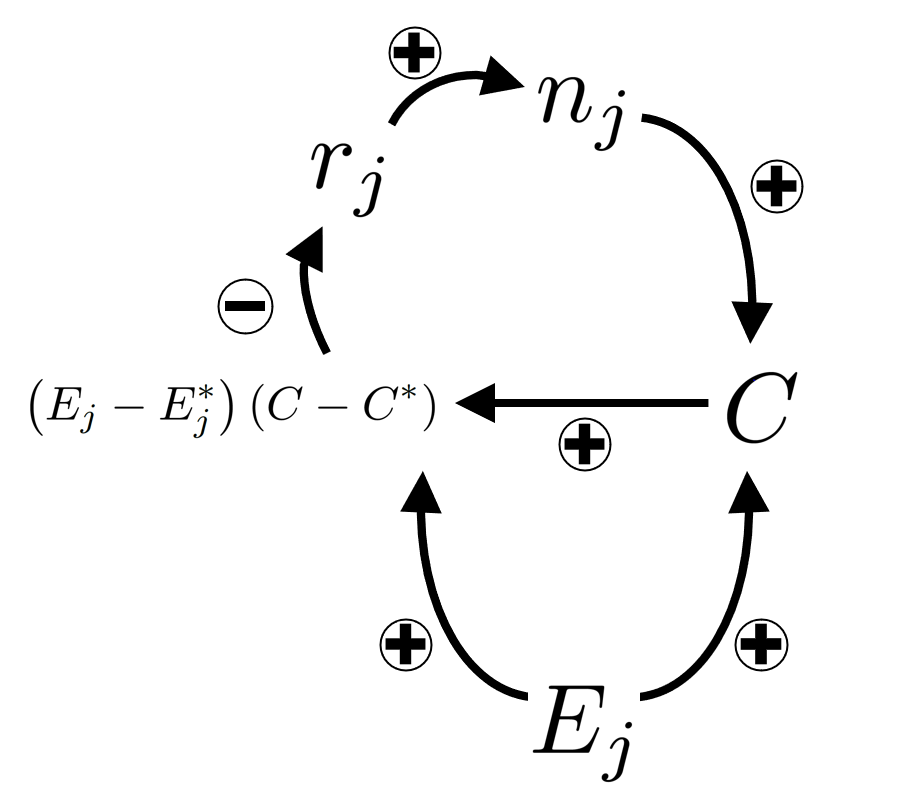}
  \caption{The storage effect. The negative feedback loop is mediated through the covariance between environment and competition. The quantity $\left(E_j - E_j^* \right)\left(C - C^* \right)$ becomes $\Cov{x,t}{E_j}{C}$ when averaged across space and time (see \eqref{taylor_decomp} and \eqref{big_decomp} in Section \ref{sec:Spatiotemporal coexistence mechanisms}). The product of fluctuations may affect the metapopulation growth rate $r_j$ either positively or negatively through the coefficient $\zeta_j$ (see \eqref{taylor_coef}, Section \ref{sec:Spatiotemporal coexistence mechanisms}); here we show a negative sign, since the archetypical storage effect involves a negative interaction effect (also known as \textit{subadditivity} or\textit{buffering}). The species-specific response to the environment, $E_j$, serves two functions. 1) $E_j$ affects $C$ (a good environment leads to high competition), thus ensuring that the covariance term is non-zero. 2) $E_j$ is \textit{species-specific}, which ensures that there is an element of \textit{specialization}. The competition parameter is generally a species-specific parameter (see Section \ref{sec:Spatiotemporal coexistence mechanisms}), but here we drop the species-specific index to emphasize that the storage effect is about specialization on environmental states. } 
    \label{fig:The storage effect}
\end{figure}

\begin{figure} \label{lottery model}
  \centering
     \includegraphics[scale = 1.2]{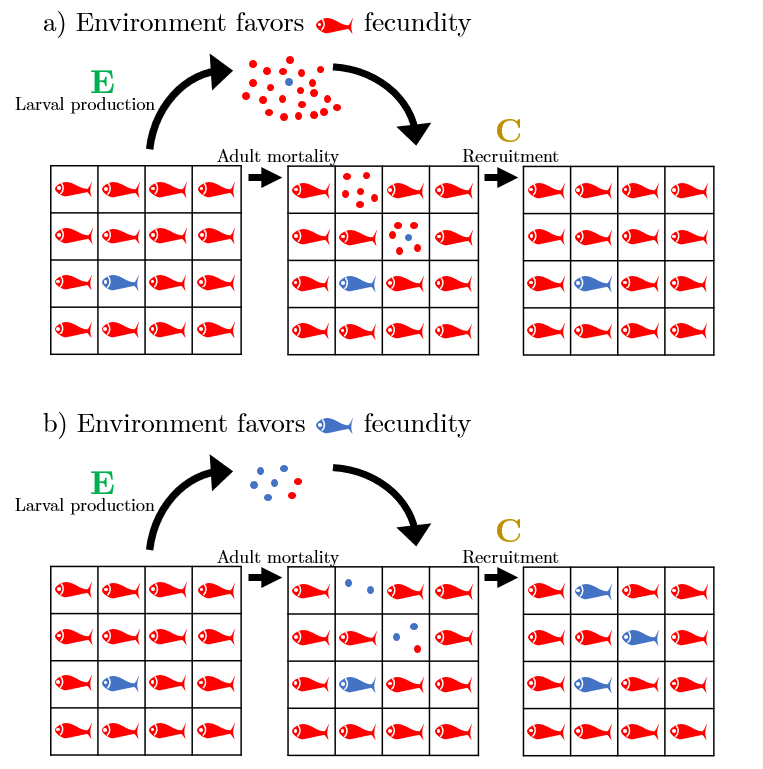}
    \caption{An illustration of the storage effect in the lottery model (from \cite{johnson2022storage}). \textit{Panel a)}: For a common (red) species, a good environment (high per capita fecundity) is undermined by the competition (total larvae per empty site) that it brings about. \textit{Panel b)}: For a rare (blue) species, a good environment does not lead to high competition. There are not many larvae produced in total because the rare species is rare, and because a good environment for the rare species is a poor environment for the common species.} 
    \label{fig:The Lottery Model}
\end{figure}

% \begin{figure}[h]
% %   \centering
% %   \hspace*{-1in}  
%   \makebox[\textwidth][c]{\includegraphics[width=1.5\textwidth]{lottery model fishes.pdf}}%
% % \includegraphics[width=1.2\textwidth, center]{lottery model fishes.pdf}
%   \caption{An illustration of the storage effect in the lottery model (from \cite{johnson2022storage}). \textit{Panel a)}: For a common (red) species, a good environment (high per capita fecundity) is undermined by the competition (total larvae per empty site) that it brings about. \textit{Panel b)}: For a rare (blue) species, a good environment does not lead to high competition. There are not many larvae produced in total because the rare species is rare, and because a good environment for the rare species is a poor environment for the common species.} 
%   \label{fig:The Lottery Model}
% \end{figure}

The second fluctuation-dependent mechanism, $\Delta\mathrm{I}_i$, is called \textit{the storage effect}. It can be interpreted as the rare-species advantage that results from environmental niche partitioning: specialization on environmental states that are distributed across space or time. Later, we will see that fitness-density covariance can be described similarly (as spatial environmental niche partitioning) but that there are substantial differences between fitness-density covariance and the spatial storage effect (see Section \ref{sec:Interpreting coexistence mechanisms:Fitness-density covariance}). The storage effect depends on three ingredients: 1) species-specific responses to the environment, 2) an interaction effect between environment and competition, also known as non-additivity ($\zeta_j \neq 0$; see \eqref{taylor_coef}), and 3) covariation between the environment and competition.

In order to gain an intuitive understanding of the storage effect, we can compare the invader and resident with respect to a good-environment scenario (Fig. \ref{fig:The Lottery Model}). For the resident, a good environment leads to increased population size, which leads to increased competition. This causal relationship between environment and competition is captured by the covariance, ingredient \#3. If the environment and competition have a negative interaction effect on growth rates  (ingredient \#2: $\zeta_j < 0$), a good environment (for the resident) will be undermined by the competition that it brings about. One the other hand, when an invader experiences a good environment, competition will not increase substantially, assuming that a the invader and resident the not perceive the environment identically (ingredient \#1). In general, the asymmetry between common species and rare species occurs because the strength of the environment's effect on competition is proportional to population density. 

What does any of this have to do with storage? The conventional but imprecise interpretation of the storage effect is that species coexist by specializing on different parts of a fluctuating environment, so species must have a robust life stage in order to "wait it out" for a favorable time period. A number of models have shown that neither stage-structure  nor overlapping-generations is necessary for the operation of the storage effect (\cite{li2016effects}; \cite{letten2018species}; \cite{schreiber2021positively}). What is essential is that at least some species experience an interaction effect between environment and competition (i.e., $\zeta_j \neq 0$), and it just so happens that a long-lived life stage is required for an interaction effect in the seminal models of MCT (i.e., the lottery model and the annual plant model). Another conventional but imprecise interpretation of the storage effect is that \textit{buffering} (operationalized as a negative interaction effect; $\zeta_i < 0$) helps the invader recover from rarity, since it protects them against the double-whammy of a bad environment and high competition. In reality, higher buffering corresponds to a lower storage effect for most species; elsewhere (\cite{johnson2022storage}), we have discussed these misinterpretations of the storage effect at-length. 

The temporal storage effect has been studied empirically in communities of zooplankton (\cite{caceres1997temporal}), desert annual plants (\cite{pake1995coexistence}; \cite{angert2009functional}; \cite{chesson2012storage}; \cite{ignace2018role}), perennial prairie grasses (\cite{adler2006climate}), protist microcosms (\cite{Jiang2007}); nectar yeast microcosms (\cite{letten2018species}), and tropical trees (\cite{usinowicz2012coexistence}). To our knowledge, there are only three empirical studies of the spatial storage effect (\cite{sears2007new}; \cite{angert2009functional}; \cite{towers2020requirements}), all of which involve herbaceous plants. 

The storage effect has also served as the basis for several theoretical advances. The temporal storage effect can be mediated through a particular life-stage (\cite{chesson1988community}, autocorrelated environmental variation (\cite{li2016effects}; \cite{schreiber2021positively}), causally-related environmental variables, and transgenerational plasticity (\cite{johnson2022towards}). The storage effect can arise in a periodic environment (\cite{loreau1989coexistence}; \cite{loreau1992time}; \cite{klausmeier2010successional}), which is just a special case of a temporally autocorrelated environment. From this perspective, it is easy to see how coexistence via phenology differences (\cite{godoy2014phenology}; \cite{rudolf2019role}) can be attributed to the storage effect. 

The temporal storage effect can also be mediated through predation, but only if predation changes concomitantly with the the the environmental parameter. This can happen through temporal autocorrelation in the predator's demographic rates (\cite{schreiber2021temporally}), or if predators exhibit behavioral responses to changes in prey density (\cite{kuang2010interacting}, \cite{chesson2010storage}). Prey species can experience a negative temporal storage effect (undermining coexistence) if their predators have type 2 functional responses (\cite{Stump2017}), such that residents can satiate their predators but invader cannot. In the absence of predator behavioral responses, we expect that the temporal storage effect due to apparent competition will generally weaker be than the storage effect due to resource competition (\cite{chesson2010storage}; \cite{johnson2022towards}). 

The spatial storage effect depends on a spatial covariance between environment and competition, which may arise either through sedentary life-stage (\cite{muko2000species}; \cite{snyder2003local}) or local retention (\cite{chesson2000general}), which generically occurs when dispersal distance is not much greater than the grain size of environmental variation (\cite{snyder2003local}; \cite{snyder2004spatial}). In either case, population buildup in good patches leads to high competition, thus satisfying ingredient \#3.

Some formulas in the literature (e.g., \cite[Eq.~81]{Chesson1994}) suggest that the storage effect can allow for the coexistence of an arbitrary number of species. While the storage effect is potentially powerful, it is not uniquely powerful, nor does "an arbitrary number of species" have much practical significance. When a regulating factor is expanded to be a continuum of density-independent factors (e.g., a density-distribution of different phenotypes), then ann arbitrary number of species can coexist via the linear effects of competition, $\Delta \rho_i$ (\cite[p.~534]{RoughgardenJoan1979Topg}). Similarly, when space is treated as continuum of environments (and not as discrete patches), an arbitrary number of species can coexist via fitness-density covariance (\cite{szilagyi2009limiting}). On the other hand, when there are $M$ \textit{discrete} environmental states and $L$ regulating factors, the storage effect can only support $L \times M$ species (\cite{miller2017evolutionary}; \cite{johnson2022coexistence}). 

Further, it likely that limits set by coexistence mechanisms (Table \ref{tab:co_max}) are  practically irrelevant, due to the fact that other forces limit biodiversity. These include developmental/physiological constraints on extreme forms of specialization (e.g., a plant can only grow so tall without falling over); structural stability (in the mathematical sense, \cite{Gyllenberg2005}); extinction-speciation balance (which putatively depends on environmental stability; \cite{krug2009generation}; \cite{shiono2018roles}); and succession towards limitation by only ground-level light (\cite{wisheu2000makes}; \cite{borer2014herbivores}).

\subsection{$\Delta\kappa_i$: Fitness-density covariance}
\label{sec:Interpreting coexistence mechanisms:Fitness-density covariance}

\begin{figure}[h]
 \centering
      \includegraphics[scale = 0.5]{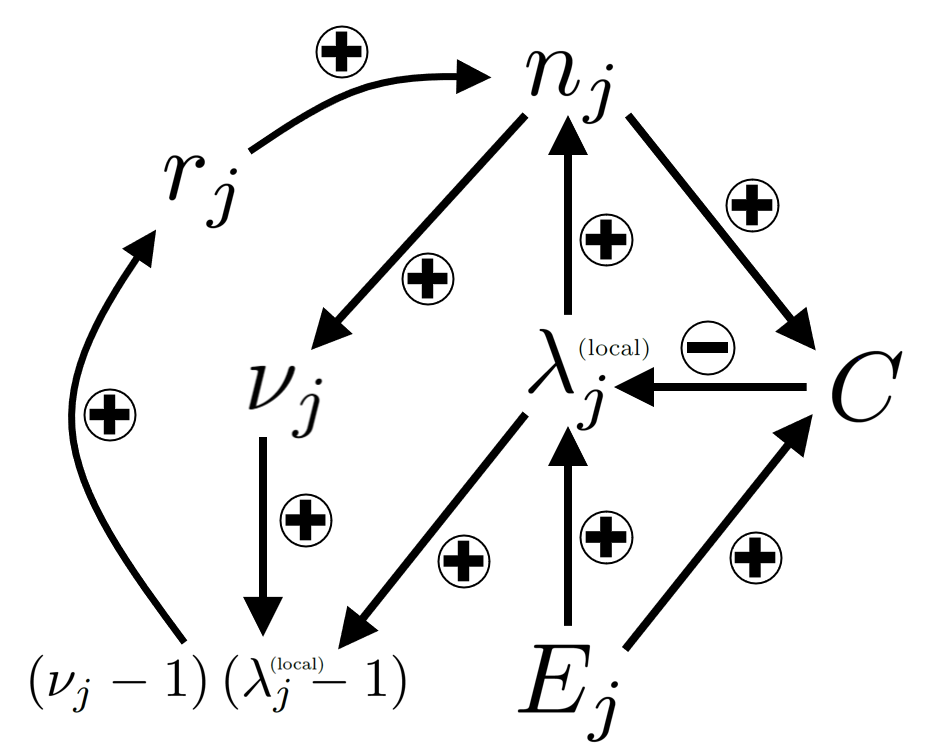}
    \caption{Fitness-density covariance. The full negative feedback loop is $n_j \rightarrow C \rightarrow \lambda_j^{\text{(local)}} \rightarrow n_j \rightarrow \nu_j \rightarrow (\nu_j - 1) (\lambda_j^{\text{(local)}} - 1) \rightarrow r_j \rightarrow n_j $. The local finite rate of increase is given to superscript "(local)" to help differentiate it from the metapopulation growth rate, $r_j$. The product of fluctuations in relative density and fitness, $(\nu_j - 1)(\lambda_j^{\text{(local)}} - 1)$ becomes $\E{t}{\Cov{x}{\nu_j}{\lambda_j^{\text{(local)}}}}$ when properly averaged across time and space (see \eqref{big_decomp} in Section \ref{sec:Spatiotemporal coexistence mechanisms}). Local population growth along with the \textit{local retention} of individuals leads to a buildup of population density in certain patches. This \textit{local} process is represented by the subset of the feedback loop $n_j \rightarrow C \rightarrow \lambda_j^{\text{(local)}} \rightarrow n_j$. Local population density affects relative density, $\nu_j$, and local fitness, $\lambda_j^{\text{(local)}}$. Then, local densities and finite rates of increase feed into the term $(\nu_j - 1)(\lambda_j^{\text{(local)}} - 1)$, which affects the metapopulation growth rate $r_j$. This process is represented by the subset of the feedback loop $n_j \rightarrow \nu_j \rightarrow (\nu_j - 1) (\lambda_j^{\text{(local)}} - 1) \rightarrow r_j \rightarrow n_j$. As in the causal diagram of the storage effect, the species-specific environmental parameter functions to imbue the negative feedback loop with some degree of specialization.} 
    \label{fig:Fitness-density covariance}
\end{figure}

The final fluctuation-dependent mechanism, $\Delta\kappa_i$, is known as \textit{fitness-density covariance} (also sometimes called \textit{growth-density covariance}). This term can be interpreted as the differential ability of an invader's individuals to end up in locations where they have high fitness (\cite{chesson2000general}; \cite{chesson2005scale}; \cite{Chesson2012}). Fitness-density covariance accounts for the tremendous potential for biodiversity that can result from the spatial partitioning of density-dependent factors. Examples include phytoplankton partitioning a light gradient in the water column (\cite{huisman1999competition}; \cite{gervais2003small}), nesting birds partitioning tree branches (\cite{macarthur1958population}), and plants partitioning microhabitats with different ratios of resource-supply rates (\cite{tilman1982resourceST}; \cite{crozier1984correlations}). Fitness-density covariance also accounts for the spatial partitioning of density-independent factors (i.e., environmental niche partitioning), the competition--colonization trade-off, and heteromyopia.

We can immediately see that $\Delta\kappa_i$ contains enormous complexity in a simple formula: relative density is the outcome of a multi-generational interplay between dispersal and local growth, and "fitness" itself can be decomposed into many parts (see \eqref{big_decomp}). To get a better grasp on this complexity, we will look at several scenarios that do not lead to fitness-density covariance, followed by several scenarios that do.

First, consider a landscape with a homogeneous environment. If per capita growth rates decrease with population density, then chance fluctuations in relative density will generate a negative $\Cov{x}{\nu_j}{\lambda_j}$ (\cite{Lloyd1966}; \cite{lloyd1967mean}; \cite{matsuda1992statistical}). However, if there are no systematic differences between species, an invader will quickly attain the same spatial correlations as the resident, resulting in $\Delta \kappa_i = 0$ (\cite{chesson1991need}). On the other hand, $\Delta \kappa_i$ will be positive if individuals aggregate semi-independently across space, as will occur naturally for invaders with herding, swarming, and/or mate-finding behavior. This scenario demonstrates that environmental niche differences are not necessary for coexistence, but that niche differences of some kind --- whether they be environmental or behavioral --- are still required.

Second, consider a landscape with a heterogeneous environment which is inhabited by organisms that have complete information, can disperse at no cost, and do not exhibit interference competition. Here, residents will attain the \textit{ideal free distribution} (\cite{fretwell1969territorial}): the spatial distribution of a resident $s$ where fitness is constant at $\lambda_s = 1$ across the landscape. If the invader has environmental niche differences, it will concentrate in its best patches without increasing competition, thus producing a positive $\Delta \kappa_i$. If the invader has the same environmental niche as the resident, then $\Delta \kappa_i$ equals zero. This scenario demonstrates that fitness-density covariance is not merely a measure of how good species are at actively dispersing to "good" patches. Since the local fitness of a patch depends on the number of competitors in that patch, the ability to \textit{end up} in location with high fintess is inextricably tied to abundance. 

For our final no-coexistence scenario, consider a heterogeneous environment with organisms that exhibit \textit{widespread dispersal}: in each time-step, all individuals disperse and rain uniformly across the landscape. Here, relative density is the same in every patch, and thus fitness-density covariance is zero. 

What does generate a positive $\Delta \kappa_i$? Environmental niche differences, environmental heterogeneity, and local retention naturally engender a positive fitness-density covariance:. A good environment for a resident leads to a buildup of resident individuals, but a good environment for an invader does not lead to a buildup of invader individuals. This asymmetry leads to reduced fitness for residents in high density patches, which translates to a positive $\Delta \kappa_i$. This whole process can also generate a storage effect if the interaction effect between environment and competition is non-zero (see Section \ref{sec:Interpreting coexistence mechanisms:The storage effect}). As \textcite{chesson2000general} points out, variation in relative density can generate a positive covariance between environment and competition, even if a negative covariance would be attained in the absence of density variation. 

Fitness-density covariance accounts for the coexistence of competitors in patches with varying resource supply rates (\cite[ch.~6]{chase2003ecological}). Resources in different patches can be treated (mathematically) as separate regulating factors (\cite{levins1974discussion}; \cite{szilagyi2009limiting}). Alternatively, this process can be thought of as a special case of the environmental niche partitioning described in the previous paragraph, where the spatially fluctuating resource supply point \textit{is} the environmental parameter $E_j$. Even though $E_j$ is typically a demographic parameter that directly affects some focal species  (e.g., the probability of germination), $E_j$ can just as easily directly affect the regulating factors (for an example, see Appendix \ref{Spatial variation in resource supply promotes coexistence}).

Fitness-density covariance also accounts for two previously proposed explanations for coexistence: the \textit{competition--colonization trade-off} (\cite{skellam1951random}; \cite{levins1971regional}) and \textit{heteromyopia} (\cite{murrell2003heteromyopia}). In the competition--colonization trade-off, the superior colonizer has a positive $\Delta \kappa_i$ (\cite{shoemaker2016linking}) because it ends up in recently disturbed patches which are devoid of competitors. In the case of heteromyopia --- the phenomenon where intraspecific competition occurs over longer distances than interspecific competition --- intraspecific competition lowers the residents' density, creating small holes in the landscape. The invader settles into theses holes and only competes strongly with a few conspecific in the near vicinity, resulting in a positive $\Delta \kappa_i$ (\cite{snyder2008does}). 

Another way to understand fitness-density covariance is to understand how it differs from other coexistence mechanisms. The spatial storage effect and fitness-density covariance seem inextricably related, since they both arise readily in models with environmental niche differences, environmental heterogeneity, and local retention. However, we can tease out differences just by looking at the mathematical definitions of the two coexistence mechanisms (\eqref{dI} \& \eqref{dkappa}). For one, the spatial storage effect requires spatial variation in the environment, whereas fitness-density covariance does not. Species may have biased dispersal based on sensory preferences of environmental conditions that otherwise do not affect growth rates (\cite{barabas2018chesson}, Appendix S5); this is "no environmental variation" in the technical sense that there is no varying demographic parameter than affects the local finite rate of increase. 

A fitness-density covariance can potentially emerge in a truly homogeneous environment if species engage in aggregating behavior (e.g., swarming, schooling, herding, "natal homing"). It is quite possible that such behavior will generate a positive fitness-density covariance: the clustered resident individuals experience lower fitness $\lambda$ (via increased competion), but continue to aggregate because there is an individual-level fitness benefit for doing so. However, aggregating behavior could also lead to a negative fitness-density covariance. If aggregation is strong enough, invader populations can be regionally rare but locally abundant, thus eliminating a rare-species advantage. Alternatively, aggregation can have positive fitness consequences (e.g., increased mating-finding ability, group-level vigilance) such that the residents benefit on-net. 

Even in the face of environmental heterogeneity, there are still notable differences between the spatial storage effect and fitness-density covariance. In Appendix \ref{The spatial storage effect vs. fitness-density covariance} we derive expressions for both coexistence mechanism in an arbitrary model with two species, permanent spatial heterogeneity, and dispersal. When species are identical except for their responses to the environment, we find that

\begin{equation}
    \Delta I = \frac{q N_{0}^* \alpha^{(1)}}{1-q(\theta N_{0}^* \beta^{(1)} +1)} \left[ \zeta \sigma^2 (\phi - 1) \right], \text{and}
\end{equation}

\begin{equation}
\begin{aligned}
    \Delta \kappa = & \frac{q N_{0}^* \alpha^{(1)}}{1-q(\theta N_{0}^* \beta^{(1)} +1)} \left[\frac{q}{1-q}  2 \alpha^{(1)} \theta \beta^{(1)} \sigma^2 (\phi - 1)\right].
\end{aligned}
\end{equation}

All of the symbols are described in Table \ref{tab:symbols} and Appendix \ref{The spatial storage effect vs. fitness-density covariance}; but notably, $\sigma^2$ is the variance in the environmental parameter $E_j$; $\phi$ is the spatial correlation between the two species' environmental parameters; $\zeta$ is the interaction effect between environmental and competition; and $q$ is the local retention fraction (a $(1-q)$ fraction of individuals disperse after each time-step).

Both the spatial storage effect and fitness-density covariance are proportional to $q \times \sigma^2 \times (1- \phi)$, respectively representing local retention, environmental heterogeneity, and spatial niche differences. Under these conditions, fitness-density covariance is all but inevitable, whereas the spatial storage effect depends on there being a substantial interaction effect $\zeta$. This may explain why some simple spatial models produce fitness-density covariance but not the spatial storage effect (e.g., \cite{amarasekare2001spatial}; \cite{muko2000species}).

Another notable difference is that the spatial storage effect is proportional to $q/(1-q)$, whereas fitness-density covariance is nearly proportional to $(q/(1-q))^2$, which is very large when the local retention $q$ is large. The discrepancy occurs because the density of a species is proportional to $q/(1-q)$; in turn, the density of species is only proportional to competition (which shows up in the spatial storage effect), but is proportional to both relative density $\nu$ and fitness $\lambda$ (the product of which shows up in fitness-density covariance). This may explain \posscite{shoemaker2016linking} finding that fitness-density covariance is more important than the spatial storage effect: the authors used large retention fractions ($q \approx 0.9$). 

It has been argued that the spatial storage effect "seems to be inevitable under realistic scenarios" (\cite{chesson2000general}), and that "...space itself is often the bet-hedging strategy that generates storage ..." (\cite{barabas2018chesson}). However, we believe that these statements are over-generalizations from particular versions of the lottery model and annual plant model, where spatially-fluctuating fecundity automatically generates an interaction effect $\zeta_j$. However, with a slight tweak to these models, we see exactly the opposite: when survival fluctuates instead of fecundity, the interaction effect is automatically zero. The spatial storage effect may indeed be prevalent in nature, but it is by no means inevitable (see \cite{towers2020requirements}), and is likely to be less widespread than fitness-density covariance.

\section{Other topics in Modern Coexistence Theory}
\label{sec:Discussion}

We have already discussed how Modern Coexistence Theory (MCT) is able to relate real coexistence to simple explanations for coexistence (see Fig. \ref{fig:edifice} in Section \ref{Introduction}). However, there remain a number of specific topics in MCT that ought to be expounded on. Some topics have been the subject of mild controversy in the MCT community; other topics have not been adequately addressed or may otherwise be confusing to newcomers. Here, we discuss these topics and attempt to provide insight when possible. 

\subsection{How do coexistence mechanisms correspond to explanation for coexistence}
\label{How do coexistence mechanisms correspond to explanation for coexistence}

In Section \ref{sec:Interpreting coexistence mechanisms}, we argued that coexistence mechanisms correspond to different classes of simple explanations for coexistence. But why does this work? What counts as a explanation for coexistence, and how is that reflected in the mathematical definitions of coexistence mechanisms.

Historically, mathematical explanations for coexistence have centered around the idea of \textit{specialization} or trade-offs or ecological differentiation. To give one example, "\ldots each species must consume \textit{proportionately} more of the resource that more limits its growth" (\cite[p.~96]{tilman1982resourceST}; emphasis added). The word "proportionately" insinuates a comparison of multiple species, and this is born-out in both the the mathematical and graphical conditions for coexistence (\cite[p.~77]{tilman1982resourceST}). At the same time, verbal/textual explanations often focus on the idea of a \textit{rare-species advantage}. For instance, one may explain resource partitioning by saying "when a species falls to low density, the resource on which it specializes will become more abundant, thus increasing the focal species' per capita growth rates". 

In the coexistence mechanisms of MCT, the rare-species advantage is captured by the invasion growth rate, and the notion of specialization is captured by the invader--resident comparison. However, the invader--resident comparison not only captures the notion of the specialization, but also intrinsic differences between species that are not relevant for coexistence. For example, consider the two species the two-species Lotka-Volterra model with the per capita growth rate $r_j = b_j(1 - \alpha_{j1} n_1 - \alpha_{j2} n_2)$. The parameters $b_j$ have no bearing on coexistence (the condition for coexistence is $\alpha_{12} / \alpha_{22} < 1 < \alpha_{11} / \alpha_{21}$), yet can greatly modulate the values of coexistence mechanisms. To correct for these sorts of between-species differences, we may re-scale resident growth rates.

\subsection{Scaling factors}
\label{Scaling factors}

The conventional way to re-scale resident growth rates is with the so-called scaling factors, a measure of the residents' relative sensitivity to competition (defined by \cite{Chesson1994}, Eq.24). The scaling factors have been the source of much confusion, as they 1) were introduced without an explicit justification, 2) are not unique when there are more residents than regulating factors, and 3) can be very difficult to compute (see \cite{ellner2016quantify}, SI.5).

The purpose of the scaling factors is to cancel the linear density-dependent effects, $\Delta \rho_i$. As we argue in \textcite{johnson2022methods}, this can be useful in certain theoretical studies, but it is not recommended for quantifying coexistence mechanisms in real communities. In their quest to cancel $\Delta \rho_i$, the scaling factors can dramatically modulate other coexistence mechanisms. This "collateral damage" can lead to incorrect inferences about how species are coexisting. 

There are several alternatives to the scaling factors (\cite{johnson2022methods}). The two best options are \textit{the simple comparison}, and \textit{speed conversion factors}. The simple comparison, first suggested by \textcite{ellner2016quantify}, gives a weight of $1/(S-1)$ to each resident (recall that $S-1$ is the number of resident species). The simple comparison has the effect of weighting each resident equally, but giving the same total weight to the low-density state (i.e., the invader) and the high density state (i.e., the sum of residents).

For an invader $i$ and resident $s$, the speed conversion factor method scales resident growth rates by $ (a_i / a_s) \times 1/(S-1)$, where $a_j$ is the reciprocal of the generation time of species $j$. The speed conversion factors hypothetically convert the residents' population-dynamical speed to that of the invader, in an attempt to correct for intrinsic differences between species that are not relevant for coexistence. In the Lotka-Volterra example provided in the previous section, the speed conversion factors should eliminate the speed parameters, $b_j$, from all coexistence mechanisms. The speed conversion factors are recommended when species have dissimilar generation times. 

\subsection{Exact coexistence mechanisms}
\label{Exact coexistence mechanisms}

Different expositions of MCT have given subtly different definitions of coexistence mechanisms. In some papers (e.g., this paper; \cite{barabas2018chesson}), coexistence mechanisms are derived via a Taylor series expansion of the function $g_j(E_j, C_j)$ (Section \ref{sec:Spatiotemporal coexistence mechanisms}). Because the truncated Taylor series only approximates the per capita growth rate, this approach results in coexistence mechanisms do not sum exactly to the invasion growth rate. We call these the \textit{small-noise coexistence mechanisms}, because their accuracy as approximations depend on the technical assumption of small environmental noise. 

In other papers (e.g., \cite{Chesson1994}), coexistence mechanisms are derived by first performing a coordinate transformation to the the so-called \textit{standard parameters}, which represent the main effects of the environment and competition on per capita growth rates. They are defined as $\mathscr{E}_j = g_j(E_j, C_j^*)$ and $\mathscr{C}_j = g_j(E_j^*, C_j)$, where $E_j^*$ and $C_j^*$ are the \textit{equilibrium parameters} (i.e., $g_j(E_j^*, C_j^*) = 1$). This approach results in coexistence mechanisms that sum exactly to invasion growth rate. Fittingly, we call these \textit{exact coexistence mechanisms}.

For reasons that are not entirely clear, authors have mixed small-noise coexistence mechanisms and exact coexistence mechanisms together in the same mathematical expression (e.g., \cite{barabas2018chesson}, Eq.19; \cite{Chesson1994}, Eq. 19--22). While the two types of coexistence mechanisms are equivalent in the limit of small environmental noise, users of MCT should be aware that they are distinct mathematical objects. A clear delineation between the small-noise and exact coexistence mechanisms is presented by \textcite{johnson2022coexistence}.

What are the pros and cons of the two types of coexistence mechanisms?  There are several situations in which small-noise coexistence mechanisms are useful. First, small-noise coexistence mechanisms are easier to compute, which may be a relevant consideration when computing coexistence mechanisms for many draws from a bootstrap or posterior distribution of model parameters. Second, small-noise coexistence mechanisms have proven useful in theoretical work: in simple models, the small-noise coexistence mechanisms sometimes permit analytical expressions of coexistence mechanisms. Finally, the small-noise coexistence mechanisms could be more interpretable than the exact coexistence mechanisms. For example, the small-noise storage effect contains the covariance term $\Cov{x,t}{E_j}{C_j}$, which succinctly captures the idea that a good environment can lead to high competition. The exact storage effect, on the other hand, contains higher-order Taylor series terms, such as $\E{x,t}{(E_j-E_j)^2(C_j-C_j)}$. If positive, this term implies that a small environmental fluctuation (whether negative or positive) leads to less competition than large environmental fluctuations. This relationship seems far less general and wholly separate from the idea that a good environment leads to high competition. In order words, the exact coexistence mechanism captures some growth that we cannot identify with the archetypal textual explanation for coexistence via the storage effect; it is in this sense that exact coexistence mechanisms may have less \textit{face validity} than small-noise coexistence mechanisms. 

Of course, the main boon of the exact coexistence mechanisms is that they sum exactly to the invasion growth rate. In practice, there can be a substantial difference between the true invasion growth rate and the sum of small-noise coexistence mechanisms (\cite{johnson2022methods}, Tables 1-4). We leave it to reader to decide which type of coexistence mechanism is more useful for their purposes. 

\subsection{The mutual invasibility criterion doesn't work}
\label{The mutual invasibility criterion doesn't work}

In our view, the greatest barrier to MCT's usage as an empirical tool is the unclear relationship between coexistence and invasion growth rates. The classic way to connect these constructs is the \textit{mutual invasibility criterion}, which states that coexistence occurs when each species (in a community of $S$ species) has a positive invasion growth rate in the context of the limiting dynamics of the $S-1$ resident species. But what if $S-1$ residents cannot persist in the absence of the invader species (say, because of mutualism or intransitive competition)? What if a species experiences a strong allee effect and has a negative invasion growth rate, despite being able to persist under normal conditions? These issues render the mutual invasibility criterion (in general) neither necessary nor sufficient for coexistence. One solution is to integrate MCT's partition of the invasion growth rate with the Hofbauer criterion for coexistence (\eqref{SPT}), which is a sufficient condition for a type of global stability. For reasons discussed in Section \ref{sec:The relationship between coexistence and invasion growth rates:Alternative coexistence criteria}, it is not obvious how this would be accomplished.

A temporary solution, though perhaps unsatisfactory, is to use invasion growth rates and coexistence mechanisms heuristically. That is, we may claim that the mechanisms by which a species invades one community (even if the community contains less than $S-1$ residents) are probably the same mechanisms which allow persistence in the full community. We may also claim that allee effects appear to be rare in the real world (\cite{myers1995population}; \cite{gregory2010limited}; \cite{sugeno2013semiparametric}), and that mutualisms, though common in the real world (\cite{risch1976ecologists}; \cite[p.~4]{bronstein2015mutualism}), are ritually ignored or abstracted away for tactical reasons (\cite[p.~164]{MittelbachGaryGeorge2019Ce}). Intransitive competition appears to be weak in real world (\cite{Godoy2017},  \cite{Friedman2017CommunityMicrocosms}), although there is a dearth of evidence overall. In light of these tenuous generalizations, it seems reasonable to say that invasion growth rates and coexistence mechanisms give \textit{some} indication about how species are coexisting. 

\subsection{Measuring the invasion growth rate in models with finite populations}
\label{Measuring the invasion growth rate in models with finite populations}

MCT assumes that species have varying densities, but an infinite number of individuals (see Section \ref{sec:The relationship between coexistence and invasion growth rates:Invasion growth rates}). Of course, infinite populations don't exist, but calculating the exact invasion growth rate in an approximate scenario (i.e., an infinite-population model) is equivalent to finding an approximate invasion growth rate in an actual scenario (i.e., a finite-population model). Infinite population dynamics will approximate finite-population dynamics when the invader's density is low enough to not affect any species' per capita growth rate, but high enough that demographic stochasticity can be ignored and that the probability of stochastic extinction is negligible. In such a scenario, the sign of the invasion growth rate will perfectly predict whether the invader will recover from rarity, at least in the medium-term (the "resident strikes back" is possible in the long-term; Fig. \ref{fig:phase}). However, it seems unlikely that environmental shocks/perturbations/disturbances will push the invader to this "goldilocks zone" of population density, but no farther. When perturbations severe enough for stochastic extinction to become a possibility, the invasion growth rate no longer perfectly predicts population recovery. Still, a positive invasion growth rate is necessary for a non-zero probability of recovery (\cite{Schreiber2011}). 

In models with finite populations, one can still measure an invasion growth rate by pretending that stochastic extinction is impossible. The general strategy is to perturb the invader to low density, and then wait for the ecological system to attain its limiting dynamics. Here, the "ecological system" includes all variables that may influence the invasion growth rate, including internal variables (e.g., resident densities, resource concentrations), external variables (e.g., temperature, density-independent fluctuations in fecundity) and the internal structure of the invader population (e.g., the quasi-steady spatial distribution, the stable age distribution), as discussed in Section \ref{sec:The relationship between coexistence and invasion growth rates:Invasion growth rates}. If the invader becomes too abundant, such that it significantly affects the residents' population dynamics on the regional scale, then the simulation is terminated and the whole process may be restarted if more simulation data is needed. If any of the species go extinct, then the simulation may be restarted with initial conditions equal to the state of a parallel simulation (this is known as the Fleming Viot Algorithm; \cite{ferrari2007quasi}; \cite{blanchet2016analysis}), or a past state from an archive of the current simulation (\cite{groisman2012simulation}). For more information on measuring invasion growth rates, including computational tricks for specific classes of models, see Section 4 in \textcite{johnson2022coexistence}.

\subsection{The probability of invasion increases monotonically with the invasion growth rate}
\label{The probability of invasion increases monotonically with the invasion growth rate}

The term "fitness" has had many uses (\cite[Ch.~10]{dawkins1982extended}), but contemporary use of the word --- whether in population genetics, ecology, or ethology --- is about predicting medium-term success. In the short-term, bad weather could be a temporary setback; this is why fitness is often measured as a temporal average or geometric mean in variable environments. In the long term, everything (e.g., populations, alleles, lineages) goes extinct; this is why the infinite-population assumption (Section \ref{sec:The relationship between coexistence and invasion growth rates:Invasion growth rates}) can be useful: it precludes stochastic extinction and thus allows an asymptotic analysis of coexistence. When stochastic extinction is a possibility, invasion is neither impossible nor certain. In this context, it makes sense to define fitness as the probability of fixation (\cite{proulx2002can}). Note that in ecology, fixation does not mean replacing other species, but rather hitting some high target population density. For our purposes, we may treat the \textit{probability of fixation} and the \textit{probability of invasion} as equivalent, since the two are nearly equal if invasion does not "reverse course" (i.e., the \textit{resident strikes back} phenomenon; \cite{Mylius2001TheAttractor}). To see this, consider $X$ individuals whose lineages each have probability $Q$ of eventual extinction. Because the individuals of a rare species are independent, the probability of fixation for the whole population is $1-Q^X$. This formula shows that once the population invades to a moderate $X$, fixation is nearly certain.

Interestingly, the probability of invasion is closely related to another common measure of invasion: the geometric mean of the finite rate of increase. For a single realization of a stochastic model, the geometric mean of the finite rate of increase is either zero or one in the medium-term: either the population goes extinct (the geometric mean is zero), or it attains its typical abundances and fluctuates around that zone of abundance for a very long time (the geometric mean is one). The expectation (across realizations of a stochastic process) of the medium-term geometric mean is thus equivalent to the probability of invasion. To our knowledge, this relationship has not yet been noted in the literature. 

How is the invasion growth rate related to the probability of invasion? To answer this question, we introduce a simple stochastic differential equation,
\begin{equation} \label{SDE}
    dN = A \: N \: dt + \sigma_D \: N \: dW_D + \sigma_E \: N^2 \: dW_E,
\end{equation}
where $N$ is population abundance, $A$ is the (arithmetic) mean per capita growth rate, $\sigma_D$ is the coefficient of demographic stochasticity, $\sigma_E$ is the coefficient of environmental stochasticity, $dt$ is the infinitesimal time-step, and finally, $dW_D$ and $dW_E$ are increments of independent Wiener processes. 

This simple SDE, with its constant coefficients, represents a linearization of population dynamics about some low abundance. Approximating the probability of fixation can be reduced to second-order linear differential equation (\cite[ch.~15.3]{karlin1981second}). In the supplementary \textit{Mathematica notebook}, {\fontfamily{qcr}\selectfont prob\_fix.nb} (\url{https://github.com/ejohnson6767/MCT_review}), we show that for a species at low density $N$, the probability of fixation is 

\begin{equation}
    Pr(\text{invasion}) \approx N \frac{2A - \sigma_E^2}{\sigma_D^2 + \sigma_E^2}. 
\end{equation}

The quantity $A-\sigma_E^2 / 2$ is precisely the geometric average growth rate (\cite{braumann2007ito}), which we know better as the invasion growth rate, $\E{t}{r}$.

Multiple studies (reviewed by \cite{lande2003stochastic}, Table 1.2) have found that $\sigma_D \gg \sigma_E$. Similarly, an analysis of the Global Population Dynamics Database found evidence that demographic stochasticity has detectable effects on large populations (\cite{reed2004relationship}). These studies have limitations (respectively, only a handful of species, simple models without observation error), but they suggest that we may approximate the probability of fixation as 
\begin{equation}
    Pr(\text{invasion}) \approx N \frac{2\E{t}{r}}{\sigma_D^2}.
\end{equation}
Here, we see that the probability of fixation from low density increases linearly with the invasion growth rate. For convenience we may call the quantity $2\E{t}{r} / \sigma_D^2$ the \textit{per capita probability of invasion}.

\subsection{Does the magnitude of invasion growth rates matter, or just the sign?} 
\label{Does the magnitude of invasion growth rates matter, or just the sign?} 

\textcite{pande2020mean} published a paper titled "Mean growth rate when rare is not a reliable metric for persistence of species", where they argued that the sign of the invasion growth rate, but not the
magnitude, is an important measure of persistence. To emphasize this point, they give examples where the mean time to extinction decreases as the invasion growth rate increases. 

We agree with \textcite{pande2020mean} that the magnitude of the invasion growth rate does not matter as much as the \textit{sign} of the invasion growth rate, since a positive invasion growth rate is a necessary condition for coexistence in the face of large perturbations (\cite{Schreiber2011}). However, as shown in the previous section, the magnitude of the invasion growth rate is not irrelevant. 

If one cares about whether a specific group of individuals will invade, then the magnitude of the invasion growth rate is a relevant metric. If one cares about whether an extirpated population will become reestablished eventually, then the sign of the invasion growth rate is the relevant metric. The unconditional probability of invasion in any given time-frame, or related measures such as the expected time to invasion, will depend on factors for which we do not usually have good data. For example, the time to fixation will depend on the rate of immigration. Less obviously, fixation depends on whether immigrating individuals arrive sequentially or in groups (\cite{haccou1996establishment}). 

\subsection{Does the magnitude of coexistence mechanisms matter?} 
\label{Does the magnitude of coexistence mechanisms matter?}

\textcite{pande2020mean} have claimed that "In the presence of stochastic environmental fluctuations, a contribution to [the invasion growth rate] is not necessarily a contribution to persistence. Therefore, one cannot quantify the contribution of a certain mechanism to persistence by comparing the value of [the invasion growth rate] in the presence and in the absence of this mechanism...". This claim was correctly rebutted by \textcite{Ellner2020}: the magnitude of the coexistence mechanisms does matter, since the magnitudes illustrate the degree to which different mechanisms contribute to the positivity of an invasion growth rate. For instance, if both relative nonlinearity is equal to 0.1 and the storage effect is equal to 0.2, then we would say both coexistence mechanisms are responsible for coexistence, but that the storage effect is more responsible. 

To add more nuance to the conversation, we comment that what really matters is the relative magnitude of the coexistence mechanisms, a quantity which can be obtained by dividing coexistence mechanisms by the absolute value of the invasion growth rate. This idea of "relative contributions to the sign of the invasion growth rates" suggests a novel way of calculating community average coexistence mechanisms (read on to Section \ref{Community-average coexistence mechanisms}).

A naive counterfactual or "but for" analysis (\cite{sep-causation-law}) is not an appropriate method for determining the importance of coexistence mechanisms. That is, it is not legitimate to say that the storage effect is the only important coexistence mechanism if it is the only mechanism whose absence would result in a negative invasion growth rate, nor is it legitimate to say that relative nonlinearity and the storage effect are necessarily equally important if the absence of either mechanism would result in a negative invasion growth rate. We want to know if coexistence mechanisms are \textit{causal factors}, and a causal factor is a \textit{Necessary Element of a Sufficient Set} (the NESS test for causation; \cite{wright1985causation}; also see \cite{mackie1980cement}). A set of several (positive) coexistence mechanisms constitute such a sufficient set (i.e., the invasion growth rate would be negative, had all the mechanisms in the set been zero), even if some mechanisms are strong enough to bring about coexistence on their own (see \cite[p.~1793--1795]{wright1985causation}). 

\subsection{Community-average coexistence mechanisms}
\label{Community-average coexistence mechanisms}

It is a common practice in MCT to divide each coexistence mechanism by the \textit{sensitivity to competition}, $\abs{\beta_i^{(1)}} = \abs{\pdv{g_i\scriptstyle{(E_i^*, C_i^*)}}{C_i}}$ (see \eqref{taylor_coef}). The rationale is that dividing by $\abs{\beta_i^{(1)}}$ is like correcting for between-species differences in population-dynamical speed (\cite{Chesson2018}), a variable that greatly modulates the values of coexistence mechanisms but is typically not relevant to coexistence . The correspondence between $\abs{\beta_i^{(1)}}$ and population-dynamical speed is not guaranteed, but it just so happens that $\abs{\beta_i^{(1)}}$ is the reciprocal of generation time in the two seminal models of MCT (the lottery model and the annual plant model; \cite{Chesson1994}).

This scaling enables a better comparison of species with slow and fast life-cycles. It is also often used in the calculation of \textit{community-average coexistence mechanisms} (\cite{chesson2003quantifying}; \cite{barabas2018chesson}). For instance, the community-average storage effect is usually defined as
\begin{equation} \label{averageSE}
    \overline{\left(\frac{\Delta I}{\beta^{(1)}} \right)} = \frac{1}{S} \sum_{i = 1}^{S} \frac{\Delta I_j}{\abs{\beta_i^{(1)}}}. 
\end{equation}
While this scaling of coexistence mechanisms is not without merit, it does not fully respect the fact the sign of the invasion growth rate is what determines eventual persistence (see Section \ref{Does the magnitude of invasion growth rates matter, or just the sign?}). Our suggestion is that when making cross-species comparisons or averages of coexistence mechanisms, all the coexistence mechanisms of species $i$ should be divided by the invasion growth rate, $\E{t}{r_i}$, to give the average relative importance of a mechanisms. For instance, $\Delta I_i / \abs{\E{t}{r_i}}$ is the quantity that best answers the question "What fraction of the ability of species $i$ to persist can be attributed to the storage effect?". With this scaling, the community-average storage effect becomes

\begin{equation}
    \overline{\left(\frac{\Delta I}{\abs{\E{t}{r}}} \right)} = \frac{1}{S} \sum_{i = 1}^{S} \frac{\Delta I_j}{\abs{\E{t}{r_i}}}.
\end{equation}

Community-average coexistence mechanisms tell us how different mechanisms benefit rare space \textit{in general}. If positive, the coexistence mechanism generally uplifts all species. If negative, the mechanism is generally destabilizing; in models without allee effects, mutualisms, or knock-on extinctions, a negative community-average coexistence mechanism is evidence of \textit{priority effects}, which are classically presented in a Lotka-Volterra framework  (\cite{amarasekare2012two}), but may be mediated through fluctuation-dependent coexistence mechanisms (\cite{chesson1988interactions}; \cite{schreiber2021positively}; \cite{schreiber2021temporally}). 

\subsection{Other measures of persistence/coexistence}
\label{Other measures of persistence/coexistence}

Invasion growth rates are the not the only way to measure stability/persistence, though we will argue that the alternatives are not empirically relevant. In the last couple decades, various authors have used measures of persistence/stability which are related to extinction times, including the mean time to extinction (MTE; \cite{grimm2004intrinsic}; \cite{Schreiber2018}), the quasi-potential (\cite{nolting2016balls}), and the scaled logarithm of the mean time to extinction (\cite{pande2020mean}; \cite{Dean2020Stochasticity-inducedSynthesis}). There are a number of reasons why these measures are attractive: 1) Theoretical work has shown that the mean time to extinction (MTE) should be long for established populations (\cite{lande1993risks}; \cite{faure2014quasi}; \cite{grimm2004intrinsic}; \cite{kamenev2008colored}; \cite{Meerson2011}). If we ignore the possibility of an early extinction for a population below its carrying capacity, the negligible length of the growth phase (the time it takes to reach carrying capacity) relative to the length of the stationary phase (the time it takes an established population to go extinct) means that effectively, MTE does not depend on initial conditions. 2) Directly operationalizing the folk-definition of coexistence --- "species co-occuring for a long time" --- would require us to define "a long time”. Such a straightforward approach has been used unabashedly in the population viability analysis literature (\cite{shaffer1981minimum}), but we imagine that defining a "long time" feels uncomfortably subjective to some. 3) The MTE can be calculated with well-known mathematical tools (\cite[ch.~15.3]{karlin1981second}, \cite{assaf2010extinction}; \cite{Chow2015}), thus circumventing the need for extensive simulations. 4) An asymptotic analysis allows the complete distribution of extinction times to be derived from the MTE (\cite[p.~253]{masoliver2018random}).

Although the MTE and related measures have led to insights in theoretical work, they empirically irrelevant because they cannot be estimated with any reasonable degree of accuracy. First, current estimates of model parameters will not be predictive of population dynamics on the long time-scales implied by MTE calculations (\cite{beissinger1998use}). Secondly, it is possible that many extinctions/extirpations are caused by catastrophes (\cite{Mangel1994}), which cannot be effectively modelled due to data limitations. There are empirical and theoretical reasons to believe that catastrophes --- also known as 'black swan' events (\cite{taleb2007black}; \cite{anderson2017black}) or events from a fat-tailed distribution --- are common on long time-scales but cannot be detected or estimated on short time-scales. A meta-analysis of the Global Population Dynamics Database estimated that only 3--5\% of animal populations experienced catastrophes, but that the probability of occurrence increased with time-series length. In addition, one can never know definitively that a distribution does not have fat-tails (\cite[ch.~3.4]{taleb2020statistical}), and showing that a distribution does have fat-tails can require thousands of data points (\cite{Weron2001}). 

\subsection{Variation is both stabilizing and destabilizing}
\label{Variation is both stabilizing and destabilizing}

Several papers have shown that the mean time to extinction (MTE) is a uni-modal hump-shaped function of environmental variability, with the maximum MTE attained at intermediate levels of variability (\cite{adler2008environmental}; \cite{pande2020mean}). This research demonstrates that variation has both stabilizing and destabilizing effects on persistence: increasing levels of variation can increase the fluctuation-dependent coexistence mechanisms ($\Delta I_i$, $\Delta N_i$, and $\Delta \kappa_i$). But, variation increases the magnitude of deviations from equilibrium, eventually increasing the probability of stochastic extinction.

These two forces were not synthesized mathematically until recently (\cite{Dean2020Stochasticity-inducedSynthesis}), most likely because they are most demonstrable in different types of models. Multi-species models of competition demonstrate that variability can enhance the invasion growth rate (\cite{armstrong1976coexistence}; \cite{chesson1981environmentalST}); single-species models often show the opposite phenomenon (\cite{Lewontin1969}; \cite{boyce1992population}). But single-species models demonstrate that variability increases the probability of stochastic extinction (\cite{lande1993risks}; \cite{kamenev2008colored}), whereas multi-species model are not analytically tractable in this context (\cite{assaf2017wkb}, Section VII).

Both temporal and spatial variability can enhance the invasion growth rate via the fluctuation-dependent coexistence mechanisms (of course, fluctuations can decrease the invasion growth rate too, depending on the particular model). However, unlike temporal environmental variability, spatial environmental variability tends to \textit{decrease} the probability of metapopulation extinction (\cite{matthews2007inflationary}; \cite{schreiber2010interactive}). To see this, note that environmental heterogeneity will (all else being equal) lead to a few high quality patches; and that, in the limit of small dispersal, the MTE of the metapopulation is determined by the MTE in the highest-quality patch (\cite{khasin2012minimizing}, Eq.14). As dispersal increases, the MTE smoothly transitions from this high MTE to a MTE corresponding to the spatial average of habitat quality: the exact same MTE that we would expect in the absence of environmental heterogeneity. Therefore, environmental heterogeneity leads to an elevated MTE, except in the limiting case where typical dispersal distances are much greater than the grain size of environmental variation.  

The persistence-hindering effects of spatial variation only emerge in the face of large disturbances:  occasional catastrophes that extirpate an entire patch, inducing classical metapopulation dynamics (\cite{levins1969some}).  If spatial variation is severe enough that certain patches are uninhabitable, then it becomes more difficult for populations to percolate across a landscape (\cite{fahrig2001much}; \cite{dewhirst2009dispersal}), thus reducing the effective recolonization rate and increasing the risk of metapopulation extinction. If, on the other hand, the primary risk to the metapopulation is a pathogen that spreads across the landscape, then environmental heterogeneity can reduce the risk of metapopulation extinction (\cite{hess1994conservation}).

\subsection{Autocorrelation is both stabilizing and destabilizing: the paradox of the plankton revisited}
\label{Autocorrelation is both stabilizing and destabilizing: the paradox of the plankton revisited}

Much like environmental variability, environmental autocorrelation has both negative and positive effects on persistence, such that the MTE is a humped-shaped function of autocorrelation. As autocorrelation increases, so too does the covariance between environment and competition, potentially leading to a large storage effect (\cite{li2016effects}; \cite{schreiber2021positively}). However, in the limit of strong autocorrelation, extinction is inevitable once the environment wanders into a regime where low-density growth rates are negative (\cite{kamenev2008colored}).

Environmental autocorrelation plays a crucial role in \posscite{hutchinson1961paradox} own solution to \textit{the paradox of the plankton of the plankton} (Table \ref{tab:paradox}). Hutchinson conjectured that if the time to competitive exclusion, $tc$, was much less than than the time to environmental change, $te$, then the community would behave as if in a constant environment, and the inferior competitor (in the context of the environmental variable's initial condition) is excluded. Conversely, Hutchinson argued that if $tc$ was much greater than $te$, then the community behaves as if at an equilibrium defined by the average environment (nonlinear averaging nonwithstanding). The idea is that individuals live longer than the timescale of environmental change, so they do not "feel" particular environmental states, but rather average over them. Here, there is still a single effective regulating factor, so the competitive exclusion principle for stochastic environments applies (\cite{chesson1997roles}; \cite{johnson2022coexistence}; \cite{hening2020competitive}). It is only when the timescales of competitive exclusion and environmental change are commensurable (i.e., $tc \approx te$) that coexistence can be attained. 

\begin{table}[h]
\caption{ Hutchinson's solution to the paradox of the plankton. $tc = $time to competitive exclusionl $te = $ time for significant change in the environment. The table is made in the likeness of Table 1 in \textcite{li2016effects}. The predictions are reproduced verbatim from \textcite{hutchinson1961paradox}, except for formatting and punctuation.}
\begin{tabular}{ p{0.2 \linewidth} p{0.4\linewidth} p{0.4\linewidth}}
\toprule
Scenarios & Prediction & Mechanism \\
\midrule
$tc \ll te$ & Competitive exclusion at equilibrium complete before the environment changes significantly &  Competitive exclusion in a functionally constant environment (thanks to time-scale separation; \cite{macarthur1970species}; \cite{gunawardena2014time}; \cite{kuehn2013mathematical}). Extinction occurs in finite-time due to finite population sizes. \\
$tc \approx te$ & No equilibrium achieved & The storage effect via environmental autocorrelation; \cite{li2016effects}; \cite{schreiber2021positively} \\
$tc \gg te$ & Competitive exclusion occurring in a
changing environment to the full range
of conditions to which individual competitors
would have to be adapted to
live alone & The environment changes so quickly that there is no covariance between environment and competition, and therefore there is only one effective regulating factor (e.g., nitrogen concentration); the competitive exclusion principle for stochastic environments (\cite{hening2020competitive}) applies here.  \\
\bottomrule
\end{tabular}
\label{tab:paradox}
\end{table}

\textcite{li2016effects} argue that Hutchinson's explanation for coexistence when $t_c \approx t_e$ can be attributed to the storage effect, but that the storage effect only increases as environmental autocorrelation (i.e., $te$) increases, and therefore, that Hutchinson's prediction of competitive exclusion in the case of $tc \ll te$ is incorrect. \textcite{li2016effects} and Chesson are correct that the storage effect increases monotonically with environmental autocorrelation, but they do not recognize the role of stochastic extinction. As discussed in Section \ref{sec:The relationship between coexistence and invasion growth rates:Invasion growth rates}, the formalism of MCT assumes infinite populations for which extinction is impossible; lengthy periods of bad conditions can always be offset "down the road". In reality, extinction becomes inevitable as environmental autocorrelation increases, even without competitive pressure (\cite{kamenev2008colored}). In other words, Hutchinson was right all along.

If we allow for more exotic population dynamics, temporal autocorrelation can have varied effects on coexistence and persistence (\cite{heino2000extinction}; \cite{ruokolainen2009ecological}; \cite{griffen2008review}). For example, in populations with overcompensatory dynamics, a positive temporally-autocorrelated environment (also known as red noise) can offset the negative growth that comes from overshooting carrying capacity. This generates a negative $EC$ covariance, which could lead to either a positive or negative storage effect a positive storage effect depending on the sign of the interaction effect (see Section \ref{sec:Interpreting coexistence mechanisms:The storage effect}). At the same time, the combination of overcompensatory dynamics and positive autocorrelation reduces extinction risk (\cite{ripa1996noise}; \cite{ripa1999linear}). It is hard to say how common overcompensatory dynamics are in the real-world: time-series analysis shows that real-world populations experience weak density-dependence (\cite{Ziebarth2010}; \cite{knape2012patterns}), but this may be an artifact of the spatial scale of observation being much larger than the spatial scale of intraspecific interactions, such that such that the statistical signal of density-dependence is lost via spatial averaging over weakly-coupled sub-populations (\cite{ray1996density}; \cite{thorson2015importance}).

In models with local dispersal (i.e., individuals disperse to nearby patches on a spatially explicit landscape), spatial autocorrelation increases the tendency of populations to build up in good environments, which promotes coexistence via the spatial storage effect and fitness-density covariance (see Section \ref{sec:Interpreting coexistence mechanisms:Fitness-density covariance}). More specifically, the buildup is attained if the spatial scale of environmental variation is greater than that of dispersal (\cite{snyder2003local}; \cite{snyder2004spatial}). Much like spatial variation \textit{per se}, spatial environmental autocorrelation can be thought of as creating isolated islands of favorable conditions. Therefore, the arguments in the previous Section (\ref{Variation is both stabilizing and destabilizing}) apply here: spatial autocorrelation has different effects on population persistence, depending on the context (e.g., the presence/absence of disturbances, infectious pathogens).

\subsection{Problems with empirical applications of MCT}
\label{Problems with empirical applications of MCT}

Increasingly, ecologists are attempting to quantify coexistence mechanisms in real communities by fitting and analyzing empirically-calibrated models. Based on our reading of the literature, there are some persistent problems with these empirical applications of MCT. 

\begin{enumerate}
    \item \textbf{No accounting for parameter uncertainty}. Most papers use a maximum likelihood framework, and only compute coexistence mechanisms corresponding to the maximum likelihood estimate (MLE). Uncertainty in parameter estimation must be propagated through to the level of coexistence mechanisms by calculating coexistence mechansims for many draws from a posterior or bootstrap distribution of model parameters. Uncertainty can be reported either through summary statistics (e.g. the standard deviation) or visualizations of the distribution of a coexistence mechanism. When information about uncertainty is not reported, it is nearly impossible for readers to determine whether estimated values of coexistence mechanisms reflect reality or sampling error. 
    
    Accounting for uncertainty can (potentially) completely alter one's inferences. Coexistence mechanism are often nonlinear functions of model parameters, so Jensen's inequality can be used to show that mean coexistence mechanisms (i.e., the coexistence mechanisms integrated over parameter space, with weights proportional to parameter probabilities) can be substantially different from the modal coexistence mechanisms (i.e., the coexistence mechanism computed with MLE or maximum a posteriori estimates). In extreme cases, the sign of a coexistence mechanism can flip from positive to negative (or vice versa). A Bayesian framework is the most natural way to compute the expected coexistence mechanisms, but bagging (\cite{hastie2009elements}) can be used in a MLE framework.
    
    \item \textbf{No accounting for model uncertainty.} All inferences are always conditional on a model, and history has shown that different models often lead to different inferences. The decades-long debate over the existence/ detectability of chaotic population dynamics is a consequence of different researchers using different models (\cite{hassell1976patterns}; \cite{ellner1995chaos}; \cite{rogers2021chaos}). Similarly, the strength / detectability of density dependence in the Global Population Dynamics Database depends on model structure (\cite{Sibly2007}; \cite{knape2012patterns}) and time-series length (\cite{hassell1989seeing}; \cite{Ziebarth2010}). Further, in the social sciences (which are not obviously more "messy" than ecology), several “many analyst, one dataset" studies consistently show high variance between the inferences of different researcher teams (\cite{klein2018many}; \cite{silberzahn2018many}; \cite{salganik2020measuring}; \cite{huntington2021influence}).

    It is not not enough to find the "best model(s)" with model selection or model averaging: with enough data, model weights based on cross-validation error or information criteria will give 100\% of the weight to a single model, but this is problematic because we know that the model is not "True" (i.e., all ecological models are \textit{M-open}; \cite[p.~385]{bernardo2000bayesian}) and we suspect that other models provide distinct information. Further, model selection and model averaging are only concerned with \textit{relative model fit} (\cite{mac2018model}), but due diligence requires \textit{absolute model fit}: that a model is plausible, compatible with previous research, performs well on common metrics of absolute fit when applicable (conditional $R^2$, goodness-of-fit, area under the curve (AUC)), and faithfully recreates important features of the data via \textit{graphical predictive checks} (\cite{gelman1996posterior}; \cite[ch.~6]{gelman2014bayesian}) and \textit{model-fitting diagnostics}. Model selection and model averaging can discourage hard thinking by giving a veneer of credibility to the best of a bad bunch.
    
    We recommend a two-part model-building strategy. In the first step, high-quality models are built built with iterative, continuous model expansion (\cite{box1980sampling}; \cite{draper1995assessment}; \cite{gelman2020bayesian}; \cite{gelman2020bayesian}). This step is repeated several times, using disparate model structures as the starting point. In the second step, inferences from several high-quality models (the end-point of iterative model expansion) are combined, either by simply discussing the differences between models, or by averaging some quantity of interest across models; for this, we recommend a simple average (each model gets equal weight; \cite{winkler1992sensitivity}; \cite{dormann2018model}) or weights based on Bayesian stacking (\cite{Yao2018}). 
    
    This model-building strategy above can be justified on frequentist grounds. It is well-known that the mean squared error of predictions can be decomposed into the sum of \textit{bias}, \textit{variance}, and irreducible error (\cite[p.~223]{hastie2009elements}). Using high-quality mechanistic / explanatory models reduces bias (\cite{shmueli2010explain}). Using models with disparate structures reduces covariation between models' predictions, which when combined with model averaging, reduces variance (\cite{dormann2018model}).
    
    \item \textbf{Models are too simple.} To reduce bias and increase credence in individual models, models should be as complex as the data allows; as the statistician Leonard Savage used to say, all models should be "as big as a house" (qtd in \cite{draper1995assessment}). One can build from a simple model to a complex model using \textit{iterative continuous model expansion} and a principled workflow (\cite{gelman2020bayesian}; \cite{gelman2020bayesian}). However, by using the data many times over, there is a danger of overfitting (\cite{ying2019overview}; \cite[p.~228]{hastie2009elements}). 
    
    Overfitting can be detected with k-fold cross-validation (\cite{hastie2009elements} and abated with \textit{regularization} (\cite{gelman2021most}), which is the general term for penalizing complexity in the parameter tuning process. Regularization manifests in many methods and model structures: the LASSO, ridge regression, prior distributions, horseshoe priors, hierarchical models, and more. Overfitting can often be avoided simply by avoiding a maximum likelihood framework (or any sort of optimization for that matter; \cite{cawley2010over}); In Bayesian inference, predictions are integrated over the posterior distribution of model parameters, so local maxima on the likelihood surface have less influence. 
    
\end{enumerate}

\subsection{Equalizing vs stabilizing mechanisms: two species theory}
\label{equalizing vs stabilizing mechanisms: two species theory}

There are two sides to MCT. One side of MCT (the primary focus of this paper) is concerned with partitioning invasion growth rates, usually into an intermediate number of terms called \textit{coexistence mechanisms}. The other side of MCT (the primary focus of this section) is concerned with explaining coexistence as the joint effect of equalizing and stabilizing mechanisms. Conceptually speaking, equalizing mechanisms reduce fitness differences between species, and stabilizing mechanisms increase niche differences between species. 

In actuality, the Equalizing vs. Stabilizing (EvS) paradigm can be further subdivided into two categories: a two-species framework and a multi-species framework. As shown by \textcite{song2019consequences}, these two frameworks deal with subtly different mathematical objects: the two-species framework is not a special case pf the multi-species framework. We review the two-species framework here and the multi-species framework in the next section.

\textcite{pianka1974niche} published a popular measure of niche overlap that was based on the products of two species' resource utilization functions. \textcite{chesson1990geometry} showed that Pianka's measure of niche overlap was a well-known measure of linear dependence: the square root of an $R^2$ statistic for a regression through the origin, where two species' resource utilizations were the predictor and response variables. Chesson also showed how the measure of niche overlap could be used to describe the conditions for coexistence. In the Lotka-Volterra model, 

\begin{equation}
    \frac{1}{N_i} \frac{dN_i}{dt} = b_j \left(1- \sum_{j = 1}^2 \alpha_{ij} N_j \right),
\end{equation}

The \textit{niche overlap} is denoted by $\rho$ and the \textit{fitness ratio} is denoted by $\kappa_1 / \kappa_2$ (Note: This notation is reserved with this section only, since it conflicts with the notation elsewhere in the paper). The quantities are defined as
\begin{equation}
    \rho = \sqrt{\frac{\alpha_{12} \alpha_{21}}{\alpha_{11} \alpha_{22}}}, \quad and
\end{equation}
\begin{equation}
    \frac{\kappa_1}{\kappa_2} = \sqrt{\frac{\alpha_{21} \alpha_{22}}{\alpha_{11} \alpha_{12}}}.
\end{equation}
The conditions for coexistence are described by the inequality, 

\begin{equation}
\rho <  \frac{\kappa_1}{\kappa_2} < \frac{1}{\rho}.
\end{equation}

The niche overlap $\rho$ is a measure of \textit{stabilizing mechanisms}, whereas $1/\abs{\log(\kappa_1 / \kappa_2)}$ is a measure of \textit{equalizing mechanisms}, i.e. equalizing mechanisms are large when $\kappa_1 / \kappa_2$ is close to 1. The two-species EvS paradigm has been used by many authors, both for theoretical work (e.g., \cite{chesson2008interaction}; \cite{Letten2017}) and empirical work (e.g., \cite{chu2015large}; \cite{kraft2015plant}). Yet, there are several problems with this paradigm, ranked in order from most problematic to least problematic.

\begin{enumerate}
    \item It only works in two-species models. In models with three or more species, one can defined analogous versions of $\rho$ and $\kappa_1 / \kappa_2$ (\cite{carroll2011niche}), but there is no simple inequality that describes the conditions for coexistence (\cite{saavedra2017structural}). 
    
    \item The EvS paradigm only works in Lotka-Volterra-like models. There is one caveat: one may derive pseudo competition coefficients by comparing a species' population density in the two-species community to its population density when it is the sole species (\cite{tilman1982resourceST}, Appendix; \cite{Letten2017}). We are unsure of how this workaround will perform (i.e. predict coexistence) in empirical applications, especially in models with fluctuation-dependent mechanisms.
    
    \item the terminology "average fitness ratio" can be misleading. It is well-known that ecological fitness --  also known as invasion fitness -- is the low-density growth rate, or more generally, the dominant Lyapunov exponent of a dynamical system (\cite{Metz1992}). While language is dynamic and semantics is context-dependent, the fear is that ecologists are mistaking the "average fitness ratio" for something that it is not.
    
    The "fitness ratio" $\kappa_1 / \kappa_2$ appears to be that which results from algebraically manipulating the conditions for coexistence so that $\rho$ is isolated. The definition for the fitness ratio suggests that the "fitness" of species $1$ is $\kappa_1 = \sqrt{\frac{\alpha_{21}}{\alpha_{11}}}$, which is the square root of the re-scaled invasion fitness of species $2$. We agree with \textcite{barabas2018chesson} that the term "competitive advantage" more accurately reflects the reality that $k_1$ is simply a ratio of species $1$'s competitive effects (on itself and species 2). 

    \textcite{Chesson2018} argues that the term "fitness" is justified because fitness typically predicts success (whether for individuals, groups, alleles, genotypes, or populations), and that $\kappa_j$ predicts success in a hypothetical world in which species share the same competition parameter (i.e., $C = \alpha_1 N_1 + \alpha_2 N_2)$. However, to attain this hypothetical world, the actual species-specific competition parameters must be partitioned in a particular way (Chesson 2018, eq. 18) that depends on knowledge of parameters ($\alpha_i$ and $\beta_j$) that would not be identifiable from data (statistically speaking;  \cite[p.~365]{gelman2014bayesian}).

\end{enumerate}

\textcite{song2019consequences} has criticized the two-species EvS paradigm on the grounds that the equalizing mechanisms and stabilizing mechanisms are interdependent. We applaud their analysis of how $\rho$ and $\kappa_1/\kappa_2$ jointly depend on the parameters of Macarthur's resource-consumer model, but this interdependence does not serve as a legitimate critique of the EvS paradigm. First, scientific abstractions are often interdependent but nonetheless meaningful. For example, ecology and evolution are interdependent, but it is still be meaningful to talk about ecology in isolation (as this paper hopefully demonstrates).

Second, it has long been known that $\rho$ and $\kappa_1/\kappa_2$ are interdependent at every level of mechanistic detail, so therefore, it is unlikely that the EvS paradigm is misleading ecologists in this way. At the most phenomenological level, both $\rho$ and $\kappa_1/\kappa_2$ depend on the same competition coefficients (\cite{chesson1990macarthur}). At a lower level of mechanistic detail, multiple competition coefficients depend on the same resource consumption rates (\cite{macarthur1970species}; \cite{abrams1998high}) or higher-order interactions (\cite{neill1974community}). At an even lower-level, it is safe to assume that multiple resource consumption rates depend on abiotic variables, like temperature. \textcite{song2019consequences} suggest that the graphical presentation of $\rho$ and $\kappa_1/\kappa_2$ as orthogonal axes of variation is evidence that the quantities are perceived as independent; but dependence or interdependence between variables is precisely the point of such 2D graphs (think of scatterplots and regression lines). 

\subsection{Equalizing vs stabilizing mechanisms: multi-species theory}
\label{equalizing vs stabilizing mechanisms: multi-species theory}

According to the multi-species EvS theory (\cite{chesson2003quantifying}), the stabilizing mechanisms are captured by the \textit{stabilizing term $A$}, which is simply the sum (across species) of re-scaled invasion growth rates. The equalizing mechanisms are captured by the \textit{average fitness difference $f_i$}, which is the difference between species $i$'s invasion growth rate and $A$. In symbols, the stabilizing term and the average fitness differences are respectively defined as

\begin{equation}
    A = \frac{1}{S} \sum_{i = 1}^S \frac{\E{t}{r_i}}{\abs{\beta_i^{(1)}}}, \quad \text{and}
\end{equation}

\begin{equation}
    f_i = \frac{1}{S} \frac{\E{t}{r_i}}{\abs{\beta_i^{(1)}}} - A.
\end{equation}

There are a few notable features about this framework: 1) The quantity $A$ is a community-level property, whereas $f_i$ is a species-level property. 2) $A$ and $f_i$ can be partitioned into contributions from individual coexistence mechanisms (see Eq.49--52 in \cite{barabas2018chesson}). 3) Unlike the two-species ES theory (Section \ref{equalizing vs stabilizing mechanisms: two species theory}), the term "fitness" is fitting here (if we ignore the division by $\abs{\beta_i^{(1)}}$). The $f_i$ is the difference between the invasion fitness of species $i$ and the average of fitnesses across the community, hence "average fitness difference". 4) The quantity $\E{t}{r_i} / \abs{\beta_i^{(1)}}$ is proportional to the \textit{per capita probability of fixation} (see Section \ref{The probability of invasion increases monotonically with the invasion growth rate}) after correcting for population-dynamical speed (see Section \ref{Community-average coexistence mechanisms}). While this is one way "\ldots to quantify average relative performance of the different species in the system." (\cite{chesson2003quantifying}), other measures are also possible depending on one's definition of "performance". For reasons given in Section \ref{Does the magnitude of invasion growth rates matter, or just the sign?} it may make sense to define $A = \sum_i sign\{\E{t}{r_i}\}$, and $f_i = sign\{\E{t}{r_i}\} - A$.

Although $A$ and $f_i$ can be interpreted as the average stabilizing force and the average fitness difference, these quantities do not have an extra predictive power: they do not predict coexistence (as in the two-species ES framekwork; Section \ref{equalizing vs stabilizing mechanisms: two species theory}) or persistence (as the invasion growth rate does; Section \ref{sec:The relationship between coexistence and invasion growth rates:Invasion growth rates}). Ultimately, $A$ and $f_i$ are just summary statistics that are useful for making within or between-community comparisons. There are other summary statistics (for instance the variance of $r_i$ across species) the could also be useful, depending on the context.

\subsection{Other theories/frameworks for understanding coexistence}
\label{Other theories/frameworks for understanding coexistence}

In addition to the equalizing vs. stabilizing mechanisms paradigm, there are several other frameworks for understanding coexistence. First, there is neutral theory (\cite{hubbell2001unified}) whose controversial assumption is that all species are demographically equivalent, and therefore, that they do not experience any stabilizing mechanisms. Neutral theory can be thought of as a special case of MCT where invasion growth rates are zero. 

It is arguable that neutral theory is not really a theory of coexistence. It was originally devised to predict the shape of species abundance distributions, and the crucial mechanism of neutral theory -- ecological drift -- generates the same predictions, regardless whether species richness is maintained by extinction-speciation balance (as in classic neutral theory; \cite{hubbell2001unified}), or by weak coexistence mechanisms (as in nearly neutral theory; \cite{zhou2008nearly}; \cite{lin2009demographic}; \cite{he2012coexistence}; \cite{kalyuzhny2015neutral}).

Metacommunity theory (\cite{leibold2004metacommunity}; \cite{leibold2017metacommunity}) is not a theory of coexistence per se, but rather a collection of paradigms (i.e., classes of models) that can be used as a mental springboard for thinking about community structure on broad spatial scales. \textcite{shoemaker2016linking} showed that in all of the non-neutral paradigms, fitness-density covariance and the spatial storage effect enabled coexistence (one exception is that the patch dynamics paradigm did not involve the spatial storage effect). This is not so surprising, since all non-neutral paradigms are qualitatively similar in the sense that they only differ in the degrees of spatial heterogeneity and local dispersal \cite{logue2011empirical}, Fig 1.), two features that readily generate fitness-density covariance and the spatial storage effect (see Section \ref{fig:Fitness-density covariance}).
 
The framework of Stochasticity-induced Stabilization (cite{Dean2020Stochasticity-inducedSynthesis}), claims that persistence ought to be measured by $A/\sigma_E^2$, where $A$ is the arithmetic temporal-mean of low-density per capita growth rates, and $\sigma_E^2$ is temporal-variance of the low-density per capita growth rate that can be attributed to environmental fluctuations (as in Section \ref{The probability of invasion increases monotonically with the invasion growth rate}, \eqref{SDE}). This measure of persistence is justified on the grounds that it increases monotonically with the mean time to extinction (MTE). While Stochasticity-induced Stabilization produces some interesting theoretical insights -- namely that the MTE can decrease as the invasion growth rate increases (\cite{adler2008environmental}; \cite{pande2020mean}) -- the quantity $A/\sigma_E^2$ is not empirically meaningful, since its main virtue is its correspondence with the MTE, which itself is not empirically meaningful (see Section \ref{Other measures of persistence/coexistence}). In addition, the measure $A/\sigma_E^2$ has only been shown to be relevant in one-dimensional diffusion approximations (or equivalently, one-dimensional stochastic differential equations), whereas the invasion growth rate does not have such limitations.

Tilman's (\citeyear{tilman1980resources}, \citeyear{tilman1982resourceST}) graphical theory of coexistence (also known as \textit{contemporary niche theory}) has been hugely influential. Originally limited to resource competition, this theory has been used to study the role of apparent competition, disturbances, environmental heterogeneity, and succession in coexistence (see \cite{chase2003ecological}). It has also been used to explain large-scale vegetation patterns (\cite{tilman1988plant}). Tilman's theory is limited in that it works best with two focal species (\cite{mcpeek2019mechanisms}), and that it can only informally demonstrate the roles of spatial and temporal variability in coexistence (e.g., \cite[ch.~6]{chase2003ecological}).

Finally, there are two frameworks based on the idea of \textit{structural stability}. The first framework (\cite{meszena2006competitive}; \cite{Song2020}) measures the robustness of an ecological system as the volume of parameter space that permits a stable equilibrium. Interestingly, the theory can be extended to include fluctuation-dependent mechanisms by including variances and covariance as regulating factors. The two major limitations of the theory is that it works for point equilibria, and that the relevance of \textit{robustness} -- the volume of parameter space permitting coexistence -- depends on information that we do not usually have: in theoretical studies where we have a model but no data, the probability of coexistence depends on how likely we are to observe a system with a particular set of parameters (see \cite[Section~4.2]{meszena2006competitive}); in empirical studies where we have estimates of model parameters, the relevancy of robustness depends on how parameters are expected to change in the future.

The second theory of structural stability (from \cite{saavedra2017structural}) is concerned with the volume of parameter space that permits \textit{feasibility}: the existence of a positive equilbirium (stable or not). Clearly, the focus on feasibility is a limitation of the theory, especially since feasibility isn't necessary for coexistence in the sense of \textit{local stability} (see Table \ref{tab:co_def}, row: \textit{Positive attractor}). 

\subsection{The scale-dependence of coexistence mechanisms}
\label{The scale-dependence of coexistence mechanisms}

The linear density-dependent effects, $\Delta \rho_j$, captures the fluctuation-dependent coexistence mechanisms ($\Delta N_i$, $\Delta I_i$, and $\Delta \kappa_i$) that operate on finer-grained time-scales and spatial scales than that of  data collection (\cite{chesson1993temporal}). To see why, consider a hypothetical model in which the spatial storage effect promotes coexistence. Say that the model is "coarse-grained" by averaging the fine-grained environmental parameter $E_j$ over $K_x$ microsites in patch $x$, with the microsites indexed by $w \in x$. If the size of patch $x$ is much larger than the scale of spatial variation, then the coarse-grained environmental parameter, $E_j'(x,t) = (1/K_x)\sum_{w \in x}^{K_x} E_j(w,t)$, will converge to $\E{w}{E_j}$, the spatial average of $E_j$ across all microsites. Therefore, there will be near-zero spatial variation in the coarse-grained environmental parameter, and consequently, a near-zero value of the coarse-grained storage effect. However, the effects of the fine-grained spatial storage effect -- a positive invasion growth -- will be phenomenologically captured by $\Delta \rho_i$. This same idea is true for other fluctuation-dependent mechanisms and for coarse-graining across time instead of space.

Of course, to "coarse-graining" a model by taking the spatial average of the environmental parameter would result in an unnecessarily loss of information. However, it is not implausible that an ecologist would measure $E_j$ (perhaps soil clay content) at one location in space, and then measure $C_j$ (perhaps plant density) a dozen centimeters away. If the environment changes significantly across this distance, then there will be no measurable $EC$ covariance despite an underlying spatial storage effect. Again, this fluctuation-dependent mechanisms may be captured by $\Delta \rho_i$.

With the exception of $\Delta E_i$, all coexistence mechanisms involve a negative feedback loop with population density (Fig. \ref{fig:Density independent effects}, \ref{fig:Linear density-dependent effects}, \ref{fig:Relative nonlinearity}, \ref{fig:The storage effect}, \ref{fig:Fitness-density covariance}). From this perspective, it is not so surprising that $\Delta \rho$, the \textit{linear density-dependent effects}, will also pick up on latent fluctuation-dependent mechanisms. That is not to say that phenomenological density-dependent models (e.g., \cite{ives2003estimating}) capture \textit{all} stabilizing mechanisms. Model misspecification goes hand-in-hand with estimation error, so one ought err on the side of building models that are complex/mechanistic (see Section \ref{Problems with empirical applications of MCT}).

\section{Conclusions}
\label{Conclusions}

Modern Coexistence Theory is a diverse body of work that is constantly changing through revision and debate. It was originally devised to generate theoretical insights (\cite[p.~288]{barabas2018chesson}, \cite[p.~6]{Chesson2019}), but it is now also a measurement tool (\cite{ellner2016quantify}; \cite{Ellner2019}; \cite{johnson2022coexistence}). It was originally based on the assumption of small environmental noise, but now exact coexistence mechanisms can be computed (Section \ref{Exact coexistence mechanisms}). The scaling factors, a seminal aspect of MCT, have been shown to be detrimental in certain contexts (Section \ref{Scaling factors}). The equalizing vs stabilizing mechanisms framework has been used by many authors, but there is a growing appreciation that this framework is problematic (Section \ref{equalizing vs stabilizing mechanisms: two species theory}). The storage effect has been conflated with bet-hedging strategies via dormancy or a robust life stage, but this is demonstrably false. Similarly, it has been shown that "buffering" typically hurts rare species (Section \ref{sec:Interpreting coexistence mechanisms:The storage effect}). Relative nonlinearity was once thought to be an impotent coexistence mechanism, but a more thorough review of the literature shows that no such generalization can be made (Section \ref{sec:Interpreting coexistence mechanisms:Relative nonlinearity}). The mutual invasibility criterion for coexistence, once the cutting edge in empirical research (\cite{Siepielski2010}), is now being recognized as insufficient (Section \ref{sec:The relationship between coexistence and invasion growth rates:Inferring coexistence from invasion growth rates: The mutual invasibility criterion}); new coexistence criteria will need to take the reins (Section \ref{sec:The relationship between coexistence and invasion growth rates:Alternative coexistence criteria}). There remain other topics that we have not covered here, such as the evolution of coexistence mechanisms (\cite{snyder2011coexistence}; \cite{snyder2011coexistence}; \cite{abrams2013evolution}; \cite{miller2017evolutionary}; \cite{yamamichi2021rapid}) and the effects of intraspecific variation (\cite{hart2016variation}; \cite{stump2021synthesizing}).

MCT's greatest strength is its generality: with sufficient effort, invasion growth rates can be partitioned into coexistence mechanisms in almost any model. The small set of coexistence mechanisms help us create a hierarchical taxonomy of explanations for coexistence (see Fig. \ref{fig:edifice}) and highlights the subtle similarities between seemingly disparate models. That being said, MCT is not a panacea. To properly understand coexistence and interpret coexistence mechanisms, particular models must be studied (or likened to models that are already well-studied). For example, relative nonlinearity is driven by exogenous environmental variation in the lottery model (\cite{Chesson1994}), but endogenous population cycles in the Armstrong-McGehee model (\citeyear{armstrong1980competitive}); and this distinction determines whether all species can coexist (Section \ref{fig:Relative nonlinearity}, Table \ref{tab:relative nonlinearity}). There are no shortcuts to understanding coexistence, though MCT may help us organize our thoughts.

\section{Acknowledgements}
We would like to thank Simon Stump and Sebastian Schreiber for discussions; and Logan Brissette for copy editing. This research is supported in part by NSF Grant DMS – 1817124 Metacommunity Dynamics: Integrating Local Dynamics, Stochasticity and Connectivity.

\section{Appendixes}

\subsection{The spatial storage effect vs. fitness-density covariance}
\label{The spatial storage effect vs. fitness-density covariance}

Consider a community with scalar populations inhabiting discrete patches (indexed by $x$), with discrete-time dynamics (indexed by $t$). In each time-step, there are two events. First is a bout of local population growth, determined by the local finite rate of increase, $\lambda_j(x,t)$. Second is a dispersal event, where in each patch a proportion of individuals, $p_j$, disperse and are distributed uniformly over all patches (including the patch of origin). It follows that a proportion of individuals, $q_j = (1-p_j)$, are retained locally; We call $q$ the \textit{retention proportion}. To simplify the expressions for coexistence mechanisms, we assume that there is no temporal variation, and that population densities $N_j$ and relative densities $\nu_j$ settle to an equilibrium in each patch. The time-evolution of population density $N_j$ at patch $x$ is given by 

\begin{equation} \label{dynamics}
  N_j(x,t+1) =  q_j N_j(x,t) \lambda_j(x,t) + \frac{1 - q_j}{K} \sum \limits_{s=1}^{K} N_j(s,t) \lambda_j(s,t).
\end{equation}

Often in MCT, the competition parameter is a function of both species densities and the environmental parameter; However, in such models, there is an implicit time-lag between the effects of environment and competition on population dynamics, such that the environment has enough time to affect competition within a time-step. For instance, in the lottery model, per capita fecundity (the environmental parameter) affects the per larva recruitment probability (the competition parameter), but recruitment occurs weeks or months after reproduction, a fact which is hidden by the simple structure of the lottery model equations. Here, we only which can be approximated by expanding the competition parameter. In this appendix, we only consider models in which the competition parameter is a function of a single residents $s$'s population density: $C_j = h_j(N_s)$. This simplifies things because it prevents the environment from affecting competition on two separate time-scales - within a time-step (as in the lottery model) and between time-steps (via inter-generational population growth).

To obtain the spatial storage effect, we must obtain the quantity $\Cov{x}{E_j}{C_j}$, which in the two-species/single-resident case can be approximated as $\Cov{x}{E_j}{\theta_{jr} N_s}$, where $\theta_{jr}$ is is a constant that converts species $s$'s density to species $j$'s competition: $\theta_{jr} = \frac{d h_j(N_s^*)}{d N_s}$.
Elsewhere (\cite{johnson2022coexistence}), we have shown that fitness-density covariance can be approximated as 
\begin{equation}
\begin{aligned}
\E{t}{\Cov{x}{v_j}{\lambda_j}} \approx = & \frac{q_j}{1-q_j} \Var{x}{\alpha_j^{(1)} (E_j - E_{j}^{*}) + \beta_j^{(1)} (C_s - C_{s}^{*})},
\end{aligned}
\end{equation}
which --- again assuming that the competition parameter is a function of only the resident's density --- can be approximated as
\begin{equation}
\begin{aligned}
\E{t}{\Cov{x}{v_j}{\lambda_j}} \approx \frac{q_j}{1-q_j} \left[ \left(\alpha_j^{(1)}\right)^2 \Var{x}{E_j} + 2 \alpha_j^{(1)} \beta_j^{(1)} \Cov{x}{E_j}{\theta_{jr} N_s} + \left(\beta_j^{(1)}\right)^2 \Var{x}{\theta_{jr} N_s} \right]. 
\end{aligned}
\end{equation}

Now, it is clear that simplifying the expressions for the coexistence mechanisms will require us to find the residents' density, $N_s$. Specifically, re-expressing $N_s$ in terms of the environmental parameter, $E_s$, will allow us to express the coexistence mechanisms in terms of spatial variation and between-species correlation in the environment.

To find $N_s$, we take a perturbative approach, where both $N_j$ and $\lambda_j$ are expanded in powers of the small parameter $\sigma$: $N_{j}(x,t) = N_{j,0}(x,t) + \sigma N_{j,1}(x,t) + ...$; and $\lambda_{j}(x,t) = \lambda_{j,0}(x,t) + \sigma \lambda_{j,1}(x,t) + ...$. With this, we re-write the population dynamics (\eqref{dynamics}) as

\begin{equation} 
\begin{aligned} \label{order1}
  N_{j,0}(x,t+1) + \sigma N_{j,1}(x,t+1) + \ldots = &  q_j (N_{j,0}(x,t) + \sigma N_{j,1}(x,t) + \ldots )(\lambda_{j,0}(x,t) + \sigma \lambda_{j,1}(x,t) + \ldots )  \\
  & + \frac{1 - q_j}{K} \sum \limits_{s=1}^{K}  (N_{j,0}(s,t) + \sigma N_{j,1}(s,t) + \ldots )(\lambda_{j,0}(s,t) + \sigma \lambda_{j,1}(s,t) + \ldots ).
\end{aligned}
\end{equation}

The zeroth-order dynamics are the same in every patch (because environmental fluctuations are $\mathcal{O}(\sigma)$). Assuming that there are no complex dynamics, the resident density reaches a stable equilibrium, denoted $N_{s,0}^*$, that is the same in each patch and which can be obtained by solving
\begin{equation}
    N_s = N_s \, g_{s}(E_s^*, h_s(N_s))
\end{equation}
for $N_s$. For example, if the population model is $\lambda_s(x,t) = s + E(x)/(1+c N_s(x,t))$, then $N_{s,0}^* = \frac{E_s^*}{c(1-s)} - \frac{1}{c}$.

Noting that $\lambda_{j,0}^* = g_{j}(E_j^*, h_j(N_{s,0}^*)) = 1$, the first-order dynamics can be written as
\begin{equation} 
\begin{aligned}
 \sigma N_{j,1}(x,t+1) = & \sigma \left[ q_j \left(N_{j,0}^* \lambda_{j,1}(x,t) + N_{j,1}(x,t) \right) + (1-q_j)\left(N_{j,0}^* \E{x}{\lambda_{j,1}(t)} + \E{x}{N_{j,1}(t)} \right) \right].
\end{aligned}
\end{equation}
MCT is based on small-noise assumptions (details in \cite{Chesson1994}; \cite{chesson2000general}) that ensure that all terms in the mathematical expression of the invasion growth rate are of commensurable magnitude. Specifically, MCT assumes that environmental fluctuations are small and that the average of fluctuations is even smaller; Or put into symbols, $E_j-E_j^* = \mathcal{O}(\sigma)$ and $\E{x,t}{E_j-E_j^*} = \mathcal{O}(\sigma^2)$. Analogous bounds can be put on population density, relative density, and the competition parameter, as all of these a ultimately functions of the environment. These small-noise assumptions can be used to match terms from the perturbative expansion above and the Taylor series expansion of $\lambda_j$ (see \eqref{taylor_decomp}), resulting in $\sigma \lambda_{j,1}(x,t) = \alpha_j^{(1)} (E_j(x,t) - E_{j}^{*}) + \beta_j^{(1)} (C_j(x,t) - C_{j}^{*})$. The small-noise assumptions also mean that $\E{x}{\lambda_{j,1}(t)}$ and $\sigma \E{x}{N_{j,1}}$ are $\mathcal{O}(\sigma^2)$, which simplifies \eqref{order1} to 

\begin{equation} 
\begin{aligned}
  \sigma N_{j,1}(x,t+1) = \sigma \left[ q_j \left(N_{j,0}^* \lambda_{j,1}(x,t) + N_{j,1}(x,t) \right) \right].
\end{aligned}
\end{equation}

Using the relationship $\sigma \lambda_{j,1}(x,t) = \alpha_j^{(1)} (E_j(x,t) - E_{j}^{*}) + \beta_j^{(1)} (C_j(x,t) - C_{j}^{*}) \approx \alpha_j^{(1)} (E_j(x,t) - E_{j}^{*}) + \beta_j^{(1)} \theta_{jr} N_{s,1})$ to substitute for $\lambda_{j,1}(x,t)$, we can solve for the equilibrium density of the resident

\begin{equation}
    N_{s}^* \approx N_{s,0}^* + \sigma N_{s,1}^*, \; \text{where}
\end{equation}
\begin{equation}
   \sigma N_{s,1}^* = \frac{q_s N_{s,0}^* \alpha_s^{(1)}(E_s - E_s^*)}{1-q_s(\theta_{ss} N_{s,0}^* \beta_s^{(1)} +1)}.
\end{equation}

Plugging the above expression into the formulas for coexistence mechanisms (\eqref{dE} - \eqref{dkappa}), and writing the covariance between the two species' environmental responses as $\phi \sigma_1 \sigma_2$ (where $\phi$ is the correlation coefficient), we find that the spatial storage effect is 
\begin{equation}
    \Delta I_i = \frac{q_s N_{s,0}^* \alpha_s^{(1)}}{1-q_s(\theta_{ss} N_{s,0}^* \beta_s^{(1)} +1)} \left[ \zeta_i \phi \sigma_i \sigma_s - \zeta_s \sigma_s^2 \right], 
\end{equation}
and fitness-density covariance is  

\begin{equation}
\begin{aligned}
    \Delta \kappa_i = & \frac{q_i}{1-q_i} \left[\left(\alpha_i^{(1)}\right)^2 \sigma_i^2 + 
    \frac{2 \alpha_i^{(1)} \alpha_s^{(1)} \theta_{ir} \beta_i^{(1)} \phi \sigma_i \sigma_s q_s N_{s,0}^*}{1-q_s(\theta_{ss} N_{s,0}^* \beta_s^{(1)} +1)} + 
    \left( \frac{\alpha_s^{(1)} \theta_{ir} \beta_i^{(1)} \sigma_s q_s N_{s,0}^*}{1-q_s(\theta_{ss} N_{s,0}^* \beta_s^{(1)} +1)} \right)^2 \right] \\
    - &\frac{q_s}{1-q_s} \left[\left(\alpha_s^{(1)}\right)^2 \sigma_s^2 + 
    \frac{2 \left(\alpha_s^{(1)}\right)^2 \theta_{ss} \beta_s^{(1)} \sigma_s^2 q_s N_{s,0}^*}{1-q_s(\theta_{ss} N_{s,0}^* \beta_s^{(1)} +1)} + 
    \left( \frac{\alpha_s^{(1)} \theta_{ss} \beta_s^{(1)} \sigma_s q_s N_{s,0}^*}{1-q_s(\theta_{ss} N_{s,0}^* \beta_s^{(1)} +1)} \right)^2 \right].
\end{aligned}
\end{equation}

Here, we can see that the invader's dispersal dynamics (i.e., the value of $q_i$) does not play a role in the spatial storage effect, but does play a role in fitness-density covariance. The resident's dispersal dynamics, on the other hand, play a role in both mechanisms. 

If the invader and resident have identical demographic parameters but partially uncorrelated environmental responses, then we can drop the species-specific subscripts, and the coexistence mechanisms simplify to

\begin{equation}
    \Delta I = \frac{q N_{0}^* \alpha^{(1)}}{1-q(\theta N_{0}^* \beta^{(1)} +1)} \left[ \zeta \sigma^2 (\phi - 1) \right], \text{and}
\end{equation}

\begin{equation}
\begin{aligned}
    \Delta \kappa = & \frac{q N_{0}^* \alpha^{(1)}}{1-q(\theta N_{0}^* \beta^{(1)} +1)} \left[\frac{q}{1-q}  2 \alpha^{(1)} \theta \beta^{(1)} \sigma^2 (\phi - 1)\right]
\end{aligned}
\end{equation}

\subsection{Spatial variation in resource supply promotes coexistence}
\label{Spatial variation in resource supply promotes coexistence}

Here, we analyze a 2-consumer, 1-resource model in which the two consumers exhibit an opportunist-gleaner trade-off. The model can also be described as a discrete-time approximation of a continuous-time chemostat model. The local finite rate of increase for the consumer is
\begin{equation}
    \lambda_j(x,t) = 1 + \left(\frac{w \mu_j R(x,t)}{K_j + R(x,t)} - d\right) \Delta t, \quad j = (1,2), 
\end{equation}
where $w$ is an efficiency constant (converts resource uptake to consumer biomass), $\mu_j$ is the resource maximum uptake rate, $K_j$ is the half-saturation constant, $d$ is the dilution/death rate, $\Delta t$ is the length of a time-step, and $R(x,t)$ is the concentration of the resource at location $x$ and time $t$.

Each time-step is split into two phases: growth and dispersal. Growth follows the equations above. The dispersal phase can be described as \textit{local retention with global dispersal}, and follows \eqref{dynamics} in Appendix \ref{The spatial storage effect vs. fitness-density covariance}. 

The resource dynamics are given by the equation,
\begin{equation} \label{resource dynamics}
    R(x,t+1) = R(x,t) + \left(d(S(x)-R(x,t))- \sum_{j=1}^{2} \frac{ \mu_j N_j(x,t) R(x,t)}{K_j + R(x,t)} \right) \Delta t,
\end{equation}
where $S(x)$ is the patch-specific \textit{resource supply point}. Here, $S(x)$ is effectively the environmental parameter, so $S(x)-S^* = \mathcal{O}(\sigma)$. Resources do not disperse. 

First we verify a claim in the main text: coexistence via spatial relative nonlinearity is not possible if there is complete local retention. With no dispersal, we can straightforwardly solve for the equilibrium resource concentration. When there is only a single resident $s$,
\begin{equation}
    R^* = \frac{d K_s}{\mu_s w - d}
\end{equation}
in each and every patch. There is no spatial variation in resource concentration, so there can be no spatial relative nonlinearity (assuming that local populations reach equilibrium and do no experience endogenously-driven population cycles).

As discussed in Section \ref{sec:Interpreting coexistence mechanisms:Relative nonlinearity}, coexistence is possible if the gleaner species tends to increase resource variation, compared to the opportunist. The opportunist-gleaner continuum is controlled by $\mu_j$ and $K_j$, with higher parameter values corresponding to more opportunism. We can fix the equilibrium resource concentration at some arbitrary value $R_0^*$, so that both species are competitively equivalent in the absence of spatial resource variation (via \posscite{tilman1982resourceST} $R^*$ rule). We can then solve for $\mu_j$, 
\begin{equation}
    \mu_j = \frac{d(K_j +R_0^*)}{R_0^* w},
\end{equation}
and substitute the right-hand-side into the dynamical equations. With this substitution, $K_j$ becomes the only parameter that controls the degree of opportunism, but modulating $K_j$ does not change the (equilibrium) competitive equivalence of species.

We will now calculate the resource variation in the case of a single-resident, using the same perturbative approach as in Appendix \ref{The spatial storage effect vs. fitness-density covariance}. Writing the dynamics of resource concentration (i.e., the right-hand-side of \eqref{resource dynamics}) as the function $\Phi(R(x,t), N_s(x,t), S(x))$, the first-order dynamics are

\begin{equation} 
\begin{aligned}
\sigma N_{s,1}(x,t+1) = &  \sigma q_s \left(N_{s,0}^* \frac{d \lambda_s(R^*)}{d R} R_1+ N_{s,1}(x,t) \right), \; \text{and}
\end{aligned}
\end{equation}

\begin{equation} 
\begin{aligned}
 \sigma R_{1}(x,t+1) = &  \frac{d \Phi(R_0^*,N_{s,0}^*,S^*)}{d R} \sigma R_1(x,t) + \frac{d \Phi(R_0^*,N_{s,0}^*,S^*)}{d R} \sigma N_{s,1}(x,t) + \frac{d \Phi(R_0^*,N_{s,0}^*,S^*)}{d S} (S(x)-S^*).
\end{aligned}
\end{equation}

After substituting in the Taylor series coefficients, we can solve for the equilibrium resource concentration:

\begin{equation}
    R_{s}^* \approx R_{0}^* + \sigma R_{s,1}^*, \; \text{where}
\end{equation}

\begin{equation}
    R_{s,1}^* = \left(S(x) - S^*\right) \frac{ (1-q_s) R_0^* (K_s+R_0^{*})}{d K_s q_s (R_0^{*}-S^*)+(q_s-1) \left(K_s S^{*}+{R_0^{*}}^2\right)}.
\end{equation}

Recall that $R_0^*$ is fixed. Using the fact that $\Var{x}{R_s} \approx \E{x}{R_{s,1}^2}$ and $\Var{x}{S} = \sigma^2$, we find that resource variation is 

\begin{equation} \label{resource var}
   \Var{x}{R_s} \approx \sigma^2 \left[\frac{ (1-q_s) R_0^* (K_s+R_0^{*})}{d K_s q_s (R_0^{*}-S^*)+(q_s-1) \left(K_s S^{*}+{R_0^{*}}^2\right)}\right]^2.
\end{equation}

In the \textit{Mathematica notebook}, {\fontfamily{qcr}\selectfont spatial\_DeltaN.nb} \url{https://github.com/ejohnson6767/MCT_review}, we prove that the resource variation decreases monotonically with the parameter $K_s$. This result confirms our earlier claim that gleaner species increase resource variation, compared to opportunist species. In the \textit{R script}  {\fontfamily{qcr}\selectfont spatial\_opportunist\_gleaner\_sims.R} (\url{https://github.com/ejohnson6767/MCT_review}), we provide a simulation example to show species can indeed coexist.

\section{References}
\printbibliography
\end{document}

\section{Recycling}

he determination of mechanisms that permits coexistence of species, both theoretically and em-5pirically, has been a central question in ecology. Since these explanations for coexistence (being6reductive by definition) are certainlypartial explanationsmore than one is operating in any situ-7ation and the real challenge is to determine the relative importance of any particular explanation8under particular circumstances. Here, we present a methodology for quantifying the relative im-9portance of different explanations for coexistence, based on an extension ofModern Coexistence10Theory. Current versions of Modern Coexistence Theory only allow for the analysis of communi-11ties that are affected by spatialortemporal environmental variation, but not both. We show how12to analyze communities with spatiotemporal fluctuations, how to parse the importance of spatial13variation and temporal variation in promoting coexistence, and how to measure everything with14either mathematical expressions or simulation experiments. This approach shows how to determine15the relative importance of spatial versus temporal explanations for coexistence rather than having16simply to attribute coexistence to one or the other

\subsection{Relative nonlinearity in models with one resource, two consumers, and an opportunist-gleaner trade-off (NOTE: incomplete)}
\label{app:Relative nonlinearity in models with one resource, two consumers, and an opportunist-gleaner trade-off}

To better understand the processes by which relative nonlinearity can promote coexistence, we will start simple. Consider a community with two consumers and one resource (fig??). Both consumer species' share a density-independent death rate (solid lines), but have different birth rates (solid lines) which are saturating functions of resource concentration. In the absence of fluctuations, species 1 will exclude species 2 - Resource concentrations will decrease as both consumers' populations grow, until the resource concentration dips below the minimum resource requirement of species 2. This is nothing more than a verbal explanation of Tilman's (1982) $R^*$ rule. What happens when we allow resource levels to fluctuate? Species 1 (the winner in the no-fluctuations-case) will be harmed more than species 2 by resource variation, due to Jensen's inequality and the relatively large concavity of its birth rate function. Is this enough to allow species 2 to coexist?

To find out, we will utilize a heuristic that originates from Tilman's graphical analysis of resource consumer models: For coexistence to occur, species must consume proportionately more of that which most limits their own growth. In our example (fig \ref{??}) Species 1 is hurt more by resource variation, so it is most limited by mean resource levels. Thus, for species 2 to coexist (i.e., invade), species 1 must \textit{increase} variation in resource levels when it is abundant. 

Species 1 can increase variation (and thus promote coexistence) by inducing cyclical resource-consumer dynamics (Armstrong and Mcgehee 1976 1980); this outcome is contingent upon model parameters, but it is not a quirk of meticulously selected parameters - species 1 has a faster resource consumption rate than species 2, and is therefore inherently more destabilizing (cite). That being said, cyclical and chaotic population dynamics appear to be rare in the real world (cite).

If resource dynamics are subject to environmental stochasticity or demographic stochasticity, then resource variation should scale monotonically with mean resource levels (lande and engen; gardiner). Here, since species 1 has a lower $R^*$ than species 2, species 1 tends to decrease resource variation, thus undermining coexistence. There is some reason to believe that this outcome is common in the real-world - Taylor's law (cite) suggests that the aforementioned relationship between the mean and variance is common, at least for biotic resources. 

The previous example of small environmental noise explicates Chesson's statement that "the limited ability for relative nonlinearity to promote coexistence when acting alone means that it is best viewed as modifying other mechanisms ... by decreasing the degree of dominance of a superior competitor with a relatively concave growth rate ... " (\cite{chesson2000general}) - When resource variation originates from environmental or demographic stochasticity, the dominant species does not tend to increase resource variation, so relative nonlinearity by itself does not promote coexistence. Relative nonlinearity is nevertheless important because it can change competitive outcomes (e.g., if resource variation is severe enough, then species 2 will exclude species 1).

Resource variation may also originate from variation in resource inputs. In this class of scenarios, the source of resource variation is exogenous, but (unlike the case of environmental or demographic stochasticity) there is no hard-wired relationship between mean resource levels and resource variation. When resource inputs fluctuate through time, we find that the dominant competitor (species 1) tends to increase resource variation, thus promoting coexistence. The reason for the increase is related to the different slopes the two consumer's birth rate curves around equilibrium resource levels (i.e., $R^{*}_1$ and $R^{*}_2$; see fig \ref{??}). If the slope is steep, a resource surplus causes a dramatic increase in consumer birth rates; The subsequently large consumer population then reduces resource levels. Conversely, when resources are scare, birth rates will dramatically decline, leading to a small consumer population which is ineffective at depleting resources further. Therefore, resource levels are regulated via a negative feedback loop with consumers (fig \ref{??}), and the strength of this negative feedback is proportional to the slope of the birth rate function. When the dominant competitor is a 'gleaner' species (which is necessary if there is to be any trade-off between the two consumers), it necessarily has a shallower slope.

Rapid fluctuations in resource inputs will not lead to large variation in resource levels. An analogy may illustrate this point: During a warm summer, rapidly turning an air conditioner on and off will not cause large variations in the temperature of a room, compared to leaving the air conditioner on for hours, and then turning it off for hours.  Fluctuations on long time scales will similarly not lead to large resource variation - when there is plenty of time for the consumer species to grow or decay to a compensatory population size, the aforementioned negative feedback between resources and consumers is maximized. Resource variation, and thus the propensity for coexistence, is maximized when fluctuations in resource inputs occur on intermediate timescales. 

Can spatial resource variation also promote coexistence via relative nonlinearity? Population cycles are necessarily a temporal phenomenon, so there is no purely spatial analogue of the endogenously generated resource-consumer cycles (armstrong 1977, 1980). If there is spatial variation in the per capita (or 'per concentration') parameters of resource dynamics, then we can apply the same argument that we used in the case of temporal environmental (or demographic) stochasticity: resource variation is proportional to mean resource levels, so the dominant competitor (species 1) tends to decrease resource variation, thus undermining coexistence. 

What happens when there is spatial variation in the resource inputs? When there is no dispersal, all patches converge to the same equilibrium resource level - relative nonlinearity cannot possibly be in effect. However, when a small proportion of the consumer species are allowed to disperse, we do find spatial variation in resource levels. Whether or not this generates coexistence is a matter of how model parameters are selected. Unlike the purely temporal case, there is nothing inherent about the opportunist-gleaner trade-off that causes the dominant competitor to induce more resource variation than the subordinate competitor.

(\eqref{??} in the main text)
want to do the Chesson and huntly thing. But you can't re-scale the invader, so you compensate by dividing everything by the invader's sensitivity to competition.

when L > S, no solution, whereas with Chesson, L > S-1 means no solution. So you can have more species and still have an exact answer. 

when di equals zero, the whole thing falls apart. Example on page

\renewcommand{\arraystretch}{1.25}
\begin{table}[h]
\caption{ In a resource-consumer model with an opportunist-gleaner trade-off, when does relative nonlinearity promote coexistence?}
\begin{tabular}{ p{0.5 \linewidth} p{0.3\linewidth}}
Source of resource variation & Promotes or No coexistence? \\
\toprule
\underline{temporal variation} & \\ 
endogenous population cycles & Promotes\\
environmental/demographic stochasticity & No \\
fluctuating resource supply rate & Promotes \\
\underline{spatial variation} & \\ 
endogenous population cycles & N/A \\
environmental/demographic stochasticity & No \\
fluctuating resource supply rate & Depends on parameters \\
\bottomrule
\end{tabular}
\label{tab:relative nonlinearity}
\end{table}

One can easily simulate a model forward in time and then check for coexistence (operationalized as co-occurrence for a sufficiently long time); But this only tells us whether species are coexisting, not \textit{how} they are coexisting, or failing to coexist. Instead, we would like a coexistence criterion which, through its simplicity, transparently relates coexistence to patterns in demographic parameters or patterns in the structure of population dynamical equations. MCT successfully isolates and quantifies these patterns via a partition of the invasion growth rate into \textit{coexistence mechanisms}. However, MCT struggles to clearly and universally relate invasion growth rates to coexistence. 

\textit{Stochastic Persistence Theory} (\cite{Schreiber2011}; \cite{Schreiber2012}; \cite{Roth2014}; \cite{hening2018coexistence}; \cite{Benaim2019}) shows how invasion growth rates can still be used to formulate a sufficient condition for coexistence in complex communities, which also becomes a necessary condition in the face of large, infrequent perturbations (\cite{Schreiber2006}). This criterion is satisfied by picking weights $p_j$ (i.e., positive constants that sum to one), one for each species $j$, such that the weighted sum of the $r_j$, the realized per capita growth rates, is positive for each and every sub-community $\mu$ in which one or more species is missing. In mathematical form, the criterion is

 \item niche differences (the mathematical object) is not niche differences (the ecological concept). In an analysis of Macarthur's resource-consumer model, Chesson (\citeyear{??}) performs a regression through the origin, using two species' resource consumption rates as the response and predictor variables. The square root of the $R^2$ statistic of this regression is $R = alpha_{ij}/\sqrt{\alpha_{ii} \alpha_{jj}}$. Because $\alpha_{ij} = \alpha_{ji}$ in Macarthur's resource-consumer model, Chesson's $R$ is equivalent to $\rho$ in the ES paradigm. However, niche differences is traditionally measured as the correlation between resource consumption rates
    
    Central problem is that chesson's definition includes resource parameters r, K, andv value w
    
    Schoener 1974,
    pianka cites macarthur and levin
    
    There is a tradition of defining niche overlap as the correlation between resource consumption rates (cite). However, using MacArthur's resource consumer model (cite) with fast resource dynamics (which is equivalent to the Lotka-Volterra model, it is easy to see that the $\rho$ is the not the correlation between resource consumption rates. 
    
    \item fitness differences are supposed to predict the winner of competition, but only do so in a hypothetical world where species have no niche overlap

resdient stationary dist. Invader density can be set to zero. However, in some models, this might not make sense. For example, consider a spatially explicit model - the invader must attain a quasi-steady spatial distribution in order to measure the invasion growth rate, but the only way to figure this out is to simulate from a model. Alternatively, consider a model where the

It is more likely that species will primarily experience short detours from their typical abundances, the product of several consecutive years of bad weather, punctuated by rare catastrophes that may extirpate the species (cite). Can invasion growth rates help us make sense of these scenarios?

The answer is "maybe". We may blindly claim that for a population which has fluctuated slightly below its equilibrium, a positive invasion growth rate is indicative of a positive rate of return to equilibrium. There is no guarantee that this is the case, and any such claim must be backed up with simulations or mathematical analysis. We are on slightly firmer ground in the case of a catastrophe-induced extirpation. blah blah blah. Although a positive invasion growth is no longer a perfect predictor of population recovery (as in infinite-population case), it is necessary. Further, the probability of population recovery increases monotonically with the invasion growth rate.

For invasion growth rates to predict whether a species can invade, we must assume that the focal species' density is brought low enough where it doesn't affect any species' per capita growth rates, but not so low the risk of stochastic extinction is non-negligible. Officially, t  shocks/perturbations/disturbance reduce the density of the focal species to the point where it has no effects on the per capita growth rates of any species (including itself)

\textcite{pande2000} et al. showed that the MTE (of a species that was coexisting via the storage effect) could have a humped-shaped relationship with environmental variation. As environmental variation increases, so too does MTE; as environmental variation increases even further, MTE decreases. This is because there are two opposing forces: the invasion growth rate increases monotonically with environmental variation, and the negative effects of the a few unlucky years increases.

Same with spatial variation, it increases spatial storach effect and EC cov. Unlike the temporal case, spurious extinction is not the culprit, but rather completely wiping patches. Metapopulation dynamics

Storage effect conflicting results with autocorrelation time.

No different from results that say that some intermediate amount of dispersal is needed - too much dispersal results in weakened EC covariance, but not enough dispersal will reduce the ability to recolonize disturbed patches.

Most research on MCT per se (as a framework/methdology) has focused on the mathematics -  specifically, different methods for partitioning the growth rate of a rare species ("the invasion growth rate") into additive terms that represent classes of explanations for coexistence ("coexistence mechanisms"). However, in order to use MCT to infer the relative importance of different explanations for coexistence, one also needs to 1) relate invasion growth rates to the construct of coexistence, and 2) relate coexistence in simple models to the coexistence mechanisms. This conceptual work has 

But, research on MCT as a framework has mainly focused on different ways to partition the invasion growth rate. What is far less appreciated how MCT is valuable in the first place. it connects simple explanations to real coexistence. Does this by relating simple explanations to coexistence, which are codified in simple models and accompanying commentary,

Does this by 1) relating invasion growths to coexistence, 2) partitioning the invasion growth rate into classes of explanations for coexistence, called coexistence mechanisms, and 3) relating simple explanations for coexistence to the coexistence mechanisms. Previous research has focused on element 2, mostly neglecting the other parts of chain that allows one understand coexitence. Here, we review 2 and 3. Main points, invasion growth rates themselves are full of assumptions, and there is no perfect way of using invasion growth rates to infer coexistence. Future work should focus on this blah. 3) Coexistecne mechansims are

jj

Invasion growth rate

TBD

Our paper is arranged so that the three types of relationships in figure 1 are discussed sequentially. New results - expressions of coexistence mechanisms in spatiotemporal models - can be found in Sections \ref{sec:Spatiotemporal coexistence mechanisms} and \ref{sec:Example: the spatiotemporal lottery model}. Other sections, namely \ref{sec:The relationship between coexistence and invasion growth rates} and \ref{sec:Interpreting coexistence mechanisms}, have a more review-like quality. Both types of scholarship are included in the paper, as they contribute to a more complete description of MCT. Naturally, any framework in community ecology which aims to be both quantitative \textit{and} general will need to contend with both technical and interpretational issues. 

\begin{equation}
    \kappa_2 = \sqrt{\frac{\alpha_{21}}{\alpha_{11}}}.
\end{equation}

 When $\Delta \kappa_i = 0$, models with only spatial variation and models with only temporal variation share the same set of coexistence mechanisms. Therefore, the case of complete dispersal can be thought of as a baseline from which spatial mechanisms of coexistence become dis-analogous from temporal mechanisms of coexistence. When we deviate from the case of complete dispersal - that is, when some individuals are retained locally - populations tend to build up in locations with good environments.

However, note that $\lambda_{j,0}^* = g_{j}(E_j^*, h_j(N_{s,0}^*)) = 1$, and that both $\sigma \E{x}{\lambda_{j,1}(t)}$ and $\sigma \E{x}{N_{j,1}}$ are $\mathcal{O}(\sigma^2)$, which simplifies \eqref{??} to

Analogous  the assumptions result in fluctuations in $E_j$, $C_j$, $N_j$, and $\nu_j$ being of magnitude $\mathcal{O}(\sigma)$; and the spatiotemporal average fluctuation in $E_j$, $C_j$, $N_j$, and $\nu_j$ being of magnitude $\mathcal{O}(\sigma^2)$

\begin{equation} 
\begin{aligned}
  N_{j,0}(x,t+1) & =  q_j N_{j,0}(x,t) \lambda_{j,0}(x,t) + \frac{1 - q_j}{K} \sum \limits_{s=1}^{K}  N_{j,0}(s,t) \lambda_{j,0}(s,t). \\ 
  & = N_{j,0}(x,t) \lambda_{j,0}(x,t) 
\end{aligned}
\end{equation}

Because fluctuations in $E_j$ or $C_j$ are

that species coexist via fitness-density covariance on the fine-grained spatial scale. This categorization is surprising at first glance, since the centrality of disturbances in the competition--colonization trade-off would seem to suggest that a fluctuation-dependent mechanism is at-play. However, because local per capita growth rates are linear with respect to whether or not a patch is occupied, fluctuations have no systematic effects at the regional scale. This underscores that fact that fluctuations \textit{per se} do not promote coexistence (\cite{chesson1997roles}) - nonlinear (i.e., $\beta_j^{(2)} \neq 0$) or non-additive (i.e., $\zeta_j \neq 0$) responses to fluctuations are also required. 

Also note that the diagrams focus on a species' capacity for self-regulation, suggesting a comparison of a single focal species at high vs low density (as opposed to the invader--resident comparison, which is integral to the conventional definitions of coexistence mechanisms). This type of comparison has been suggested previously (\cite{??}) and analyzed by \textcite{JohnsonScalingFactors}.

During an invasion analysis, after perturbing a species into the invader state, we let the residents attain their limiting dynamics. In the context of finite-population model, these "limiting dynamics" are defined by a unique asymptotic time-averaged distribution (\cite{glynn1998independent}) of resident densities and other variables, conditioned on non-extinction. This distribution can be computed via simulations. If one or more of the residents go extinct, the simultion may be restarted with initial conditions equal to the the state of a parallel simulation (the Fleming Viot Algorithm; \cite{ferrari2007quasi}; \cite{blanchet2016analysis}), or a past state from an archive of the current simulation (\cite{groisman2012simulation}). In models without periodic forcing, seasonality, or secular trends, the time-average distribution is a \textit{quasi-stationary distribution}, which can often be conveniently computed without simulations (see \cite{JohnsonSpatiotemporal} Appendix).

In some simple finite-population models, it is possible to set the invader's abundance to zero and still calculate the invader's per capita growth rates. However, in complex models, the invader's per capita growth rate depends on aspects of the invader population that must estimated via simulation. For instance, in a spatially-explicit individual-based model, the invader must attain a quasi-steady spatial distribution before the invasion growth rate can be measured. In age-structured population models, the invader must attain its stable-age distribution before the invasion growth rate can be measured. The general strategy is the inoculate the invader at low density, track If the invader's regional abundance becomes too great, such that the invader affects the dynamics of the residents (in spatial models with finite populations, an invader may become locally abundant and thus affect its own dynamics), then the simulation must be restarted. To counteract extinctions, the Fleming Viot algorithm (see previous paragraph) can be employed.

However, $\abs{\beta_i^{(1)}}$ may not be a proxy for population-dynamical speed in all models, so it may be more 
The purpose of this scaling is to enable a fair comparison of species whose life-cycles have different speeds.

Interestingly, dividing by $\abs{\beta_i^{(1)}}$ is equivalent to multiplying by generation time in the seminal models of MCT (the lottery model and annual plant model; \cite{Chesson1994}) and thus enables a better comparison of species with slow and fast life-histories. With this "tortoise-hair adjustment" (\cite{Chesson2018}), one can compare species 

For example, the community-average storage effect is

Unsurprisingly then, coexistence in mechanistic resource-consumer models can also be attributed to $\Delta \rho_i$. An influential heuristic emerged from Tilman's (\citeyear{tilman1980resources}, \citeyear{tilman1982resourceST}) analysis of two-species resource-consumer models - to coexist  "\ldots each species must consume proportionately more of the resource that more limits its growth" (\cite[p.~96]{tilman1982resourceST}). This heuristic later became the basis for an influential definition of the ecological niche (\cite{leibold1995niche}; \cite{chase2003ecological}), wherein a species' total niche could be decomposed into the \textit{impact niche} (i.e., resource consumption rates), and the \textit{requirement niche} (i.e., species' essential resource requirements). It is not straightforward to relate this conception of the niche to the small-noise expression for $\Delta \rho_i$. The sensitivity to competition, $\beta_j^{(1)}$, is proportional the impact niche. However, the deviation from 'equilibrium' competition, $\E{x,t}{C_j} - C_j^*$, is generated by the interplay between the impact niche and the sensitivity niche.

It is not easy to isolate the effects of intransitive competition with a partition of the invasion growth rate, since the average level of competition (and the average level of the constitutive regulating factors; see appendix \ref{app:Generalization of MCT to different classes of models:Continuous-time models:Multiple regulating factors}) is an emergent property of all density-dependent effects, intransitive or otherwise. \textcite{Gallien2017} suggest a measure of intransitive competition wherein the true invasion growth rate is compared to the invasion growth rate in a hypothetical world (a reference state) where a single resident species is eliminated (the resulting growth rate is averaged over all residents in succession). Intransitive competition has historically been studied in Lotka-Volterra-like models, where $\Delta \rho$ is necessarily the only coexistence-promoting mechanism. However, we note that it is possible for the effects of intransitivity on coexistence to be mediated through other coexistence mechanisms. For example, species A is excluded by species B when there is no resource variation; But, if species C induces cyclical resource dynamics, species A can coexist with species B via relative nonlinearity. 

and so, it has been claimed that feedback loops involving an odd-number of species are required for coexistence (\cite{allesina2011competitive}; \cite{Gallien2017})

A unique challenge in spatiotemporal models is determining the \textit{quasi-steady spatial distribution} of the invader (not to be confused with the quasi-stationary distribution of resident densities; Appendix \ref{app:Extended discussion of invasion growth rates, including computational tricks}). To accurately measure the invasion growth rate, one must inoculate the invader and then wait until it has attained its 'natural' spatial distribution. However, the longer one waits for the the invader attain this distribution, the larger the invader population becomes (assuming a positive invasion growth), leading to inaccurate measurements of the invasion growth rate. One hopes that the dynamics of spatial correlations operate on a much faster timescale than the dynamics of total density, such that a quasi-steady spatial distribution (i.e., a second-order stationary and isotropic process; \cite{cressie2015statistics}) is attained long before the total density changes too much. The requisite time-scale separation can be verified by plotting spatial correlations against total density (as in \cite{le2003adaptive}, Fig. 7). Analytical expressions for the quasi-steady distribution are only available in simple spatially implicit models (see Appendix \ref{app:Deriving the small-noise fitness-density covariance for the spatiotemporal lottery model} for a worked example) or in simple spatially explicit models with the help of 'pair approximation' (\cite{ferriere2001invasion}). 

In more complex models, simulation experiments are needed to compute the quasi-steady spatial distribution of the invader. After virtually inoculating the invader species and waiting through a sufficiently long burn-in period, one can begin measuring the invasion growth rate. If the regional invader population density exceeds a user-specified ceiling (i.e., the invader becomes common), then the simulation can be restarted. In finite population models, the invader may go extinct. To circumvent this problem, one may run several simulations in parallel, restarting extinct simulations using (as initial conditions) the state of another simulation. This approach is known as the Feller-Voit algorithm (\cite{ferrari2007quasi}; \cite{blanchet2016analysis}).

For a more detailed discussion of how to calculate invasion growth rates, including tricks that apply to specific classes of models, see Appendix \ref{app:Extended discussion of invasion growth rates, including computational tricks}.

The additive terms in the above equation (\eqref{big_decomp}), which we may call \textit{average growth rate components}, can be conceptualized as distinct processes. For example, the second term $\beta_j^{(1)} \E{x,t}{(C_j - C_{j}^{*})}$ is the effect of the mean level of competition on the average growth rate.

\begin{equation} \label{local_lambda}
n_j(x,t+1) = n_{j}(x,t) \;  \lambda_{j}(E_j(x,t), C_j(x,t)) + m_j(x,t) - e_j(x,t) \qquad j = (1, 2, ..., S),
\end{equation}

where $n_{j}(x,t)$ is the local density of species $j$, $\lambda_j$ is the local finite rate of increase of species $j$, $x$ is a discrete patch in space, $t$ is a discrete point in time, and $S$ is the number of species in the community. The local finite rate of increase is a function of the effects of the environment $E_j$ and competition $C_j$, which themselves may fluctuate over space and time. The terms $m_j$ and $e_j$ represent immigration and emmigration respectively, in units of population density. We require that the sum of $c_j$ and $e_j$ across space (i.e., net dispersal) vanishes (Appendix \ref{app:Deriving small-noise coexistence mechanisms:Spatial averaging and fitness-density covariance}), which occurs generically when either 1) the system is \textit{closed} (i.e., no individuals can enter or leave the system of patches), or 2) that the system of patches is representative or a larger metacommunity, such that it receives roughly as many immigrants as it loses emigrants.

A few notes on notation are necessary. For convenience, we will often write-out random variables without the explicit dependence on space and time; for example, we will write $E_j$ instead of $E_j(x,t)$, or simply $\lambda_j$ instead of $\lambda_{j}(E_j(x,t), C_j(x,t))$. We use the operator $\E{}{Z}$ to denote the averaging (i.e., the sample mean) of some random variable $Z$, with a subscript to denote whether the average is being taken across space, time, or both. For example, in a system with $K$ patches that has been observed for $T$ time-steps, $\E{x}{Z} = (1/K)\sum_{x = 1}^{K} Z(x,t)$, $\E{t}{Z} = (1/T)\sum_{t = 1}^{T} Z(x,t)$, and $\E{x,t}{Z} = (1/(T K)) \sum_{t = 1}^{T} \sum_{x = 1}^{K} Z(x,t)$. This notation is unorthodox for two reasons. First, the spatial dependence of temporal average, $(1/T)\sum_{t = 1}^{T} Z(x,t)$, is suppressed by the notation $\E{t}{Z}$ (a similar thing can be said of the spatial average). Second, the expectation operator conventionally denotes the average across an infinite number of instantiations of the stochastic population process at one point in time; not the temporal average of one instantiation (though they are asymptotically equivalent if the stochastic process is stationary and ergodic; see Section \ref{sec:The relationship between coexistence and invasion growth rates:Invasion growth rates}). We define $\Var{}{.}$ and $\Cov{}{.}{.}$ in a similar fashion, to the denote the sample variance and sample covariance respectively. 

The \textit{local finite rate of increase}, $\lambda_j$, is the discrete-time analogue of a per capita growth rate. The \textit{metapopulation finite rate of increase}, $\widetilde{\lambda}_j = \E{x}{(n_j / \E{x}{n_j}) \lambda_j}$, is the density-weighted average of $\lambda_j$ across space. The \textit{average growth rate} rate, $\E{t}{\log(\widetilde{\lambda}_j)}$, is the quantity which is predictive of long-term growth (\cite{Lewontin1969}; \cite{Dempster1955}; \cite{Stearns2000}). The average growth rate of the invader species is called the \textit{invasion growth rate}. The subscript $i$ references an invader species, the subscript $s$ references a resident species, and the subscript $j$ references a generic species whose status as a resident or invader is impertinent.

with spatial variation and \textit{local retention} (i.e., not all individuals disperse after every time-step), that spatial relative nonlinearity behaves in much the same way as temporal relative nonlinearity (Table \ref{tab:relative nonlinearity}). Some local retention is necessary because it allows for populations to build-up in good locations, which is critical component of local negative feedback loop (here, between consumer density and resource concentration) that is needed for spatial coexistence mechanisms; local retention in spatial models plays a similar role to temporal autocorrelation in temporal models, in the sense that both allow population build-up when/where conditions are favorable (see \cite{JohnsonHeuristic}).

This result can be derived by first recognizing that the covariance term, $\Cov{}{E_j}{C_j}$, is itself an effective regulating factor, and then straightforwardly applying the competitive exclusion principle (\cite{miller2017evolutionary}). 2) The Storage effect is potentially powerful, but it is not uniquely powerful.  3) There are other factors which limits biodiversity, so "an arbitrary number of species" is broadly irrelevant.

On the other hand, 
These results are the the consequence of treating the environment as a continuous variable. When there are $M$ discrete environmental states and $L$ regulating factors, the storage effect can only support $L \times M$ species. This result can be derived by first recognizing that the covariance term, $\Cov{}{E_j}{C_j}$, is itself an effective regulating factor, and then straightforwardly applying the competitive exclusion principle (\cite{miller2017evolutionary}). 

Finally, it is worth noting that storage effect holds a special place in the history of community ecology. Prior to the storage effect, the 'environment' was by-and-large ignored in theoretical studies of coexistence, since a constant environment can be abstracted away in constant demographic parameters, and a fluctuating environment was historically viewed as  a deleterious force on population persistence and community stability (\cite{Lewontin1969}; \cite{May1974}). This milieu was disrupted the storage effect, which was first discovered by \textcite{chesson1981environmentalST}, although the phenomenon had been previously studied in the context of population genetics (\cite{Haldane1963}; \cite{Lloyd1966}; \cite{Harper1971}; \cite{Gillespie1977}). More recent research has attempted to synthesize both the positive and negative effects of environmental variation on species coexistence (\cite{adler2008environmental}; \cite{schreiber2019rarity}; \cite{pande2020mean}, \cite{Dean2020Stochasticity-inducedSynthesis}).

As discussed in the Section \ref{sec:Interpreting coexistence mechanisms:Relative nonlinearity} \& \ref{sec:Interpreting coexistence mechanisms:The storage effect}, \textit{local retention} plays an important role in spatial coexistence mechanisms. 

As discussed in Sections \ref{sec:Interpreting coexistence mechanisms:Relative nonlinearity} \& \ref{sec:Interpreting coexistence mechanisms:The storage effect}, local retention plays an important role in all spatial coeixstence mechanisms

As discussed in Sections \ref{sec:Interpreting coexistence mechanisms:Relative nonlinearity} \& \ref{sec:Interpreting coexistence mechanisms:The storage effect}, local retention can also 

These features -- environmental heterogeneity and local retention -- may enable additional coexistence mechanisms. If environment and population density covary, then it is likely that environment and competition covary, so the spatial storage effect may result if species experience an interaction effect between environment and competition (i.e., $\zeta_j \neq 0$). As \textcite{chesson2000general} points out, variation in relative density can generate a positive covariance between environment and competition, even if a negative covariance would be attained in the absence of density variation. Additionally, as spatial environmental heterogeneity or local retention increases, the spatial variation in competition increases, which may modulate relative nonlinearity (Section \ref

$N_{s,0}$ is the equilibrium abundance of the resident; $\alpha^{(1)}$ is effect of environmental fluctuations in the per capita growth rate $g$ (see Section \ref{sec:Spatiotemporal coexistence mechanisms});  $\beta^{(1)}$ is effect of competition in the per capita growth rate $g$; and $\zeta$ is interaction effect of environmental and competition fluctuations on the per capita growth rate.

though it appears that real-world populations experience weak density-dependence (\cite{Ziebarth2010}; \cite{knape2012patterns}). That being said, it is hard to make any generalizations about the strength and functional form of density dependence: Microcosms exhibit selection bias (ecologists only want to study microcosm setups with interesting population dynamics), and real-world data is collected on large spatial scales,

To reiterate ourselves: while the invasion growth rate is not irrelevant (Section \ref{??}), what really matters for persistence/coexistence is the sign of invasion growth rates. From this lens, it may make more sense to define 

the probability of fixation after correcting for the speed of population dynamics (see Section \ref{??}).

If one cares about the raw probability of fixation, it is reasonable to not divide by $\beta_i^{(1)}$ (see Sections \ref{The probability of invasion increases monotonically with the invasion growth rate}, \ref{Does the magnitude of invasion growth rates matter, or just the sign?}).

Chesson 2003 says "relative fitness measures
because they quantify average relative performance of the different species in the system"

Carroll 2011 says "so... models can be exactly comparable"

A notable heuristic that emerged from the theory is that for coexistence to occur, "\ldots each species must consume proportionately more of the resource that more limits its growth" (\cite[p.~96]{tilman1982resourceST}; see p.77 for the mathematical formulation). 

Notably, Tilman's theory inspired a popular definition of the the ecological niche (\cite{leibold1995niche}; \cite{chase2003ecological}) wherein a species' total niche could be decomposed into the \textit{impact niche} (i.e., resource consumption rates), and the \textit{requirement niche} (i.e., species' minimal resource requirements). Though coexistence in Tilman's theory can be attributed to the $\Delta \rho_j$ coexistence mechanism, there is not a one-to-one relationship between the mathematical constituents of $\Delta \rho_i$ and the components of the total niche. The sensitivity to competition, $\beta_j^{(1)}$, is proportional the impact niche. However, the deviation from 'equilibrium' competition, $\E{x,t}{C_j} - C_j^*$, is generated by the interplay between the impact niche and the sensitivity niche.

There an increasing interest in "measuring coexistence": analyzing an empirically-calibrated model with MCT in order to infer the mechanisms of coexistence. But real communities plausibly contain complexities (e.g., intransitive competition, allee effects, "the resident strikes back" phenomenon) that pose problems for MCT. Therefore, the number one priority for future research in "measuring coexistence" should be integrating MCT (the partition of invasion growth rates) with better coexistence criteria, namely the presence of a positive attractor, which tests for local stability, and the Hofbauer condition, which tests for global stability.

This challenge can be decomposed into three sub-challenges, depicted in figure \ref{fig:edifice}: relating coexistence to invasion growth rates (Section \ref{sec:The relationship between coexistence and invasion growth rates}), partitioning invasion growth rates into coexistence mechanisms  (the focus of previous research; \cite{Chesson1994}; \cite{Chesson2000}; \cite{Ellner2019}; \cite{JohnsonSpatiotempora}), and relating coexistence mechanisms to simple explanations for coexistence (Section \ref{sec:Interpreting coexistence mechanisms}). Each sub-challenge has been discussed at length, but there remain a number of relevant topics in MCT. Here, we discuss them and provide solutions to controversies when possible.

\begin{equation}
\begin{aligned}
    \Delta \kappa_i = & \frac{q}{1-q} 
    \frac{2 \left(\alpha^{(1)}\right)^2 \theta_{ir} \beta^{(1)} (\phi - 1) \sigma^2 q N_{s,0}^*}{1-q(\theta_{ss} N_{s,0}^* \beta^{(1)} +1)} 
\end{aligned}
\end{equation}

\begin{equation}
    \Delta I = \frac{q N_{s,0}^* \alpha^{(1)}}{1-q(\theta_{ss} N_{\cdot,0}^* \beta^{(1)} +1)} \left[ \zeta \sigma^2 (\phi - 1) \right], \text{and}
\end{equation}

\begin{equation}
\begin{aligned}
    \Delta \kappa = & \frac{q N_{s,0}^* \alpha^{(1)}}{1-q(\theta_{ss} N_{s,0}^* \beta^{(1)} +1)} \left[\frac{q}{1-q}  2 \alpha^{(1)} \theta_{ir} \beta^{(1)} \sigma^2 (\phi - 1)\right]
\end{aligned}
\end{equation}

\begin{equation}
    \Delta I = \frac{q N_{s,0}^* \alpha^{(1)}}{1-q(\theta_{ss} N_{s,0}^* \beta^{(1)} +1)} \left[ \zeta \sigma^2 (\phi - 1) \right], \text{and}
\end{equation}

\begin{equation}
\begin{aligned}
    \Delta \kappa_i = & \frac{q N_{s,0}^* \alpha^{(1)}}{1-q(\theta_{ss} N_{s,0}^* \beta^{(1)} +1)} \left[\frac{q}{1-q}  2 \alpha^{(1)} \theta_{ir} \beta^{(1)} \sigma^2 (\phi - 1)\right]
\end{aligned}
\end{equation}

The invasion growth rate is always calculated in the context of the invader's environment, which naturally includes the effects of residents. Previous expositions of MCT have required that the invader's environment be an \textit{ergodic stationary stochastic process}(\cite[p.~236]{Chesson1994}). \textit{Ergodicity} means that the time-average of one realization of the stochastic process (that describes the invader's environment) is equal to the ensemble-average  (i.e., the mean across realizations) of the stochastic process (\cite{Nisbet1982Modelling}; \cite{Schreiber2011}). \textit{Stationary} means that the probability distribution of environmental states does not change over time (\cite{KarlinSamuel1981Asci}; \cite{allen2010introduction}). The assumption of an ergodic stationary process is convenient for theoretical work, because ergodicity implies that initial conditions are irrelevant, and stationarity allows the long-term average (inherent in the invasion growth rate) to be replaced with the expectation over the stationary distribution of resident densities, which can be calculated analytically in some models. However, the requirement of a stationary distribution is overly restrictive, since it excludes any models where parameters change over time, including models with seasonality and models that track climate fluctuations. Instead, we only a require that the invader's environment (again, this may include resident densities) has a \textit{unique, asymptotic, time-average distribution of resident densities} (see Appendix \ref{app:Extended discussion of invasion growth rates, including computational tricks}); this requirement only excludes models with long-term trends.

Unsurprisingly then, coexistence in mechanistic resource-consumer models can also be attributed to $\Delta \rho_i$. An influential heuristic emerged from Tilman's (\citeyear{tilman1980resources}, \citeyear{tilman1982resourceST}) analysis of two-species resource-consumer models - to coexist  "\ldots each species must consume proportionately more of the resource that more limits its growth" (\cite[p.~96]{tilman1982resourceST}). This heuristic later became the basis for an influential definition of the ecological niche (\cite{leibold1995niche}; \cite{chase2003ecological}), wherein a species' total niche could be decomposed into the \textit{impact niche} (i.e., resource consumption rates), and the \textit{requirement niche} (i.e., species' essential resource requirements). It is not straightforward to relate this conception of the niche to the small-noise expression for $\Delta \rho_i$. The sensitivity to competition, $\beta_j^{(1)}$, is proportional the impact niche. However, the deviation from 'equilibrium' competition, $\E{x,t}{C_j} - C_j^*$, is generated by the interplay between the impact niche and the sensitivity nich

$N_{r,0}$ is the equilibrium abundance of the resident; $\alpha^{(1)}$ is effect of environmental fluctuations in the per capita growth rate $g$ (see Section \ref{sec:Spatiotemporal coexistence mechanisms});  $\beta^{(1)}$ is effect of competition in the per capita growth rate $g$; and $\zeta$ is interaction effect of environmental and competition fluctuations on the per capita growth rate

Admittedly, the interpretation of the fitness ratio, $\kappa_1 / \kappa_2$, is a tricky issue. The E\&S framework is valuable precisely because it connects intuitive ecological concepts in coexistence criterion. One one hand, the framework loses value if the concepts aren't what we think they are. On the other hand, if we cal

It is hard to say how much the interpretation of $\kappa_1 / \kappa_2$ bears on the value of the E\&S framework. The framework is valuable precisely because it connects intuitive ecological concepts in coexistence criterion. Therefore, it would seem as though the framework loses value if "the average fitness ratio" isn't what people think it is. On the other hand, The $\kappa_1 / \kappa_2$